%% file: ms.tex
\documentclass[preprint,12pt]{aastex}
\usepackage{epsfig}
\usepackage{natbib}
\usepackage{verbatim}

\begin{document}

\title{Cosmic Streaming Field at Low Redshift}

\author{Lifan Wang}

\affil{Physics, Texas A\&M University}


\begin{abstract}

We study the expansion of the nearby Universe using a sample of Type Ia supernovae at redshifts below 0.08. Both the conventional supernova standardization method weighing heavily on the magnitude at optical maximum, and the color based CMAGIC method are  employed to put the supernovae on the Hubble diagram. These supernovae allow peculiar velocities of nearby galaxies to be measured at unprecedented precision. We have investigated in detail the possibility of a varying Hubble constant with redshift and found no evidence of a monopole term for the nearby Universe.  A large scale streaming motion is found at an amplitude of about $340^{63}_{-71}$ km/sec, aligned in the direction of $(l_0,\ b_0)\ = \  (312^{\rm o}.0^{13.5}_{-7.4}, 25^{\rm o}.7^{8.0}_{-9.2})$, which is close to the direction of the center of Shapley supercluster of galaxies. This streaming motion is best fit by a function involving a strong bipolar term. The streaming velocity field extends from the lowest redshift ($\sim\ 0.007$) to beyond 0.025 and likely out to even higher redshifts. The velocity field at redshift below 0.01 can be equally well described by a dipole field or by the same bipolar streaming velocity field that reaches out to beyond $z\ \sim\ 0.025$. We are also able to deduce a robust estimate of the random velocity component of the peculiar velocity field. Within the volume of redshift below 0.01 (weighted average redshift of $\sim$ 0.067), this thermal component is found to be about 270 km/sec. After correcting this smooth streaming motion, we are able to significantly improve the Hubble expansion fits of these supernovae. The CMAGIC method gives a dramatic decrease of $\chi^2$ from 90 to 63 for 69 degrees of freedom, and yields a residual scatter of only 0.12 magnitude; the 
correction also gives a moederate improvement to Hubble fit for the maximum light method. A whole sky Type Ia supernova search with carefully controlled systematic error is the key to confirm the large scale velocity patterns identified in this paper. The large scale streaming motion of the local Universe is of fundamental importance as it probes quantitatively the statistical isotropy of the local Universe. The bipolar pattern identified in this paper, if born out with future observations, will shed light on the formation of the structures of the Universe from the smallest to the largest scale. 

\end{abstract}
  
\keywords{Supernovae -- Cosmology -- Distance scales}

\section{Introduction}

Type Ia supernovae (SNIa) form a uniform class of objects that are precise cosmic distance indicators. The conventional method, established by \cite{Phillips:1993} and \cite{Hamuy:1996b}, is based on the empirical correlation between the brightness at optical maximum of SNIa and the decline rate of the light curve after optical maximum (see also \cite{Riess:1996, Perlmutter:1999}). Multi-color light curves of  SNIa are required to correct for the extinction caused by dust on the line of sight to the supernovae in the host galaxies. Dust extinction correction uncertainty usually dominants the error budget, not only because of the technical difficulties in acquiring multi-color light curves,  but also because of the lack of  knowledges of the extinction properties of dust particles in extragalactic environment. Even with the simplified assumption that the extinction law derived for dust particles in the Galaxy also applies to extragalactic environment, the coefficients of absolute extinction to reddening are known to vary from galaxies to galaxies. For example, the value $R_{B} \ = A(B)/E(B-V)$ has been deduced to be in a range of as low as slightly below 2 for some galaxies - a value which is significantly lower than the canonical value of 4.3 as has been found for the average of interstellar dust in the Galaxy \citep{Elias-Rosa:2006, Krisciunas:2007}. Furthermore, extinction correction may be complicated by the presence of circumstellar dust particles in which scattered light may affect the observed light and color curves \citep{Wang:2005}. Because these methods rely heavily on the magnitude at optical maximum, we will refer to this conventional method as maximum light method (MLM).

The Color-Magnitude Intercept Calibration (CMAGIC) of SNIa provides an alternative approach that is less sensitive to dust extinction \cite{Wang:2003CMAGIC, Conley:2006, Wang:2006}. In CMAGIC, the magnitude at a given color index is used as the primary distance indicator. The empirical CMAGIC linear relation can be employed to deduce the magnitude at a given color index. The CMAGIC linear relation shows that on the color-magnitude diagram, from about one week to one month past optical maximum (the exact epochs of the linear relation is a function of the light curve width, see \cite{Wang:2003CMAGIC, Wang:2006}), the color-magnitude diagram of SNIa can be well approximated by a straight line with a universal slope. For example, for $B$ and $V$ band data, this relation is given as $B\ = \ B_0-\beta(B-V)$, with $\beta\ = \ 1.94$. This simple relation reduces the degree of freedom of the light curve fits, and when compared to the MLM method, yields significantly reduced statistical errors \citep{Wang:2006, Conley:2006} in the final distance modulus. In \cite{Conley:2006}, application of CMAGIC to high redshift  SNIa shows that with a small sub-set of SNe with enough data for the application of CMAGIC, it is possible to achieve constraints on cosmological parameters that are tighter than those derived from a bigger sample of SNe but using the conventional method.

The velocity field of the local universe has been explored previously by several groups \citep{Zahavi:1998, Haugboelle:2006} using distance moduli of SNIa. These studies are based mostly on the MCLS approach of the MLM method \citep{Riess:1998, Jha:200644SNe}. A tentative monopole field is identified by \cite{Zahavi:1998} and \cite{Jha:2007}. \cite{Haugboelle:2006} recently modeled the distance moduli published in \cite{Jha:200644SNe, Jha:2007} and identified dipole and quadruple components of the local peculiar velocity field. 

Large scale streaming patterns of the local universe can be a source of systematic errors for supernova cosmology if not well understood. A streaming motion of about 300 km/sec can introduce a magnitude bias larger than 0.02 mag out to redshift of 0.1. Since all high redshift SNIa surveys need to anchor the Hubble diagram to a low redshift sample, a streaming peculiar velocity field can introduce a noticeable bias for future precision cosmology studies targeting accuracies of the order of 0.01 mag.

In this study, we apply the MLM and CMAGIC methods to a sample of nearby SNIa to study the expansion of the local universe at $z\ < \ 0.08$. Our goal is to study the angular distribution of the peculiar velocities of the host galaxies to probe the kinematics of the local universe. In this paper, \S\ref{S:DATA} describes the data sample and the CMAGIC Hubble diagram, \S\ref{S:FITS} analyzes the streaming motion and decomposes them into monopole, dipole, and quadruple components, \S\ref{S:NEWH} gives the Hubble diagram after corrections of the streaming velocity flow, and a brief discussion of the current findings is given in \S\ref{S:DIS}.

\section{The Data Sample and the CMAGIC Hubble diagrams}

\label{S:DATA}
The details of the SN sample, the light curve fit and the CMAGIC method are described in Table~2 of \citep{Wang:2006}, but this study adds a few additional SNe: SN~2003du \citep{Anupama:2005, Stanishev:2007}, SN~2004eo \citep{Pastorello:2007}, 2004S \citep{Krisciunas:2007}, SN~2005am \citep{Li:2006}, and SN~2005cf \citep{Pastorello:20072005cf}. 

As in \cite{Wang:2006}, we have used the the SuperStretch algorithm to fit the supernova light curves. A major advantage of both SuperStretch and CMAGIC methods is that the light or color curve models can be applied to the raw observed data before extinction and filter K corrections. Unlike other light curve fitting methods which assume that all SNIa light curves can be described by a one parameter family of functions, the SuperStretch method is a multi-paramter light curve fitting algorithm with a maximum of 6 free parameters of each light curve of which 4 are relevant to the light curve shapes. The SuperStretch method is built upon the success of the conventional stretch method \citep{Perlmutter:1999, Goldhaber:2001}, but allows the stretch parameter to vary with time from optical maximum, and introduces a secondary component to compensate the light curve bumps in $R$ and $I$, and sometime bumps in $B$ and $V$ bands for some peculiar SNe.  As such, most of the light curve parameters such as magnitudes at $B$ maximum, $\Delta m_{15}$, and CMAGIC parameters are identical to those published in \cite{Wang:2006}. As shown in \cite{Wang:2003CMAGIC, Wang:2006}, very frequently the CMAGIC linear fits give $\chi^2$/DoF that are smaller than 1, suggesting that the statistical noise of the original data is often very conservative, or with unknown correlations. In this study, the errors of light curve and CMAGIC fits are scaled to $\chi^{2}$ per degree of freedom to be one, whereas a slightly more sophisticated approach is adopted in \cite{Wang:2006}. This only introducs some small differences between the errors in this paper and in \cite{Wang:2006}. Such differences are not expected to significantly alter the statistical properties of the data set.

\begin{figure}[tbh]
\epsscale{1}
\includegraphics[angle=90,scale=0.65]{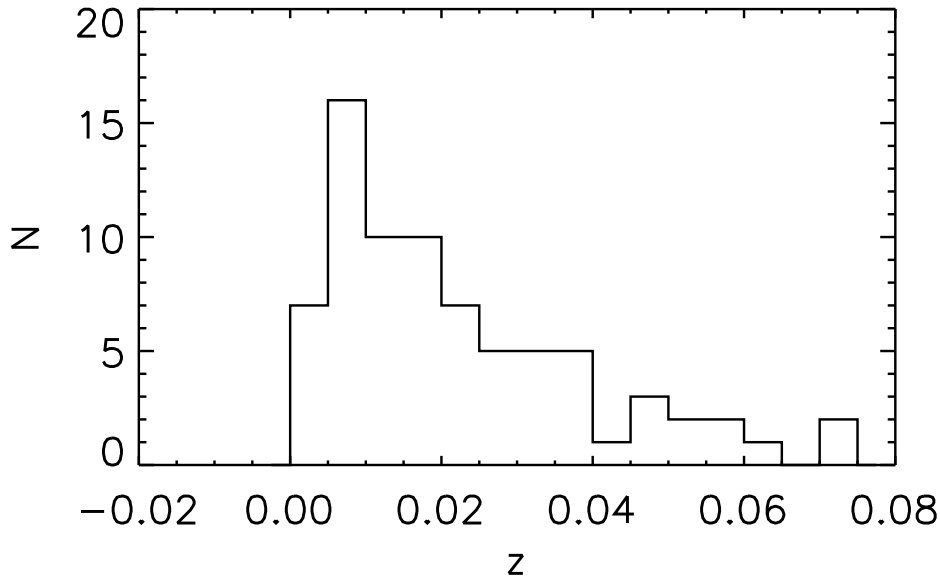}

\caption{Histogram of the redshift distribution of the SNe included in this study. }
\label{Fig:redshift}
\end{figure}

This paper studies only SNIa with sufficient observations to allow for our multi-parameter light curve and CMAGIC fits (\cite{Wang:2006}). In addition, the following criteria are applied to select the final data sample: (a) Only SNIa with $\Delta m_{15}$ smaller than 1.7 are chosen - this practically excludes all SN~1991bg-like fast decliners \citep{Filippenko:1992}; (b) SNIa with $B_{max}-V_{max}\ >\ 0.5$ are excluded - this reduces potential errors in extinction correction of heavily reddened SNe; and (c) only SNe at redshifts below 0.1 are selected. The original data set has a total of 125 SNIa of which 76 satisfy the above criterions. The redshift distribution of the sample is shown in Figure~\ref{Fig:redshift}. Most of the SNe are at redshift below 0.05. 

The magnitudes deduced from SuperStretch and CMAGIC fits need to be corrected by the light curve shapes. We will analyze in parallel both the MLM and CMAGIC method in this paper.  For the MLM in $B$ and $V$ bands, the distance modulus to a supernova is given by
\begin{equation}
\mu \ = \ B_{max} - \mathcal{M}_{1}- a_{1} (\Delta m_{15}-1.1) - b_{1} (B_{max}-V_{max} ), 
\label{eq:mlm}
\end{equation}
where $\mathcal{M}_{1}$ is the absolute $B$ magnitude at optical maximum, $a_{1}$ and $b_{2}$ are correction coefficients to be determined together with parameters of cosmological interest, $B_{max}$, $V_{max}$, and $\Delta m_{15}$ are the magnitude at $B$ and $V$
maximum, and the $B$-band magnitude decline during the first 15 days past $B$ maximum, respectively.
 The CMAGIC method does not make use of the magnitude at $V_{max}$. As shown in an our earlier studies, it is equivalent to 
 \begin{equation}
 \mu \ = \ B_{BV} - \mathcal{M}_{2}-a_{2} (\Delta m_{15}-1.1) - (b_{2}-\beta)[\frac{B_{max} - B_{BV}}{\beta} +0.6 + 1.2(\frac{1}{\beta}-\frac{1}{<\beta>})], 
\label{eq:cmagic}
 \end{equation}
where $\mathcal{M}_{2}$ is the absolute $B_{BV}$ magnitude,  $a_{2}$ and $b_{2}$ are coefficients that needs to be determined simultaneously with other cosmological parameters, $B_{BV}$ and $\beta$ are the CMAGIC intercept magnitude and slope \citep{Wang:2003CMAGIC}, the number 0.6 inside the square bracket is 
because $B_{BV}$ is calculated at $B-V\ = \ 0.6$.
As shown in previous studies, $b_{1}$ and $b_{2}$ has a range of values from 1.5-4 depending on details of the subsets of supernovae but is around 2 in average \citep{Tripp:1997, Astier:2006, Wang:2006}, and $\beta\ = \ 1.94$.  It is readily seen from equation (\ref{eq:cmagic}) that the CMAGIC method is essentially independent of the measured colors of the SNe.

In this study we also deduce the $R_B$ values for Galactic extinction of SNe based on a library of well observed SN spectra. The $R_B$ value of SNIa evolves with time $t$ after explosion due to the time evolution of SN spectra. This leads to a small correction to the  $\Delta m_{15}$ values of reddened SNe, of the order of $(R_B(0) - R_B(15))\times E(B-V)$, where $R_B(0)$ and $R_B(15)$ are the absolute to relative extinction ratio at day 0 and day 15 for Galactic dust.  For the standard extinction law \citep{Cardelli:1989}  with $R_B\ =  \ 4.1$, typical values are: $R_B(0) \ = \ 4.15\pm0.06$, $R_V(0)\ = \ 3.14\pm0.02$, $R_B(15) \ = \ 4.05\pm0.11$,  and $R_V(0)\ = \ 3.12\pm0.16$. Note that in general $R_B-R_V\ \neq\ 1 $. We also deduce estimates of the errors of $R_B$ and $R_V$ based on the various spectra of SNe we have in the library. Typical values of $R_B$ and $R_V$ uncertainties are 0.06 and 0.02 at optical maximum, and are 0.11 and 0.16 at day 15, respectively, Although not a dominant source of errors, the uncertainties in the extinction coefficients are propagated to the errors of the distances we deduced from these SNe. The second term on the right hand side of equations (1) and (2) corrects the extinction due to extragalactic dust or any intrinsic magnitude dependence on the measured color. This approach, first proposed in \cite{Tripp:1997} and \cite{Tripp:1999}, has also been employed in recent high redshift supernova studies \cite{Astier:2006}. The method has the obvious shortcomings of not being able to correctly separate potential systematic effects of the magnitude-intrinsic color dependence from the errors of dust extinction. However, this is still the best  one can do with SNIa considering the limitations on the current  understandings of the physical properties of SNIa and the properties of extragalactic dust. Other approaches for handling dust extinction include: (a) assumption of a correlation between intrinsic supernova color and the light curve shape, and a theoretical prior on the distribution of the extragalactic dust reddening \citep{Riess:1998, Jha:2007}; (b) assumption that the late time color of SNIa follows a linear relation as shown in \cite{Phillips:1999}; (c) using the SN color at 12 days past optical maximum \citep{WangX:2005}; and (d) using near-IR observations to pinpoint the reddening and dust extinction law \citep{Krisciunas:2004,Krisciunas:2007,Elias-Rosa:2006}. Not all there methods are applicable to the current data set. Some of these methods introduce additional assumptions that are difficult to justify.

A random thermal peculiar velocity component of 350 km/sec, assumed to be Gaussian, is added to each supernova. This is again a poorly known quantity. However, the assumed peculiar velocity is compatible with pairwise velocity distribution functions determined from galaxy redshift surveys \citep{Davis:1997, Landy:2002, Hawkins:2003}, and is likely to be a good guess of the thermal component of the galaxy peculiar velocity. This will be discussed further in the next  sections.

The quantities that are relevant to this study are shown in Tables~\ref{Tab:mlm} and \ref{Tab:cmagic}, where the columns are: (1) the identification of the supernova, (2) the redshift in the rest frame of the CMB, (3) and (4) the $B-$band maximum magnitude (Table~\ref{Tab:mlm}), or the color magnitude (Table~\ref{Tab:cmagic}), and error, (5) and (6) the $\Delta m_{15}(B) $ value derived using SuperStretch fit and error, respectively, (7) and (8)  the color term as in Equations~(\ref{eq:mlm}) or (\ref{eq:cmagic}), and error, respectively, (9)  the deduced distance modulus, (10) and (11) the residual on the Hubble diagram, and error, respectively, (12) and (13) the deduced peculiar velocity and error, respectively, (14) and (15) show the R.A. and Decl. of the supernovae in ecliptic coordinates. For column (7) and (8), the color index is $B_{max}-V_{max}$ and $\mathcal{E}(B-V)\equiv \frac{B_{max} - B_{BV}}{\beta} + 0.6 + 1.2 (\frac{1}{\beta}-\frac{1}{<\beta>})$. This definition of $\mathcal{E}(B-V)$ is different from that in \cite{Wang:2003CMAGIC} by an extra last term which helps to reduce the covariance between $\mathcal{E}(B-V)$ and $\beta$.  It is important to note that the quantities such as magnitude at maximum, $\Delta m_{15}$, and CMAGIC magnitude and color shown in Tables~\ref{Tab:mlm} and \ref{Tab:cmagic}  are in general correlated. It is impossible to tabulate the covariance matrix here but they are included in the analyses throughout this paper.

\begin{figure}[tbh]
\epsscale{1.}
\includegraphics[angle=90,scale=0.33]{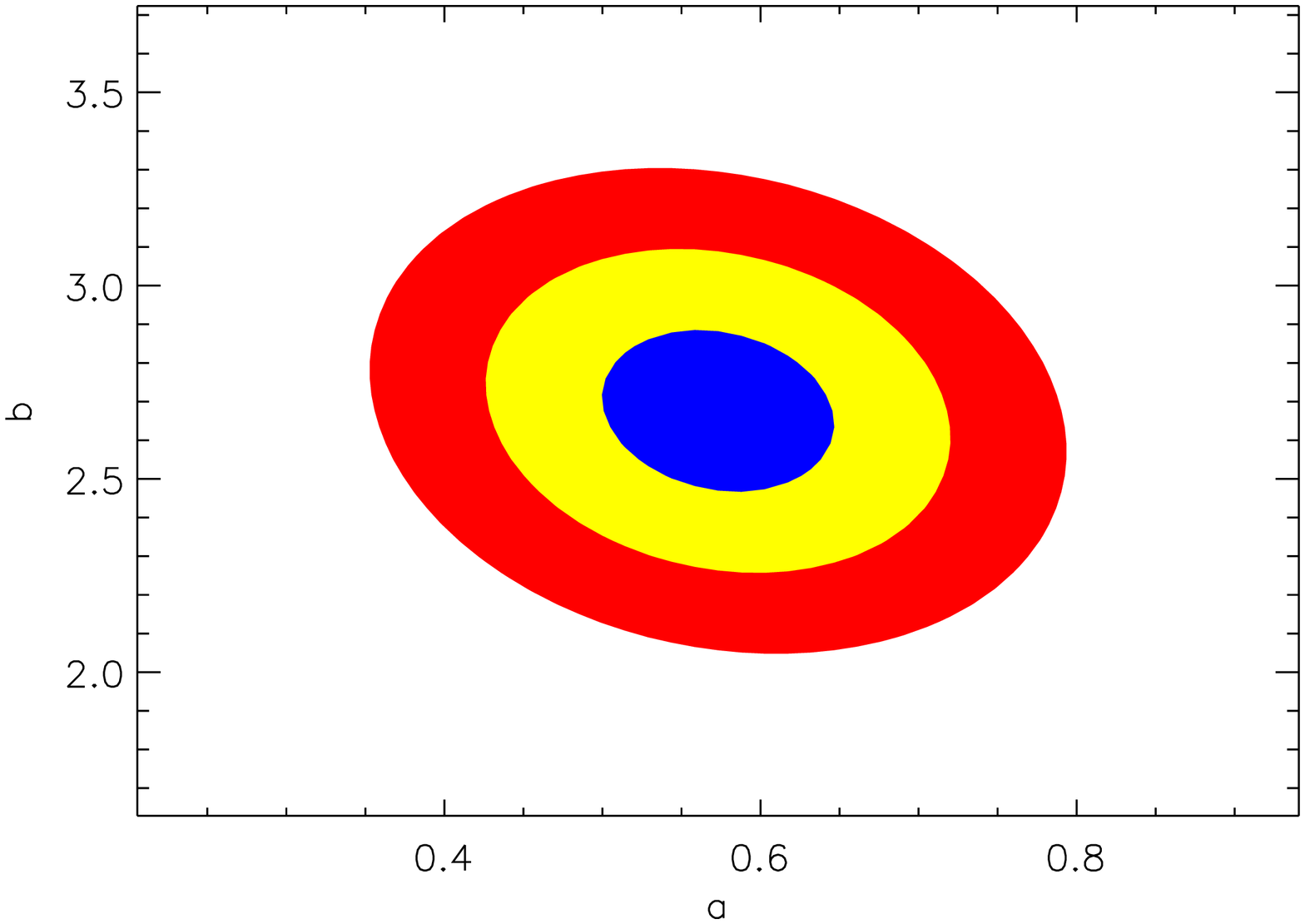}
\includegraphics[angle=90,scale=0.33]{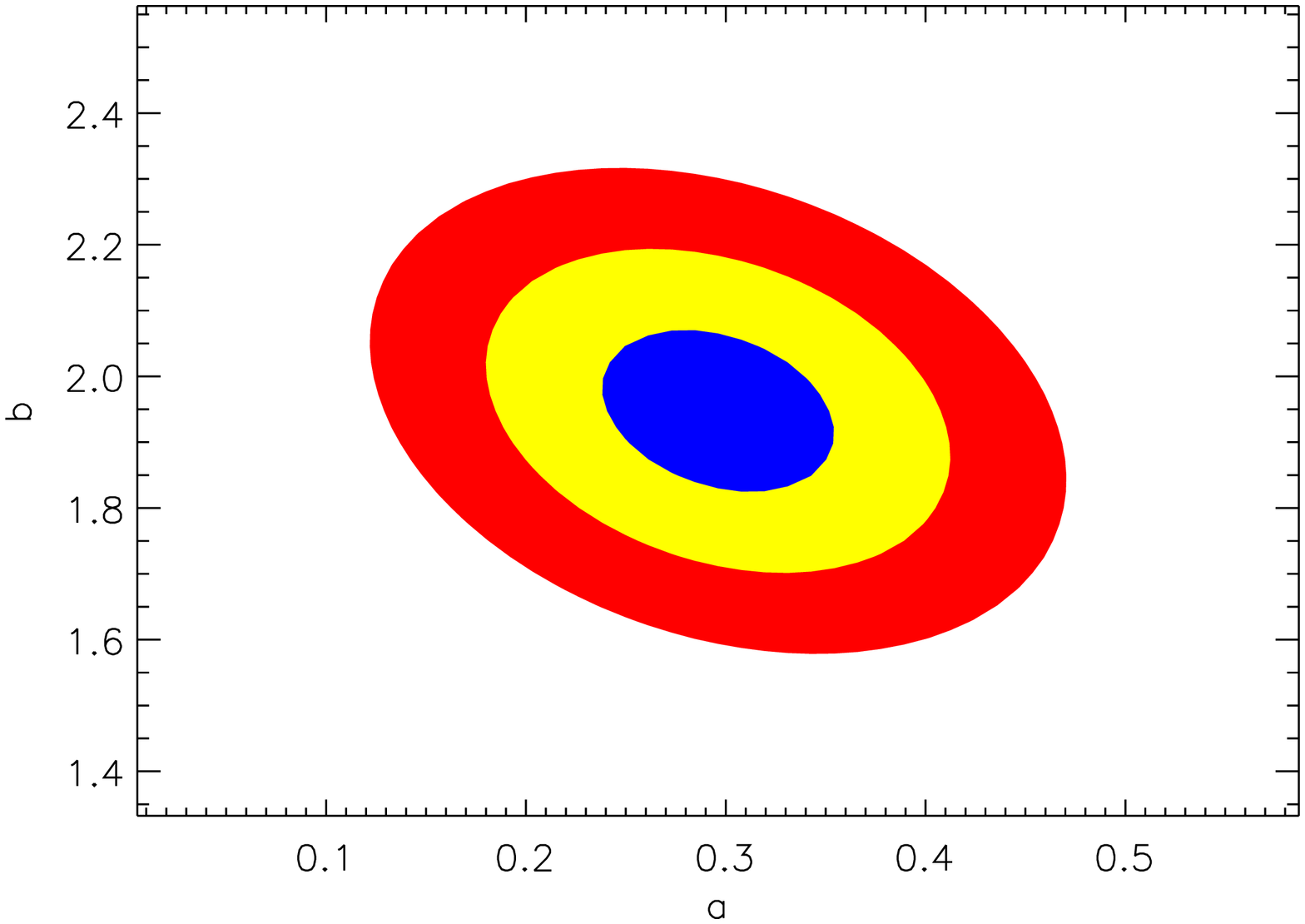}
\caption{(a, left): The correction coefficients $a_1$ and $b_1$ of Equation~(\ref{eq:mlm}), and (b, right): The correction coefficients $a_2$ and $b_2$ for Equation~(\ref{eq:cmagic}). The contours are 1, 2, and 3-$\sigma$ levels from the inner most to the outer most. This shows that errors of the light curve shape correction coefficients are anti-correlated for both the MLM and CMAGIC method.
}
\label{Fig:ab}
\end{figure}

 By minimizing the magnitude scatter on the Hubble diagram, we can set constraints on the coefficients $a_1$, $b_1$, $a_2$, and $b_2$ of Equations (\ref{eq:mlm}) and (\ref{eq:cmagic}).  The results are shown in Figure~\ref{Fig:ab}. We found that $b_{2}$ is around 2 which is very close to being equal to the mean slope $<\beta>$ of the CMAGIC linear fits \citep{Wang:2003CMAGIC}, and $a_{2}$ around 0.26. With these values, $B_{BV}$ is an excellent standard candle requiring only a very small $\Delta m_{15}$ and color correction. The errors listed in columns (11) and (13) of Table~\ref{Tab:cmagic} are in general a few times smaller than those given in the gold sample of \cite{Riess:2004} and in \cite{Jha:2007}. This is mainly due to the introduction of the CMAGIC method which reduces significantly the light curve shape and color dependence of the derived distance scales.  
 
 It is frequently assumed that SNIa are not perfect standard candles but with some intrinsic magnitude dispersion \citep{Perlmutter:1999, Riess:1998}. The exact level of intrinsic dispersion of SNIa magnitude is of fundamental interest in applying SNIa to cosmological studies.  Earlier studies have shown that the total (intrinsic+measurement) dispersion deduced for the sample of nearby supernovae (e.g. \cite{Riess:1996,Phillips:1999}) is about 0.18 mag. The fact that the $\chi^{2}$ per degree of freedom is generally well around 1 on the Hubble diagram in the fits of \cite{Riess:1996} and \cite{Phillips:1999} suggests that 0.18 mag is only an upper limit of the intrinsic dispersion. The actual intrinsic dispersion should be much smaller than 0.18 mag as it is highly possible
 that observational or dust extinction errors are the dominant sources of 
errors in those analyses. The intrinsic magnitude dispersion of SNIa has been studied in detail in \cite{Wang:2006} where the observed magnitude scatter around the best fit Hubble diagram is attributed to the combination of a random component and a coherent correlated component. \cite{Wang:2006} found that after systematic corrections are performed, MLM and CMAGIC exhibit mutual rms intrinsic variation equal to 0.074 $\pm$ 0.019 mag, of which at least an equal share likely belongs to CMAGIC.  In this study, we probe the streaming velocity field of the local universe which is likely to be a dominant source of correlated noise on the Hubble diagram of the nearby Universe. 

\input{BmaxBmaxVmaxBipolar.tex}

\input{BmaxmathcalEBVBipolar.tex}

\begin{figure}[tbh]
\epsscale{1.}
\includegraphics[angle=90,scale=0.5]{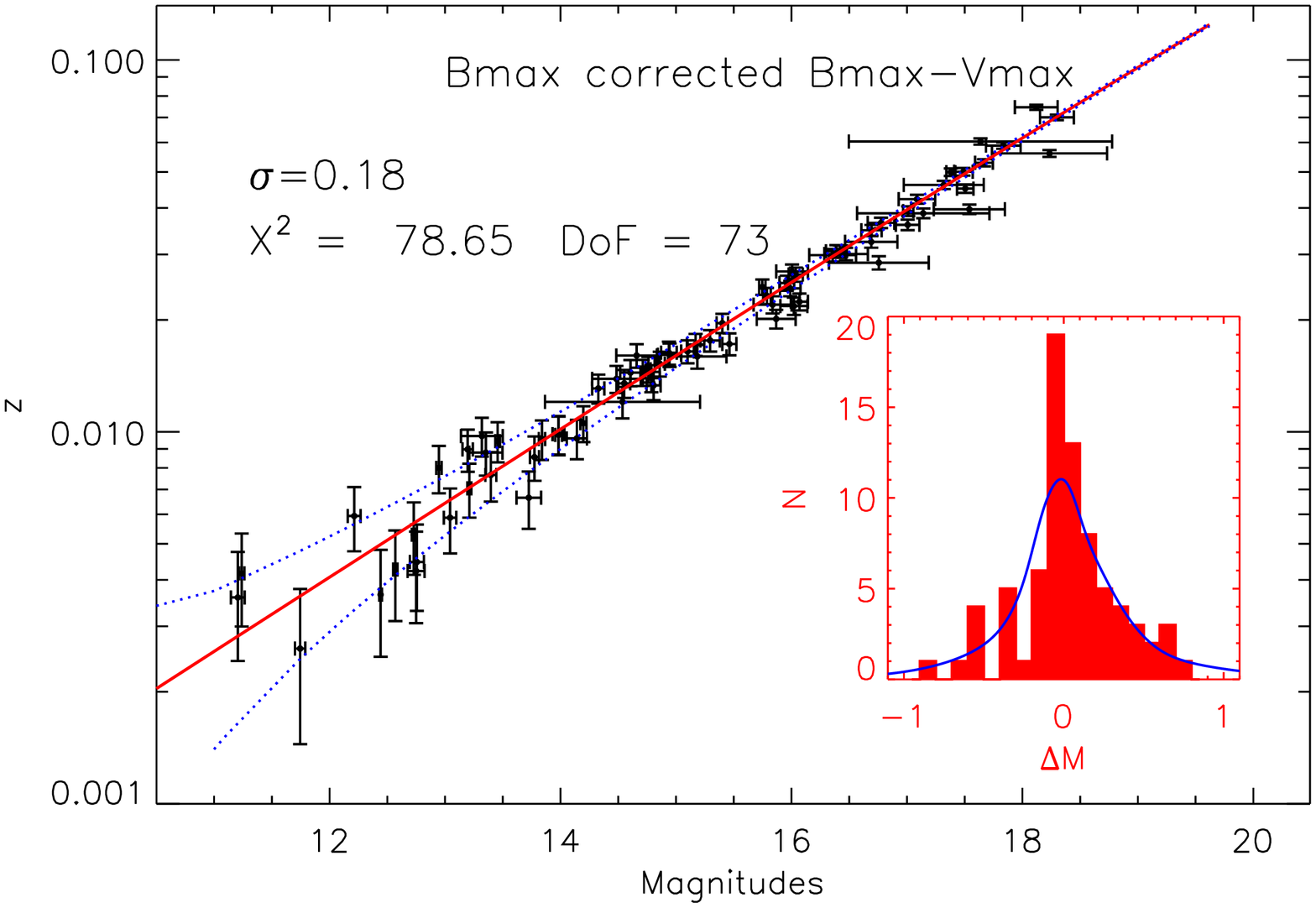}

\includegraphics[angle=90,scale=0.5]{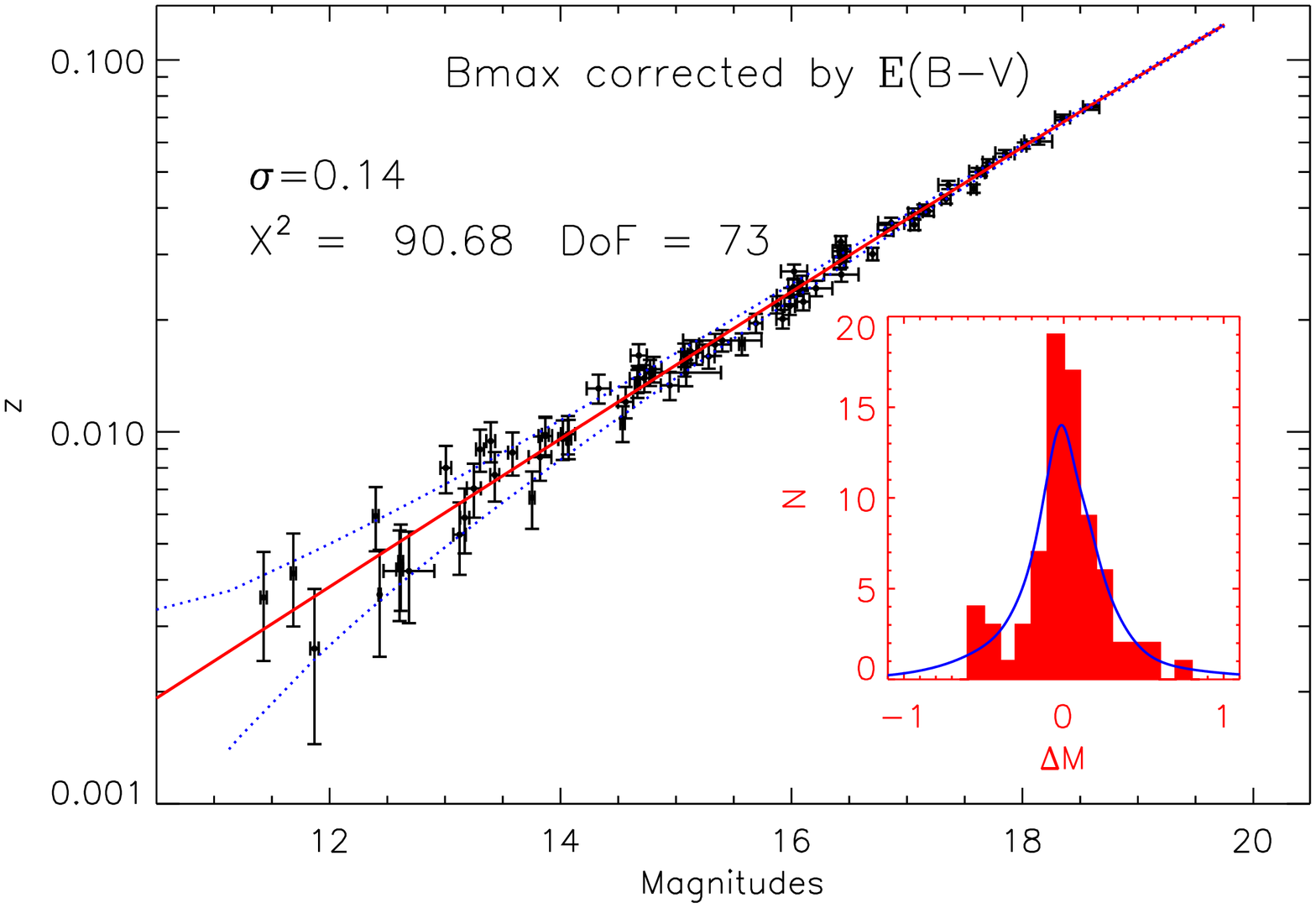}
\caption{Hubble diagrams from the conventional MLM method (a, upper panel) and from the CMAGIC (b, lower panel). Notice that the error bars for the CMAGIC are significantly smaller than for MLM. The CMAGIC method also yields a significantly smaller magnitude dispersion. The solid lines show the best fit Hubble diagram. The dotted lines show the magnitude deviation introduced by a random peculiar velocity of 350 km/sec. The inserts show the histogram of the magnitude deviations from the best fit Hubble diagram. The solid lines in the inserts show the ideogram of the residual distribution. The baseline cosmology model assumes $\Omega_M\  = 0.3$ and $\Omega_\Lambda\ = \ 0.7$. 
}
\label{Fig:hubble}
\end{figure}

\clearpage

Figure~\ref{Fig:hubble} (a) and (b) shows the resulting Hubble diagrams for the MLM and CMAGIC methods, based on  Equations~(\ref{eq:mlm}) and (\ref{eq:cmagic}), respectively. The solid lines are from cosmological models of a flat universe with $\Omega_{M}\ = \ 0.3$ and $\Omega_{M}\ = \ 0.7$. A Hubble constant of 71 km/sec/Mpc is assumed in these fits, but the exact value of that does not affect the goodness of the fits. It is seen that the CMAGIC Hubble diagram has a magnitude dispersion of 0.14 mag which is smaller than the 0.18 mag for the MLM method. This indicates that the CMAGIC method does help tightening up the dispersions on the Hubble diagram. However, the CMAGIC method yields a significantly worse Hubble fit, with a  $\chi^{2}$ = 90.68 for 73 degrees of freedom, which is remarkably larger than the $\chi^{2}\ = \ 78.65$ for the same degrees of freedom but with MLM. The probability of get $\chi^2$ larger than 90.68 from a random sample with 73 degrees of freedom is about 8\%, suggesting that either the CMAGIC method has underestimated the errors or unidentified systematic effects such as extinction errors, or large scale streaming motions may be important. If so, such effects have shown up strongly with CMAGIC but the MLM method may be less sensitive to those effects. 

With the assumed peculiar velocity of 350 km/sec, the Hubble diagram of the MLM method allows very little room for a significant intrinsic dispersion. Adding a small intrinsic dispersion to the CMAGIC method will reduce the $\chi^2$ to an acceptable level, however, we will see below that this is unlikely to be the case; the failure of a satisfactory Hubble fit with CMAGIC actually carries important information about the streaming motion of the local universe.

\section{The Peculiar Velocity Field}

There is little physical motivations of the functional form of the streaming velocity field. Conventionally it has been modeled in terms of spherical harmonic expansions and the monopole, dipole, and quadruple terms are studied. 

\label{S:FITS}
\subsection{The Monopole Flow}

It has been suggested that the nearby Universe may have an expansion velocity that is different from the Hubble expansion at higher redshift. Such a configuration may arise if the Galaxy is are inside a low density void. \cite{Kim:1997}  compared supernova distances at redshifts around 0.4 and 0.1 and set a 2-$\sigma$ upper limit of 10\% for an outflow
within the local 300 h$^{-1}$ Mpc. \cite{Zahavi:1998}, based on a larger sample of well measured supernovae, studied the redshift dependence of the Hubble constant, and found a putative void extending to about 70 h$^{-1}$ Mpc.  The existence of such a void is confirmed recently at a 2-$\sigma$ level by \cite{Jha:2007} based on an even larger sample of SNe.

To investigate the nature of the monopole field, we performed a joint Hubble diagram fit using the following model for the expansion of the Universe

\begin{equation}
\mathcal{D}_L(\hat{H}, q, z_0, z, \Omega_M,\Omega_\Lambda)\ = \ [1/q+(1-1/q)\mathcal{H}(z-z_0)] D_L(\hat{H}, z, \Omega_M,\Omega_\Lambda),
\label{eq:mono}
\end{equation}
where $D_L$ is the luminosity distance in the Friedmann-Robertson-Walker Universe, $\hat{H}$ the Hubble constant at redshift above $z_0$, $\mathcal{H}(z)$ the Heaviside function, $z_0$ the redshift cut above which the flow is consistent with the Hubble flow with Hubble constant $\hat{H}$, $\Omega_M$, and $\Omega_\Lambda$ are the density parameters for matter and dark energy, respectively. We assume $\Omega_M\ = \ 0.3$, and $\Omega_\Lambda\ = \ 0.7$ in these studies. 

The observed supernova magnitude is then fit by minimizing the quantity

\begin{equation}
\chi^2\ = \ \sum_i[m_i^c(z_i) - 5log\mathcal{D}_L(\hat{H}, q, z_0, z_i, 0.3,0.7)-25]^2/\sigma_i^2. 
\label{eq:jchi}
\end{equation}

where $z_i$ is the observed redshift of the supernovae, $m_i^c$ is the observed SN magnitude corrected by Equations (\ref{eq:mlm}) or (\ref{eq:cmagic}), $\sigma_i$ the errors of the corrected magnitudes.
The three additional free parameters are $\hat{H}$, $q$, and $z_0$ and the 
free parameters are determined simultaneously through $\chi^2$ minimization. In such a fit, the parameter $\hat{H}$ actually represents the combined effect of the Hubble constant and the zero point of the absolute magnitude of the supernovae. The final constraint on $z_0$ and $q$ is derived by marginalizing over nuisance parameters such as $\hat{H}$, and the light curve shape and color correction coefficients.

\begin{figure}[tbh]
\epsscale{1.}
\includegraphics[angle=0,scale=0.45]{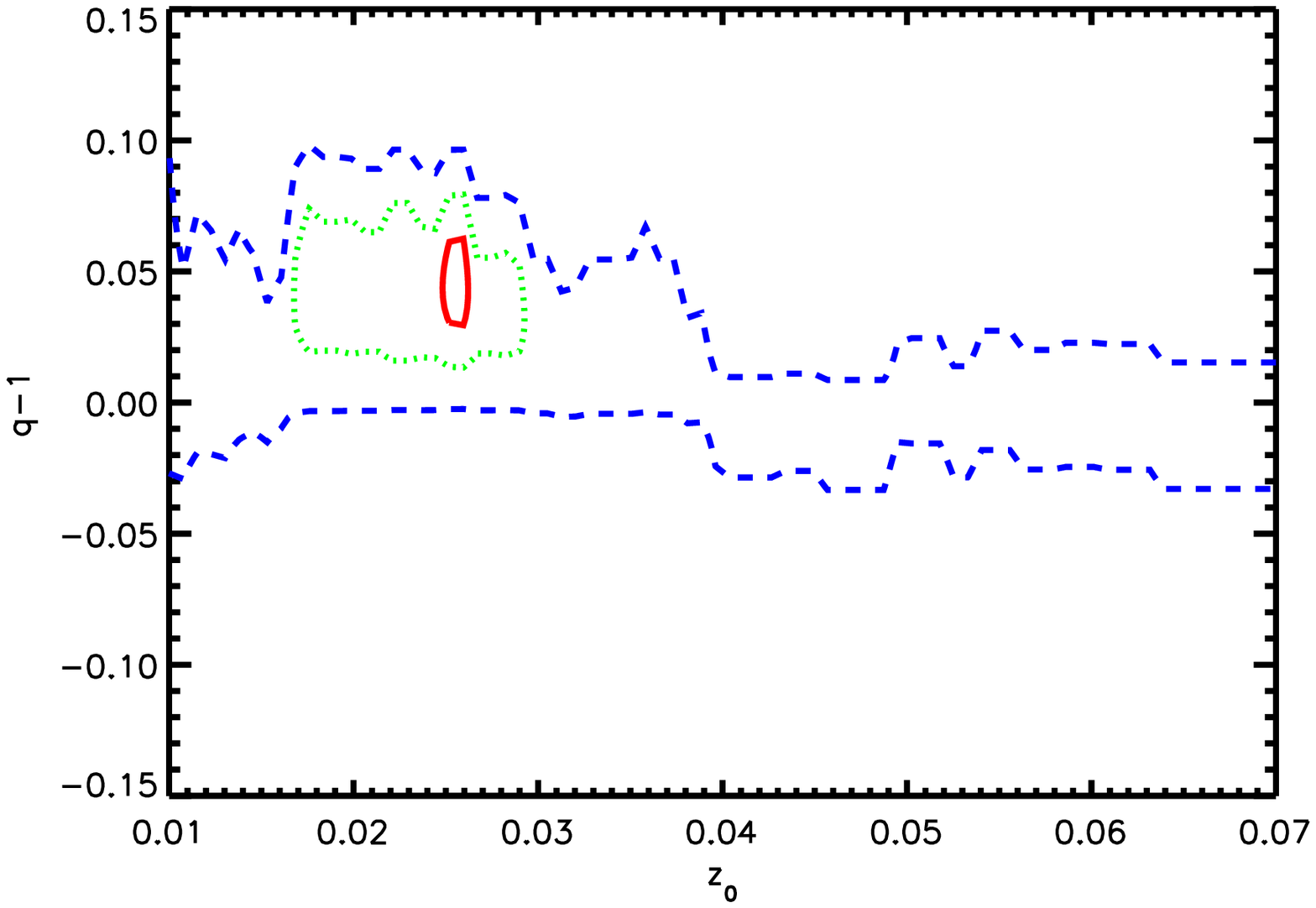}
\includegraphics[angle=0,scale=0.45]{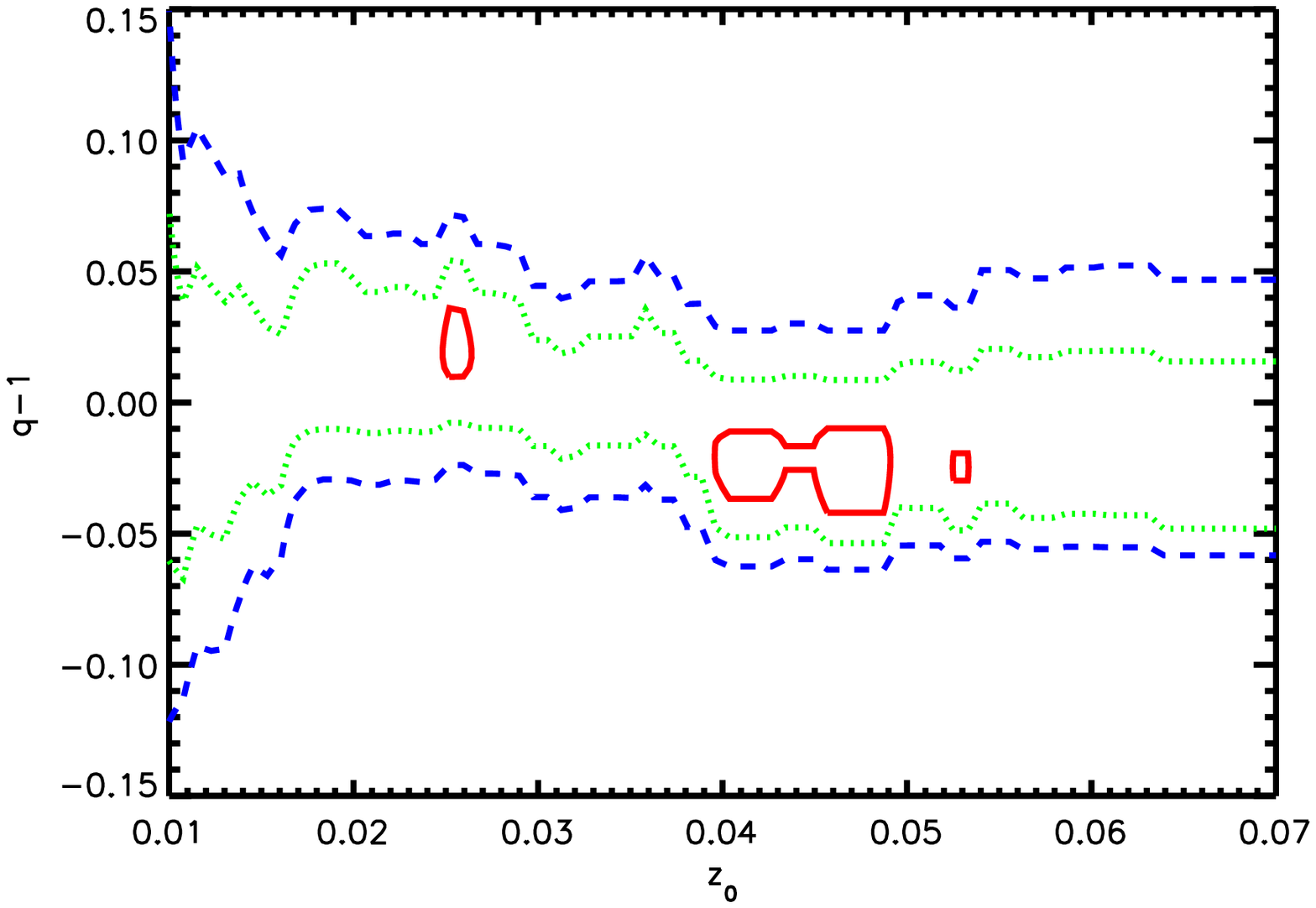}
\caption{(a) Left hand panel, confidence level on the monopole field using the distance moduli given in \cite{Jha:2007}. (b) Right hand panel, the confidence level calculated using the same data as in (a) but after correcting the linear relation shown in Figure \ref{Fig:Jhaext}. The horizontal axes are the cut-off redshift, and the vertical axes the ratio of Hubble constant below and above $z_0$ minus 1,  as defined in Equation~(\ref{eq:mono}). The contours show the 1-$\sigma$ (solid lines), 2-$\sigma$ (dotted lines), and 3-$\sigma$ (dashed lines) significance levels. 
}
\label{Fig:mono}
\end{figure}

First, we tried to reproduce the monopole field shown in \cite{Jha:2007}. We used the distance moduli given in \cite{Jha:2007} and applied the $\chi^2$ minimization procedure described above. The sample of the SNe was chosen to match exactly the 95 Hubble flow ($zc \ > \ 2,500 km/sec$) SNe used by \cite{Jha:2007}. The resulting constraint on $z_0$ and $q$ is shown in Figure \ref{Fig:mono}. Indeed, at a level slightly above 2-$\sigma$ the fit suggest $q\ = \ 1.046$, at $z_0\ = \ 0.026$, which is in agreement with the Hubble bubble reported in \cite{Jha:2007}.

\begin{figure}[tbh]
\epsscale{1.}
\includegraphics[angle=0,scale=0.8]{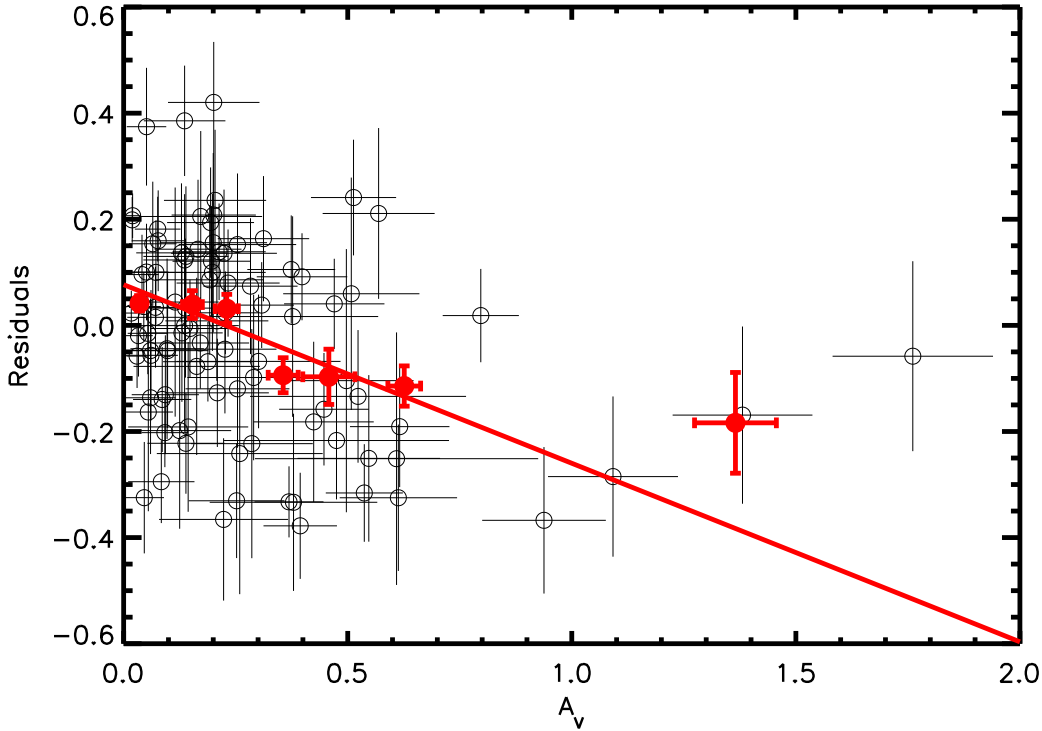}
\caption{Correlations between the supernova magnitude  residuals on the Hubble diagram and the host galaxy extinction as given by \cite{Jha:2007}. 
The linear fit results in a significant reduction of the $\chi^2$/DoF and implies a correlation between the distance moduli and the extinction.  Note that the linear correlation is determined mostly by the majority of the data with 
extinction below 0.5 mag, not those deviant points at larger than 1.0 mag. For clarity, the weighted average of the data points for $A_V$ in the intervals [0, 0.1), [0.1, 0.2), [0.2, 0.3), [0.3, 0.4), [0.4, 0.5), [0.5,1.0), and [1.0, 2.0) are shown as solid circles. The open circles are from \cite{Jha:2007}.
}
\label{Fig:Jhaext}
\end{figure}

The presence of a Hubble bubble at such a large distance scale is of sufficient cosmological significance that it deserves careful examination using different approaches. As is well known, the uncertainties in extinction correction is by far the dominant source of errors in supernova cosmology. To further investigate the authenticity of the Hubble bubble, we have analyzed the extinction correction of \cite{Jha:2007}. Figure~\ref{Fig:Jhaext} shows the magnitude residuals on the Hubble diagram and the extinction in $V$-band ($A_V$) as given in \cite{Jha:2007}. A linear fit to these data points reveals a significant correlation, as shown in Figure~\ref{Fig:Jhaext}. The $ \chi^2$ reduced from 117.3 to 101 after the introduction of the linear fit which suggests a 87\% confidence level. The deduced relation is given by $Residual \ = \ 0.077(\pm0.025)+0.336(\pm0.085)A_V$, where the numbers in parenthesis are the errors of the corresponding coefficients. This residual correlation with $A_V$ suggests 
that the extinction correction in \cite{Jha:2007} is systematically incorrect 
for SNe with different amount of reddening. Such a correlation violates the 
requirement that the distance to an SN be independent of the reddening 
to the SN.

As there is not yet a reliable model to determine the color-excess of SNIa, the
 alternative approach is to completely remove the correlation of the distance and any non-distance related parameters such as $\Delta m_{15}$ and the measured color. This is the approach used in equations (\ref{eq:mlm}) and (\ref{eq:cmagic}). As an exercise based on the same principle, we may correct the distance moduli in \citep{Jha:2007} with the linear correlation shown in Figure~\ref{Fig:Jhaext}. The resulting distance moduli are again fitted by the broken Hubble law given in Equation (\ref{eq:mono}). The confidence levels of the ratio of the Hubble constants below and above the cut-off redshift $z_0$ are shown in Figure~\ref{Fig:mono}(b). This shows a dramatic change as compared to Figure~\ref{Fig:mono} (a). No significant deviation of $q$ from 1 is seen at 2-$\sigma$ level. The best fit has in fact moved to $z_0 \ = \ 0.045$ and $q \ = \ 0.97$, but is consistent with $q$ been 1. This indicates that the putative Hubble bubble is perhaps an artifact of the data analysis. 

\begin{figure}[tbh]
\epsscale{1.}
\includegraphics[angle=90,scale=0.35]{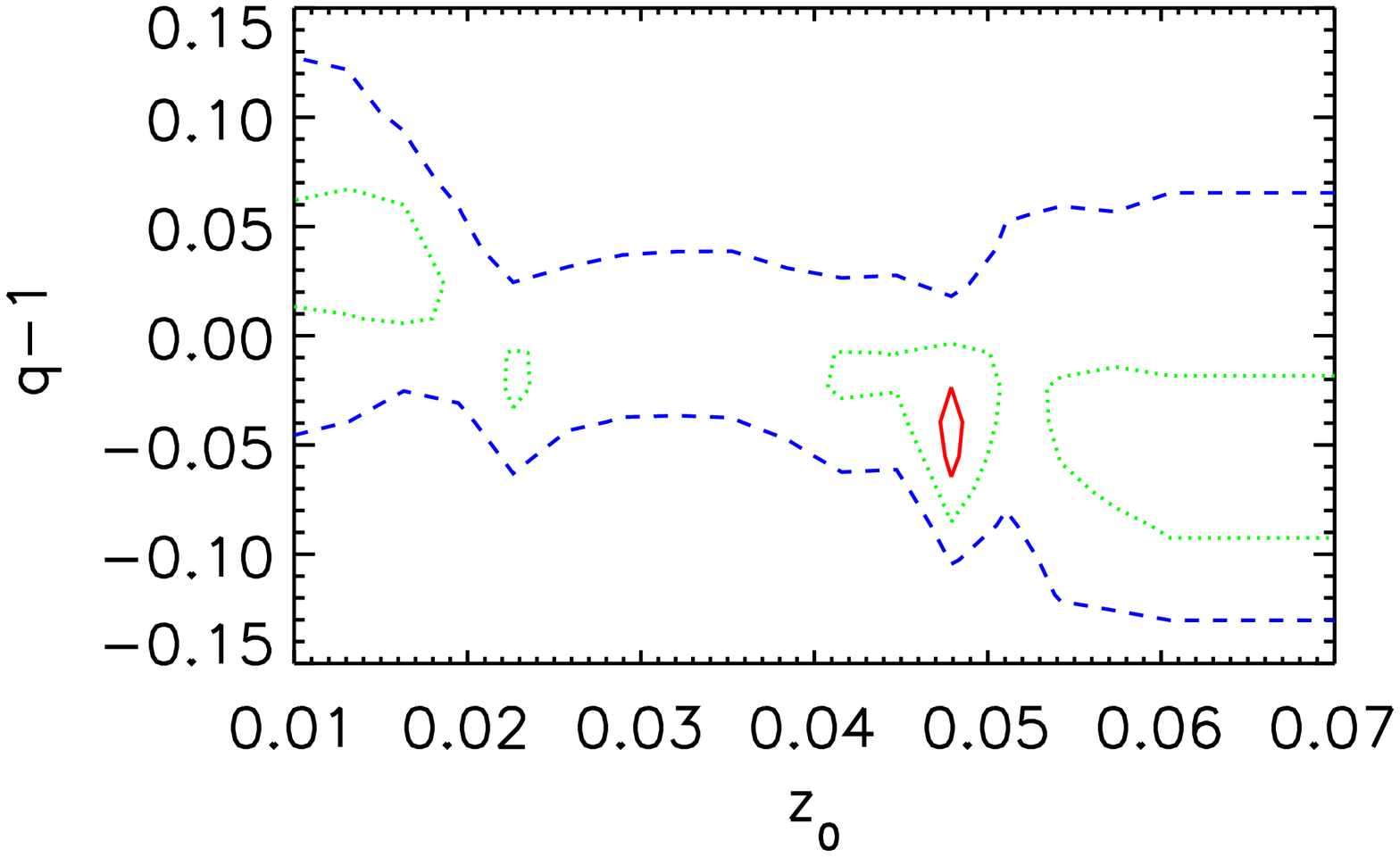}
\includegraphics[angle=90,scale=0.35]{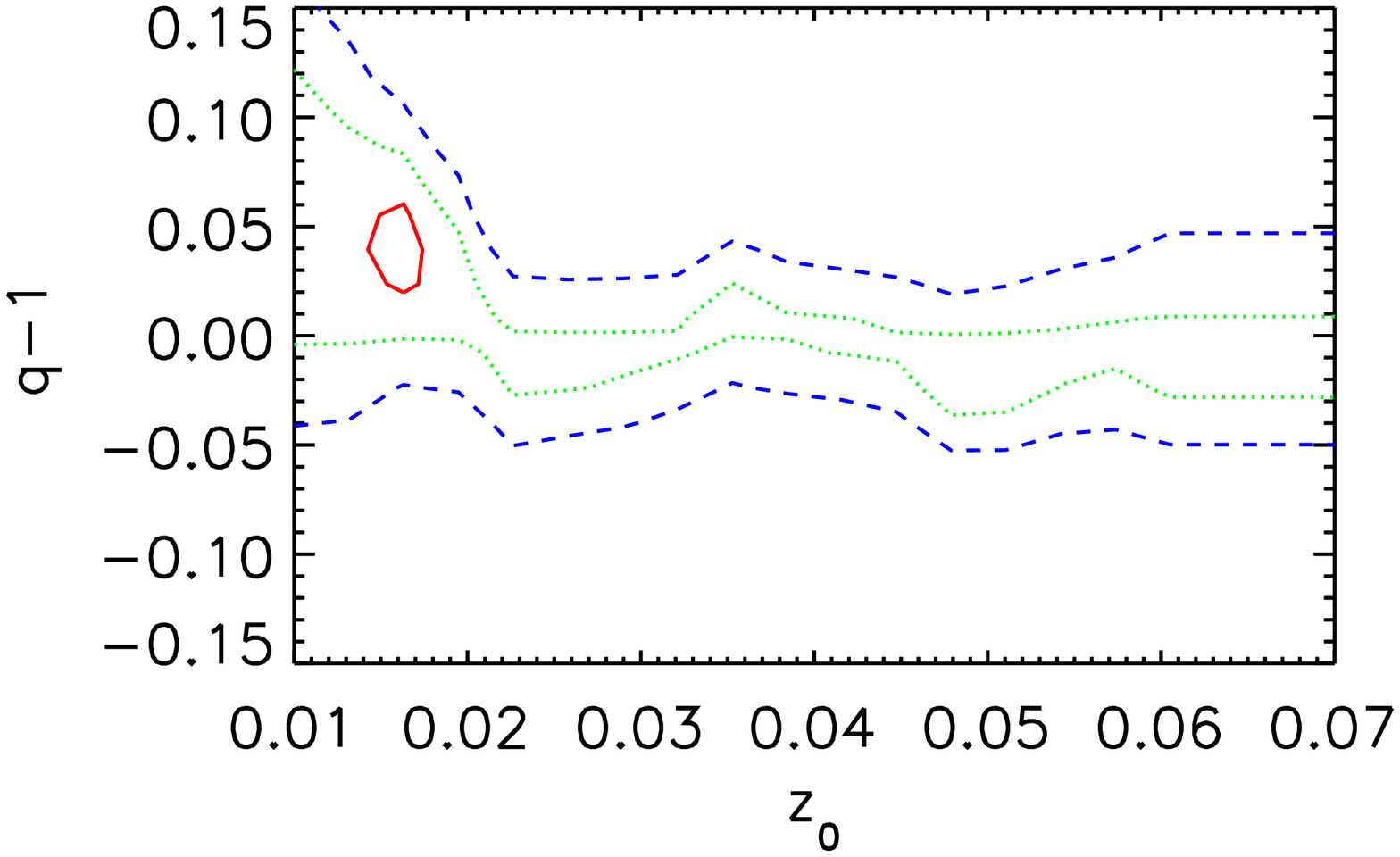}
\caption{ 
Confidence levels of the monopole field from the conventional MLM method as given by equation (\ref{eq:mlm}) (a, left panel), and  from the CMAGIC method (b, right  panel).  The horizontal axes are the cut-off redshift, and the vertical axes the ratio of Hubble constant below and above $z_0$ minus 1,  as defined in equation \ref{eq:mono}. The contours show the 1-$\sigma$ (solid lines), 2-$\sigma$ (dotted lines), and 3-$\sigma$ (dashed lines) significance levels. 
}

\label{Fig:mono1}

\end{figure}

The CMAGIC method can shed more light on  the monopole field. Not only because of its improved statistical errors, but also because it has a different sensitivity to systematic errors from extinction and light curve correction. The distance moduli of the sub-sample of SNIa derived in \S~\ref{S:DATA} are fit with Equation~(\ref{eq:mono}). Here we did not apply a redshift cut to restrict the sample to only SNe at $z\ > \ 2,500$ km/sec as done in \cite{Jha:2007}. Our analyses show very little difference with or without this redshift cut. The results are shown in Figure~\ref{Fig:mono1}. The conventional MLM yields contours that are very similar to the $A_V$ corrected case shown in Figure~\ref{Fig:mono}(b). In both MLM and CMAGIC, no monopole field more significant than 2-$\sigma$ is detected. 

We thus conclude that our analyses do not support the presence of a monopole field in the local Universe. This brings the supernova data into consistency with peculiar velocity measured from clusters of galaxies \citep{Giovanelli:1999}. A similar result was found independently by Conley et al. (2007, in prep.) 

\subsection{The Dipole Flow}

It is interesting to test wether there is a significant dipole field after the observed redshifts are corrected to the reference frame at rest with the Cosmic Microwave Background. Such a dipole field could signify a global streaming motion of the local Universe, and has been found by previous studies \citep{Riess:1995, Jha:2007, Haugboelle:2006}. The CMAGIC data, with the improved precision offer another probe of the local velocity field.

\begin{figure}[tbh]
\epsscale{1.}
\includegraphics[angle=90,scale=0.55]{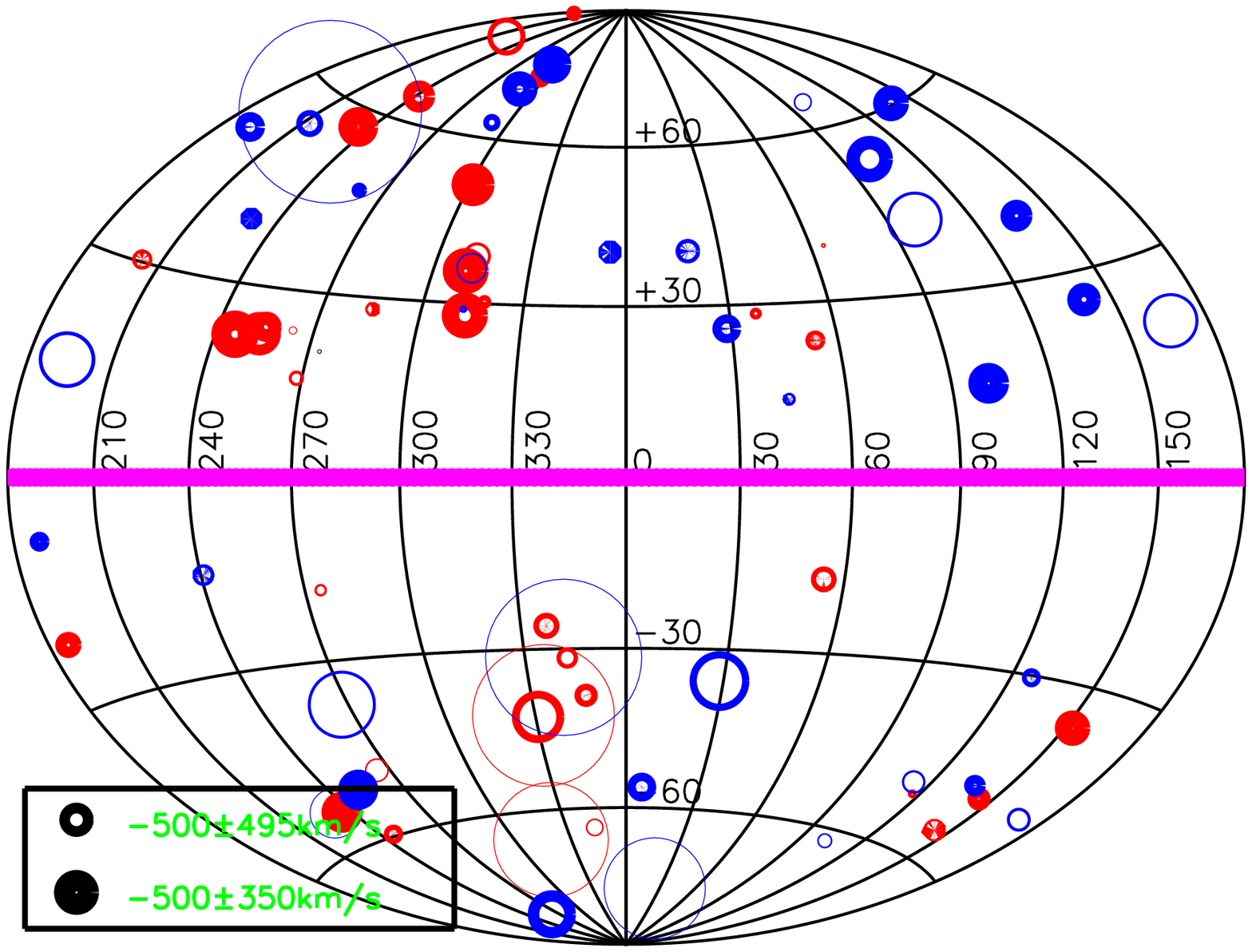}\\
\includegraphics[angle=90,scale=0.55]{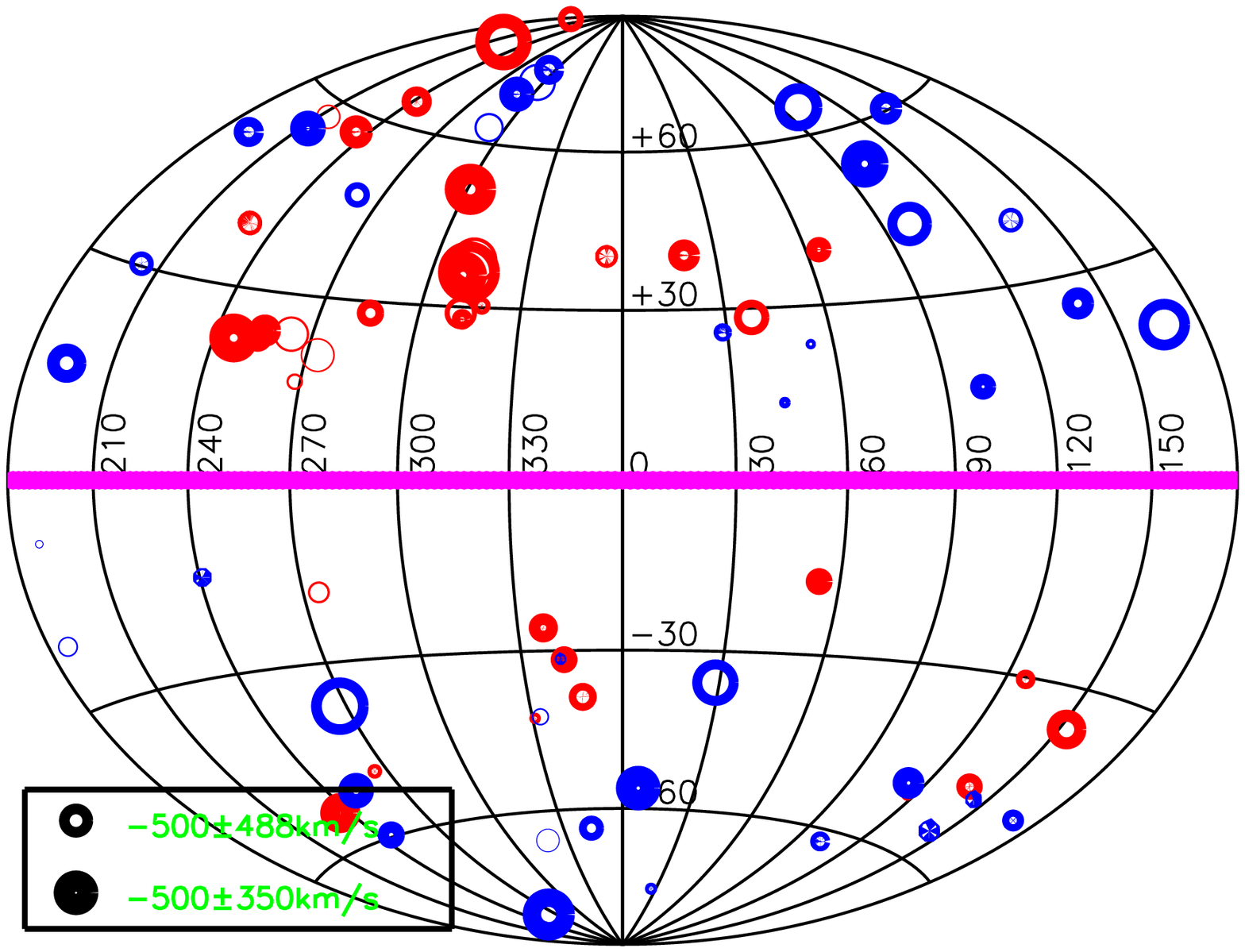}
\caption{The peculiar velocity in the CMB frame on the Aitoff projection for the conventional MLM (a, upper panel) and CMAGIC (b, lower panel) method. The Galactic coordinate system is used. The red and blue symbols show positive and negative peculiar velocities, respectively. The size of the symbols are linearly proportional to the velocity and the thickness of the circles is inversely proportional to the velocity errors. The thick solid lines show the location of the Galactic plane. The inserts illustrate typical velocities in the figure.
}
\label{Fig:Aitoff}
\end{figure}

Assuming that SNIa are perfect standard candles, the residual on the Hubble diagram is converted to peculiar velocity $V_{pec}$ using the formula 
$V\ = \ zc (1-10^{0.2\delta M})$, 
where $z$ is the redshift of the supernova, and $\delta M$ is the residual magnitude on the best fit Hubble law. In Figure~\ref{Fig:Aitoff} we show the peculiar velocity field derived with this equation on the Aitoff diagram. 

Visual inspections show overall clustering of positive and negative values. The lumpiness of the velocity field is obvious in both Figure~\ref{Fig:Aitoff}(b) for the CMAGIC Hubble diagram  and Figure~\ref{Fig:Aitoff}(a) for the MLM. Since none of the methods actually employs the position of the SN on the sky as input parameters, any angular correlation of the residuals are likely to be associated with unknown physical processes or systematic errors that are not yet accounted for in the analysis. 

\clearpage

\begin{figure}[tbh]
\epsscale{1.}
\includegraphics[angle=90,scale=0.45]{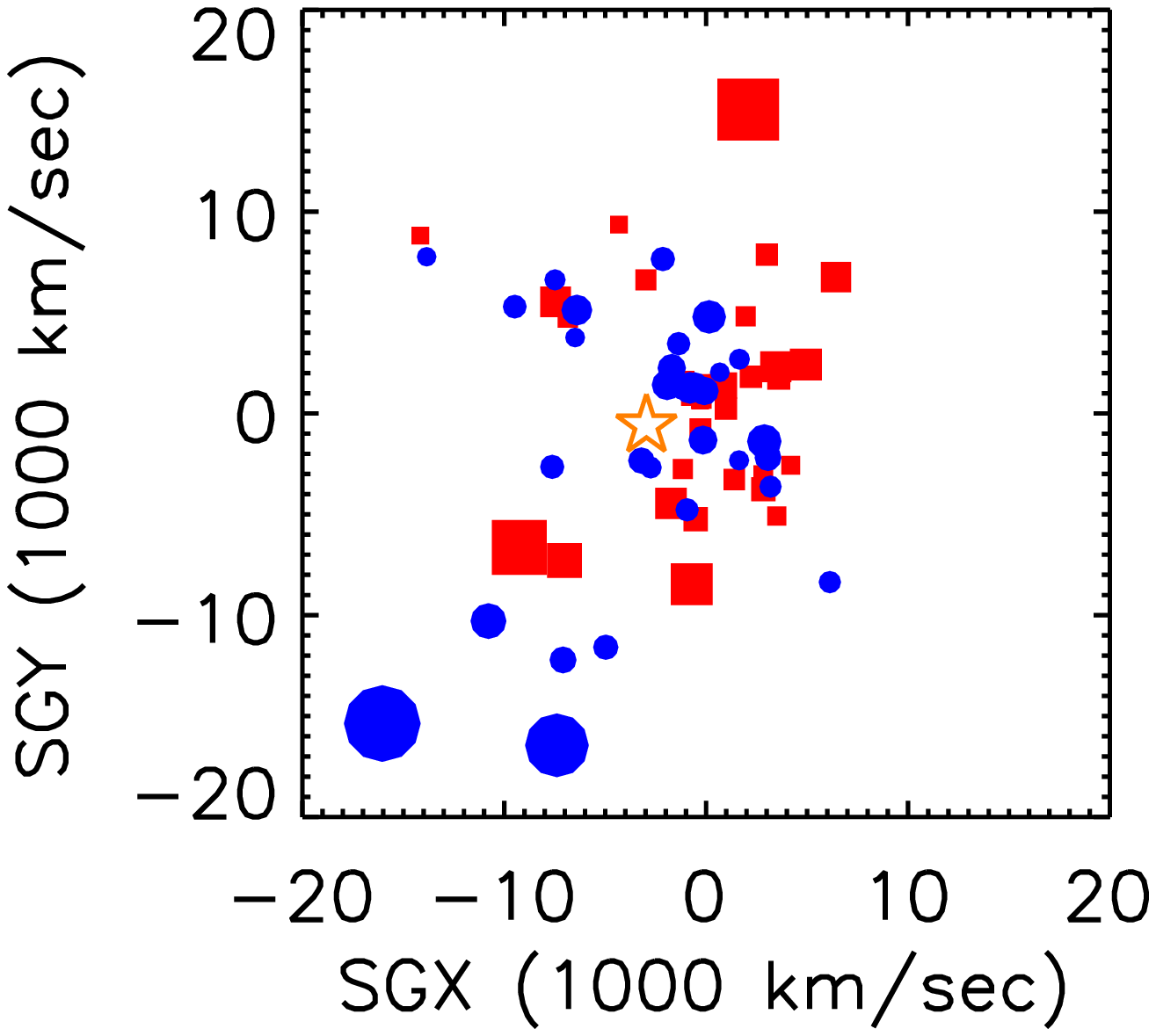}
\hspace{-4cm}
\includegraphics[angle=90,scale=0.45]{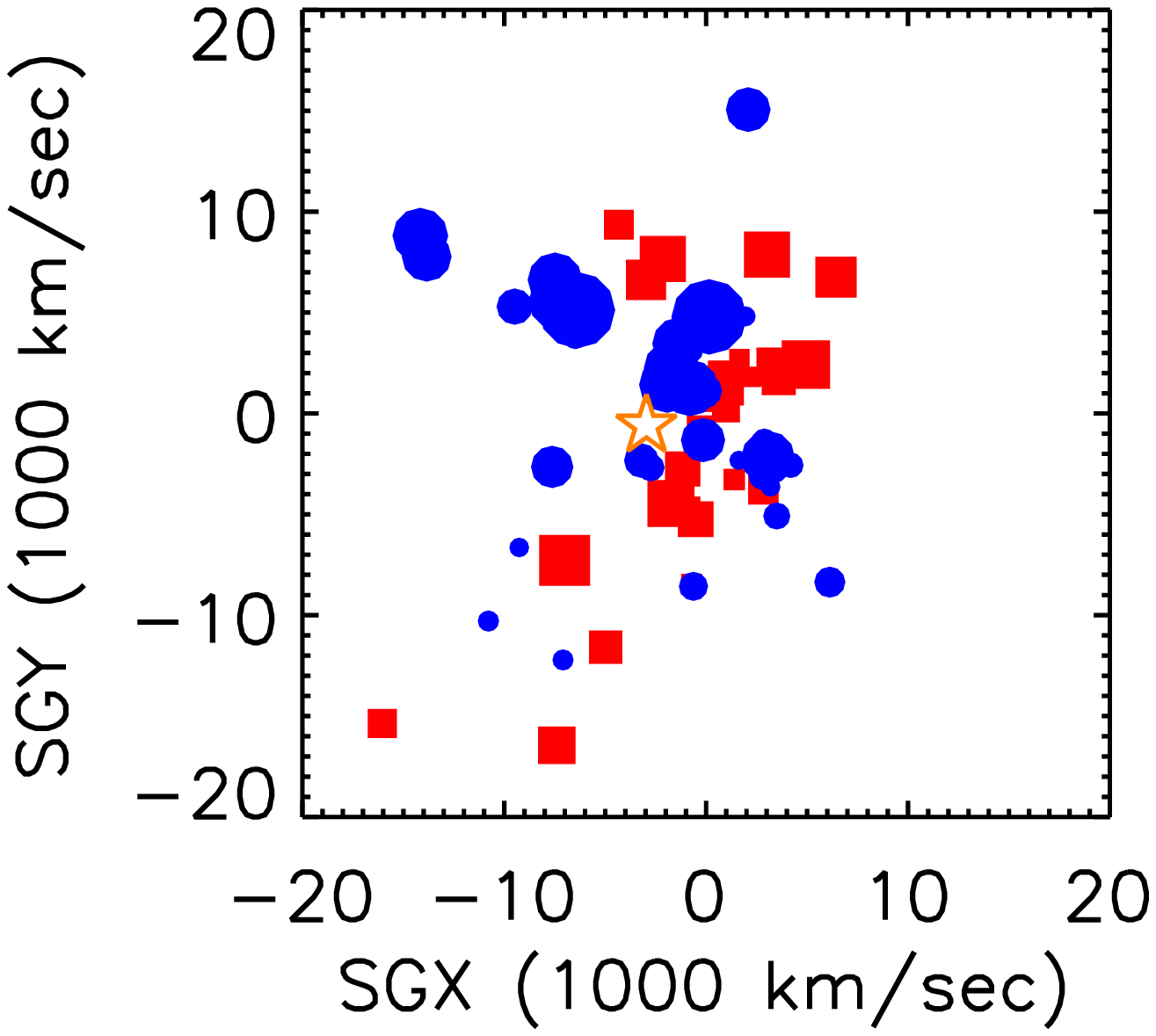}

\caption{The peculiar velocity on the supergalactic X-Y plane for the conventional MLM (a, left panel) and CMAGIC (b, right panel) methods. The star marks the location of the Great Attractor. The circles show positive peculiar velocities and the squares negative. The size of the symbols are proportional to the amplitude of the peculiar velocities. Note that in (a), the errors are large for the SNe at $z\ > \ 0.025$.
}
\label{Fig:Gal}
\end{figure}

The projected distribution on the supergalactic X-Y plane, shown in Figure~\ref{Fig:Gal} provides another view of the clustering of positive and negative peculiar velocity points. Figure~\ref{Fig:Gal}  marks also the location of the hypothesized Great Attractor (GA). The supernova data do not reveal a convergence of the velocity field toward the GA. The results from CMAGIC, shown in Figure~\ref{Fig:Gal}(b) seems to indicate excessive expansion velocity reaching far beyond the redshift of the GA.  The positive and negative velocities appears to be distributed in two orthogonal  directions. This pattern is less clear in Figure~\ref{Fig:Gal}(a) for the conventional MLM method, due apparently to its the higher level of noise than the CMAGIC method. Nonetheless, as we shall see that the two methods actually reveal consistent information about the local peculiar velocity field.

The data sparsely cover the entire sky, but with huge gaps that complicates spherical harmonic decomposition of the velocity field. To constrain the model parameters to a meaningful level, we impose further a constraint that the velocity field is axially symmetric. This is a rather arbitrary constraint but it reduces considerably the mathematical uncertainties and enables quantitative descriptions of the velocity field with simple mathematical formulae. We will consider both the dipole and quadrupole terms 
in this study. This is a natural step above the monopole field studied in the previous section.

For the axially symmetric dipole model, we model the velocity field in terms of the spherical harmonics $Y_1^0$, 
\begin{equation}
V_{pec}(l, b) \ = \ p_0 + p_1 \mu\ = \ p_0 + p_1(\cos b_0 \cos b \cos(l_0-l)\ +\ \sin b_0 \sin b),
\label{Eq:dip}
\end{equation}
where $(l,\ b)$ is the longitude and latitude of the supernovae, $p_0$ and $p_1$ are the fitting coefficients of the monopole and quadruple field, $(l_0,\ b_0)$ are the coordinates of the symmetry axis, and $\mu\ = \ sin(b')$ with $b'$ being the latitude measured in the polar coordinates with the polar axis pointing at  $(l_0,\ b_0)$, {\it i. e.} the symmetry axis in Galactic coordinates.

The $\chi^2$ of the model is then
\begin{equation}
\chi^2 \ = \ \sum_i (V_i - V_{pec}(l_i, b_i))^2/(\delta V_i^2),
\label{Eq:chi2dip}
\end{equation}
where the errors $\delta V$ of the peculiar velocity measurements are derived from the distance measurement errors. There are three major contributions to the velocity uncertainties: the measurement error, the error due to an unknown random velocity of the peculiar velocity field, and the error due to the intrinsic magnitude dispersion of supernovae.

Applying equations (\ref{Eq:dip}) and (\ref{Eq:chi2dip}) to the observed data, we found that a dipole fit results in a reduction of $\chi^2$ for both the MLM and the CMAGIC method. The resulting dipole for the entire sample is found to be at an amplitude of 293$\pm100$ km/sec and  289$\pm$94 km/sec for the MLM and CMAGIC respectively. The direction of the dipole field is at $(l_0, b_0) \ = \ (288^{\rm o}.0^{+19.3}_{-10.5}, -5^{\rm o}.7^{+16.2}_{-9.7})$ and $(l_0, b_0) \ = \ (306^{\rm o}.0^{+19.3}_{-16.4}, 17^{\rm o}.5^{+11.3}_{-15.4})$, for the MLM and CMAGIC, respectively. These results are consistent with each other, and are also consistent with earlier studies \citep{Riess:1995, Jha:2007, Haugboelle:2006}.
This success is encouraging, and motivates further explorations of the  peculiar velocity field from the CMAGIC method. 

\begin{figure}
\epsscale{1.}
\includegraphics[angle=90,scale=0.45]{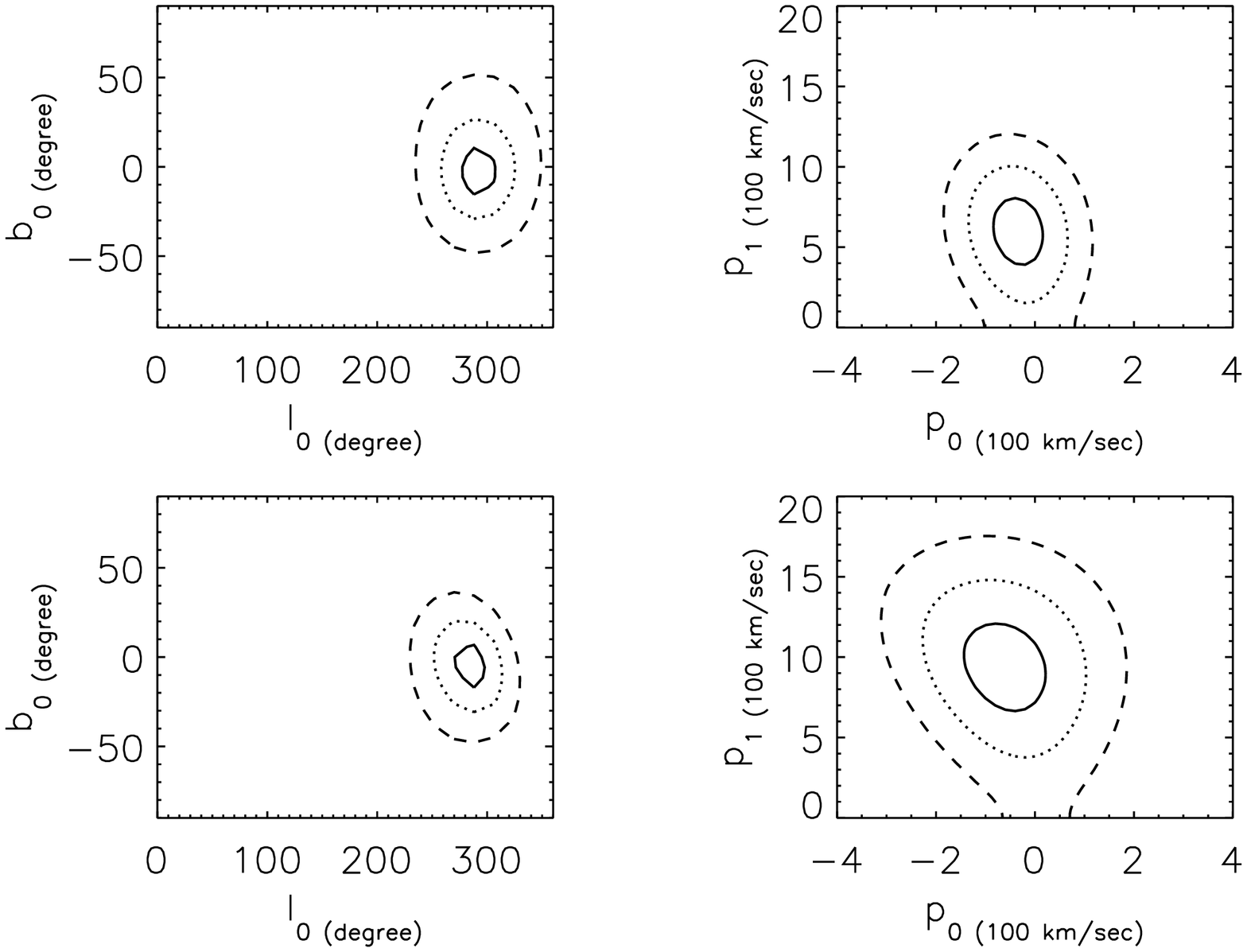}\\
\includegraphics[angle=90,scale=0.45]{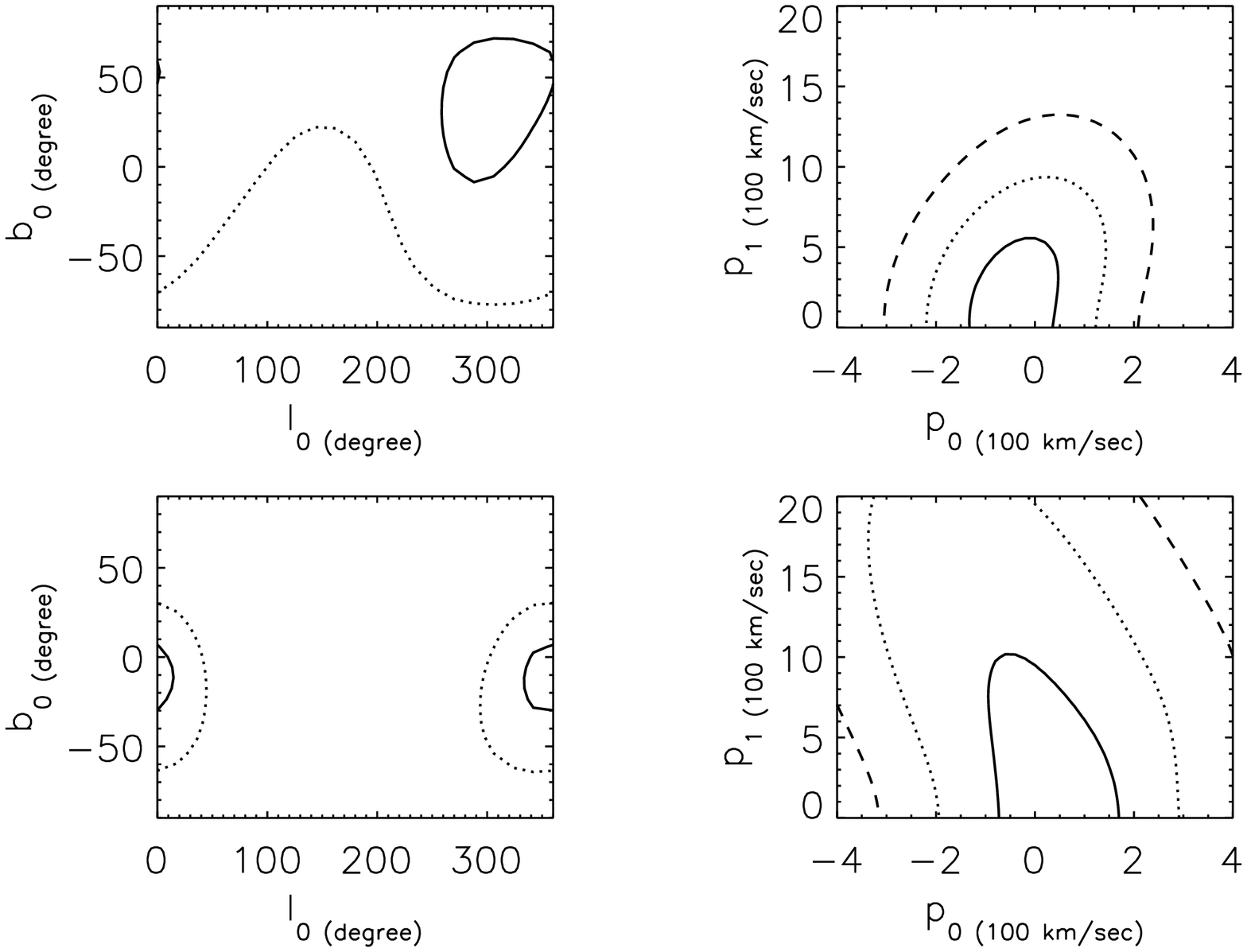}

\caption{The error contours of the parameters for the dipole fits using the MLM of SNIa. The left column 
shows the axis of symmetry $(l_0, b_0)$, and the right column shows the fitted coefficients of 
the monopole and dipole components. From the top to the bottom, the panels show the fits to the entire 
sample, sample $A$, sample $B$, and sample $C$. For the contours of $(l_0, b_0)$, the probability 
distributions are marginalized over $p_0$ and $p_1$; for those of $p_0$ and $p_1$, they are 
marginalized over $l_0$ and $b_0$. The $\chi^2$/DoF were normalized to unity to make these contours.The solid, dotted, and dashed lines are 1-, 2-, and 3-$\sigma$ confidence levels.
}
\label{Fig:fitsmlm}

\end{figure}

\begin{figure}
\epsscale{1.}
\includegraphics[angle=90,scale=0.45]{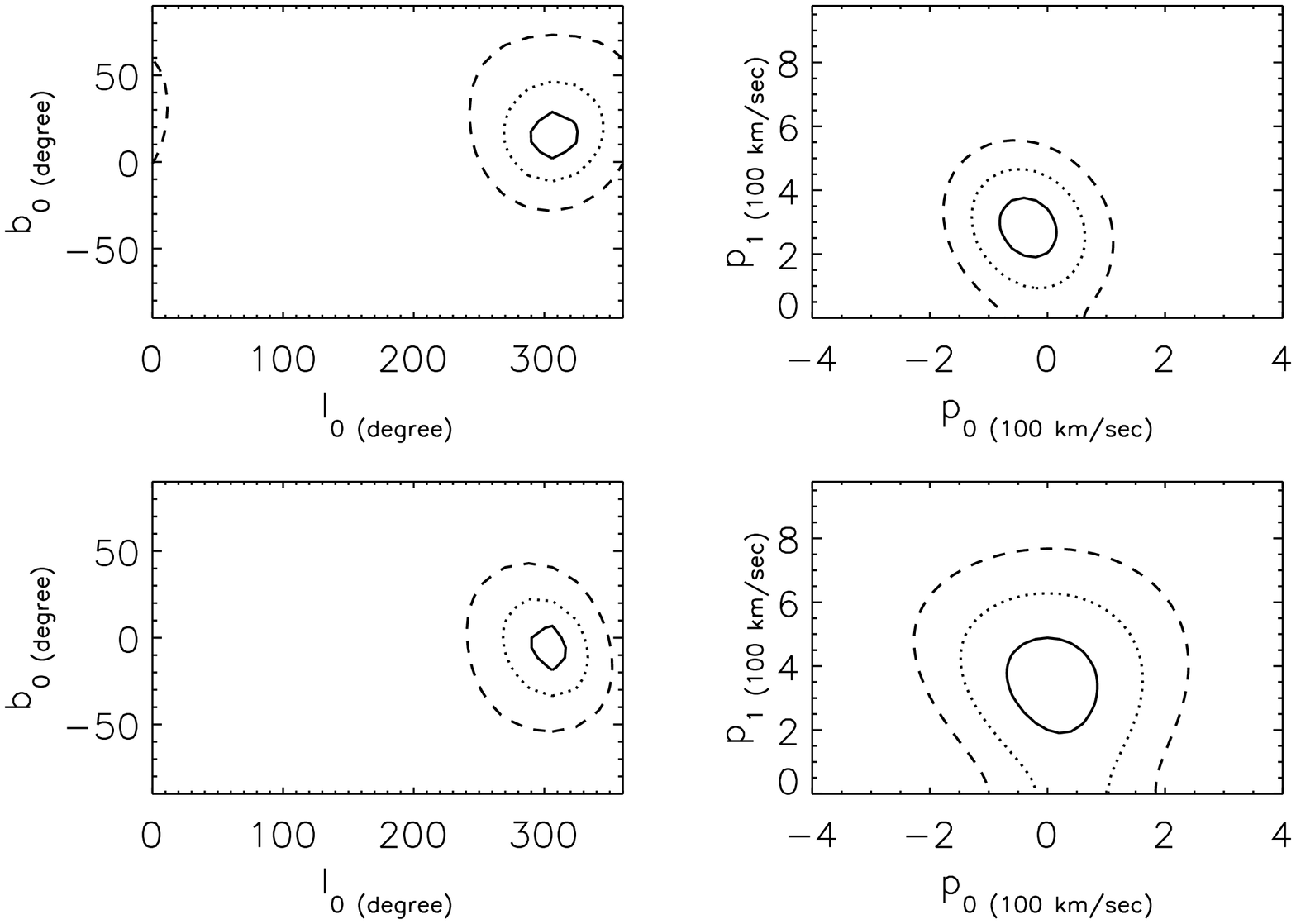}\\
\includegraphics[angle=90,scale=0.45]{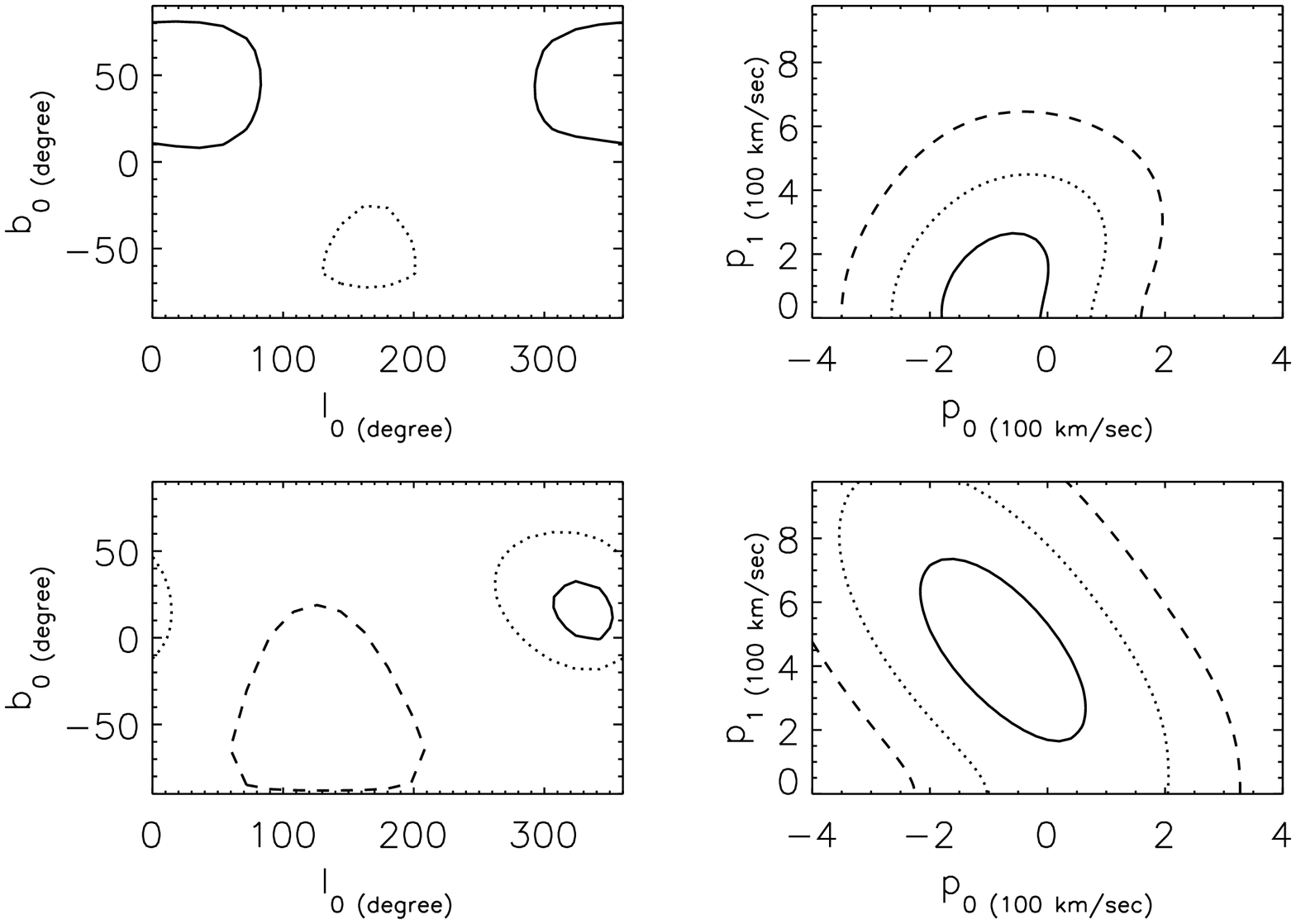}

\caption{Same as Figure~\ref{Fig:fitsmlm}, but for the CMAGIC method.
}
\label{Fig:fitscmagic}

\end{figure}

A question of particular  importance is the depth of the dipole pattern in the redshift space. If the Universe is homogeneous and isotropic on average, we should see a decreasing dipole amplitude as we average over larger and larger volume. To quantitatively  investigate this problem, we have divided the supernova sample into three different sub-samples $A$, $B$, and $C$ according to their redshifts. Each sample has roughly the same number of SNe. The samples A, B, and C are for SNe at redshifts below 0.01, between 0.01 and 0.025, and above 0.025, 
respectively. The weighted mean redshifts (standard deviations) are 0.00675 (0.00239), 0.01653 (0.00378), and 0.03450 (0.00887), for samples $A$, $ B$, and $C$ respectively. The optimal estimates of the parameters of these fits are given in Table \ref{Tab:fits}, and are shown in Figure~\ref{Fig:fitsmlm} and Figure~\ref{Fig:fitscmagic}. The dipole directions for the lowest redshift sample $A$ is remarkably consistent with the direction of Norma cluster at (l, b) = (325$^{\rm o}.3308$, $-7^{\rm o}.2557$) - the suspected center of the 
Great Attractor (see for example, \cite{Radburn-Smith:2006} for a recent study of the GA). 
At higher redshift with sample $B$, neither the MLM data nor the CMAGIC data show a prominent dipole. 
At even higher redshift sample $C$, the MLM data reveal no dipole, but the CMAGIC method does show a 
rather strong dipole whose direction is at (l, b) = (342$^{\rm o}.0^{+10}_{35}$, 11$^{\rm o}.5^{+21}_{-12.5}$), 
at an amplitude of  452$^{+284}_{-287}$ km/sec (Figure~\ref{Fig:fitscmagic}). This is only a 2-$\sigma$ 
detection, but it is a very intriguing possibility considering the fact that its direction is consistent with 
the dipole direction of sample $A$, and points 
right at the center of Shapley supercluster of galaxies located at (l, b) = (327$^{\rm o}$, 32$^{\rm o}$) 
and a redshift of about 0.048 \citep{Smith:2004}.

\linespread{1.7}
\begin{deluxetable}{l|rrrrrrrrr}
\tabletypesize{\scriptsize}\tablecolumns{9}\tablewidth{0pt}\tablecaption{Dipole and Quadruple Fits to the Observed Peculiar Velocity Field\tablenotemark{*}}
\tablehead{\colhead{Model\tablenotemark{\dag}}      & \colhead{$<z>$\tablenotemark{\ddag}}                          & \colhead{$p_0\ (\sigma)$}                    & \colhead{$p_1\ (\sigma)$}                    & \colhead{$l_0\ (\sigma)$}                    & \colhead{$b_0\ (\sigma)$}                    & \colhead{$p_2\ (\sigma)$ }                   & \colhead{$\chi^2_o$}                    & \colhead{$\chi^2$}                    & \colhead{DoF}                   \cr}

\startdata
D$^M$   & &  -40.0$_{ -43.9}^{  +55.9}$  &   293.2$_{-102.1}^{ +100.3}$   &    288.0$_{ -10.5}^{  +19.3}$  &    -5.7$_{  -9.7}^{  +16.2}$   &   --- ---  &  78  &  66  &  72  \\
D$^M_A$ & 0.00675 &  -60.0$_{ -83.4}^{  +81.3}$  &   458.1$_{-133.4}^{ +132.3}$   &    288.0$_{ -17.5}^{   +9.7}$  &    -5.7$_{ -11.3}^{  +12.6}$   &   --- ---  &  24  &  12  &  19  \\
D$^M_B$ & 0.01653 &  -40.0$_{-112.6}^{ + 67.0}$  &    61.1$_{ -210.4}^{+ 210.4}$   &    288.0$_{-12.0}^{  +72.0}$  &    23.6$_{ -32.2}^{  +48.3}$   &   --- ---  &  29  &  27  &  23  \\
D$^M_C$ & 0.03450 & 40.0$_{-114.3}^{ +149.8}$  &     0.0$_{   -497.7}^{ +497.7}$   &      0.0$_{  -25.0}^{+15.0}$  &   -11.5$_{ -18.3}^{  18.4}$   &   --- ---  &  23  &  19  &  22  \\
\hline
D$^C$              & &  -40.0$_{ -61.2}^{  +35.4}$  &   280.9$_{ -90.8}^{  +95.2}$   &    306.0$_{ -16.4}^{  +19.3}$  &    17.5$_{ -15.4}^{  +11.3}$   &   --- ---  &  90  &  76  &  72  \\
D$^C_A$ & 0.00675 & 0.0$_{ -69.2}^{  +84.7}$  &   342.0$_{-151.8}^{ +146.6}$   &    306.0$_{ -15.8}^{  +10.3}$  &    -5.7$_{ -12.8}^{  +12.6}$   &   --- ---  &  19  &  11  &  19  \\
D$^C_B$ & 0.01653 & -100.0$_{-130.0}^{  +51.3}$  &    48.9$_{ -265.3}^{ +216.3}$   &     36.0$_{ -104.0}^{ +34.0}$  &    36.9$_{ -28.8}^{  +44.1}$   &   --- ---  &  33  &  29  &  23  \\

D$^C_C$ &  0.0345 & -80.0$_{-176.4}^{ +104.7}$  &   452.0$_{-287.4}^{ +283.6}$   &    342.0$_{ -35.0}^{  +10.2}$  &    11.5$_{ -12.5}^{  +21.1}$   &   --- ---  &  37  &  28  &  22  \\
\hline
\hline
DQ$^M$ & & --- ---  &   244.3$_{-101.5}^{ +106.7}$   &    300.0$_{ -17.7}^{  +10.2}$  &    11.5$_{ -12.6}^{  +10.2}$   &    189.2$_{-168.1}^{ 115.4}$  &  78  &  65  &  72  \\
DQ$^M_A$ & 0.00675 & --- ---  &   305.4$_{-147.0}^{ +125.1}$   &    276.0$_{ -13.1}^{  +11.1}$  &   -47.2$_{  -6.2}^{  +11.6}$   &   -378.5$_{-156.2}^{ 170.4}$  &  24  &  10  &  19  \\

DQ$^M_B$ & 0.01653 & --- ---  &   280.9$_{-161.4}^{ +157.6}$   &    348.0$_{ -27.9}^{  +10.5}$  &    36.9$_{ -13.1}^{   +9.0}$   &    307.5$_{-207.2}^{ +213.2}$  &  29  &  22  &  23  \\
DQ$^M_C$ & 0.003450 & --- ---  &     0.0$_{-393.9}^{+493.9}$   &      0.0$_{   -20.0}^{ +20.0}$  &   -11.5$_{ -17.2}^{  +13.5}$   &     23.7$_{-557.1}^{ +415.3}$  &  23  &  19  &  22  \\
\hline
DQ$^C$ & & --- ---  &   244.3$_{ -74.5}^{  +82.7}$   &    312.0$_{ -15.8}^{   +6.0}$  &    23.6$_{  -7.8}^{   +7.3}$   &    331.2$_{ -87.7}^{  +96.6}$  &  90  &  63  &  72  \\

DQ$^C_A$ & 0.00675 & --- ---  &     0.0$_{   -477.6}^{ +447.6}$   &    300.0$_{  -5.0}^{  +9.1}$  &     3.8$_{  -7.9}^{   +9.9}$   &   -402.1$_{-114.7}^{ +644.9}$  &  19  &  10  &  19  \\
DQ$^C_B$ & 0.01653 & --- ---  &   317.6$_{-163.5}^{ +156.1}$   &    336.0$_{-41.0}^{  +24.0}$  &    36.9$_{ -14.0}^{  +12.4}$   &    331.2$_{-212.3}^{ +195.4}$  &  33  &  24  &  23  \\

DQ$^C_C$ & 0.03450 &  --- ---  &   232.1$_{-232.1}^{ +177.1}$   &    324.0$_{ -17.8}^{  +12.3}$  &   -53.1$_{  -8.6}^{  +13.8}$   &   -686.0$_{-198.1}^{ +216.9}$  &  37  &  20  &  22  \\

\hline
\hline
Bip$^M$ & & --- ---  &  -260.0$_{ -72.1}^{  +73.2}$   &    300.0$_{ -11.4}^{  +17.0}$  &    19.5$_{  -9.2}^{   +8.0}$   &    -27.1$_{  -6.6}^{   +8.7}$  &  78  &  63  &  72  \\
Bip$^M_A$ & 0.00675  & --- ---  &  -340.0$_{-105.0}^{ +104.2}$   &    288.0$_{ -15.1}^{  +12.8}$  &    15.5$_{ -14.7}^{  +11.0}$   &    -21.1$_{ -10.2}^{  +11.4}$  &  24  &  13  &  19  \\
Bip$^M_B$ & 0.01653 & --- ---  &  -320.0$_{-137.1}^{ +139.9}$   &    336.0$_{ -12.4}^{  +16.3}$  &    36.9$_{  -9.8}^{   +8.9}$   &    -24.1$_{ -11.7}^{   +7.9}$  &  29  &  20  &  23  \\
Bip$^M_C$ & 0.03450 & --- ---  &     0.0$_{-255.9}^{+255.9}$   &    348.0$_{-20.0}^{  +12.0}$  &     0.0$_{   -16.7}^{  +16.7}$   &    -60.3$_{   -60.3}^{  +60.3}$  &  23  &  19  &  22  \\
\hline
Bip$^C$ & & --- ---  &  -340.0$_{ -71.3}^{  +62.4}$   &    312.0$_{  -7.4}^{  +13.5}$  &    25.7$_{  -5.2}^{   +7.7}$   &    -24.1$_{  -3.7}^{   +6.4}$  &  90  &  62  &  72  \\
Bip$^C_A$ & 0.00675 & --- ---  &  -260.0$_{-108.4}^{ +105.7}$   &    300.0$_{  -9.1}^{  +14.2}$  &     1.9$_{  -13.1}^{  +14.9}$   &    -27.1$_{ -12.5}^{  +10.5}$  &  19  &  11  &  19  \\
Bip$^C_B$ & 0.01653 & --- ---  &  -440.0$_{-110.2}^{ +103.1}$   &    324.0$_{  -6.6}^{   +7.7}$  &    36.9$_{  -5.1}^{   +6.1}$   &    -30.2$_{  -5.6}^{   +6.0}$  &  33  &  16  &  23  \\
Bip$^C_C$ & 0.03450 & --- ---  &  -400.0$_{-146.3}^{ +150.3}$   &    336.0$_{ -18.2}^{   +9.5}$  &    25.7$_{ -12.4}^{  +13.3}$   &    -18.1$_{ -11.4}^{  +11.1}$  &  37  &  25  &  22  \\

\hline
\enddata
\linespread{1.0}
\tablenotetext{*}{The errors are generally non-Gaussian, check Figures~\ref{Fig:fitsmlm}, \ref{Fig:fitscmagic}, \ref{Fig:fits12mlm}, and \ref{Fig:fits12cmagic} for the error contours.}
\tablenotetext{\dag}{The superscripts $^M$ and $^C$ in the Model column stand for fits to MLM and CMAGIC data, respectively. The subscripts $_A$, $_B$, and $_C$ stand for sample $A$, $B$, $C$, respectively. The Model D, DQ, and Bip refer to the streaming motions as given in Equations~\ref{Eq:dip}, \ref{Eq:quad}, and \ref{Eq:mquad}, respectively. }
\tablenotetext{\ddag} {The redshift given in column $<z>$ are the weighted average redshifts for the SNe in each sample.}
\label{Tab:fits}
\end{deluxetable}

\clearpage

\clearpage

It is also remarkable that the fits to the whole sample and samples $A$, $B$, and $C$ are all consistent with the monopole component $p_0$ being zero. This is a direct confirmation of the previous section where we found no measurable monopole components with a different approach.

For sample $A$, the dipole fits reduce the  $\chi^2$ from 24 to 12 and 19 to 11, for the MLM and CMAGIC, respectively. Such a significant improvement of the goodness of fit suggest that the hypothesized model is indeed valid. After the dipole correction, the $\chi^2$/DoF are well below unity for both MLM and CMAGIC. This can be understood if we have assumed a thermal peculiar velocity component that is too large. To make the $\chi^2$/DoF to unity, the thermal peculiar velocity should be around 270 km/sec, assuming no
intrinsic dispersions due to supernova standardization. At these low redshifts, a thermal peculiar velocity of 
270 km/sec introduces a magnitude dispersion of about 0.27 magnitude, which is well above $\sim$ 0.08 - the upper limit of the
intrinsic dispersions of CMAGIC calibrated supernova distances  \citep{Wang:2003CMAGIC, Wang:2006}. 
It thus appears that 270 km/sec is a reasonable estimate of the thermal velocity of the peculiar 
velocity field within the volume of z $\sim$ 0.00675. This estimate is only weakly dependent on the 
intrinsic magnitude dispersions of SNIa - subtracting the intrinsic dispersion of 0.08 mag \citep{Wang:2006} 
results in an estimate of the level of  thermal component of $\sim$ 258 km/sec.

The absence of a consistent dipole field for sample $B$, however, does not imply the absence of a 
coherent streaming motion at larger scales. 
It is possible that the data are still noisy, or the assumed mathematical model is wrong and is 
incapable of fully capturing the large scale streaming motions.
Although it is true that future wide area supernova surveys will help in determining the high 
redshift bound of the dipole field, and its extension to higher redshift,   we will investigate further with the current data set 
the nature of the dipole velocity component in the next sections by introducing other 
formulations of the coherent streaming  flow.
\label{S:dip}

\subsection{The Quadruple Flow}
\label{S:quad}

The data allow for studies of higher oder components of the peculiar velocity field. To restrict the number of free parameters to minimal level, we will again assume axial symmetry so that the dipole and quadruple field point to the same direction. There is no physical motivation to model the velocity field this way other than for its simplicity. With a large enough data set, it should be possible to determine the complete quadruple components without the assumption of spherical symmetry. As we learned from the previous sections, the monopole field is small and non-detected. We will assume in this section that the monopole field is zero at all redshifts.  

\begin{figure}
\epsscale{1.}
\includegraphics[angle=90,scale=0.45]{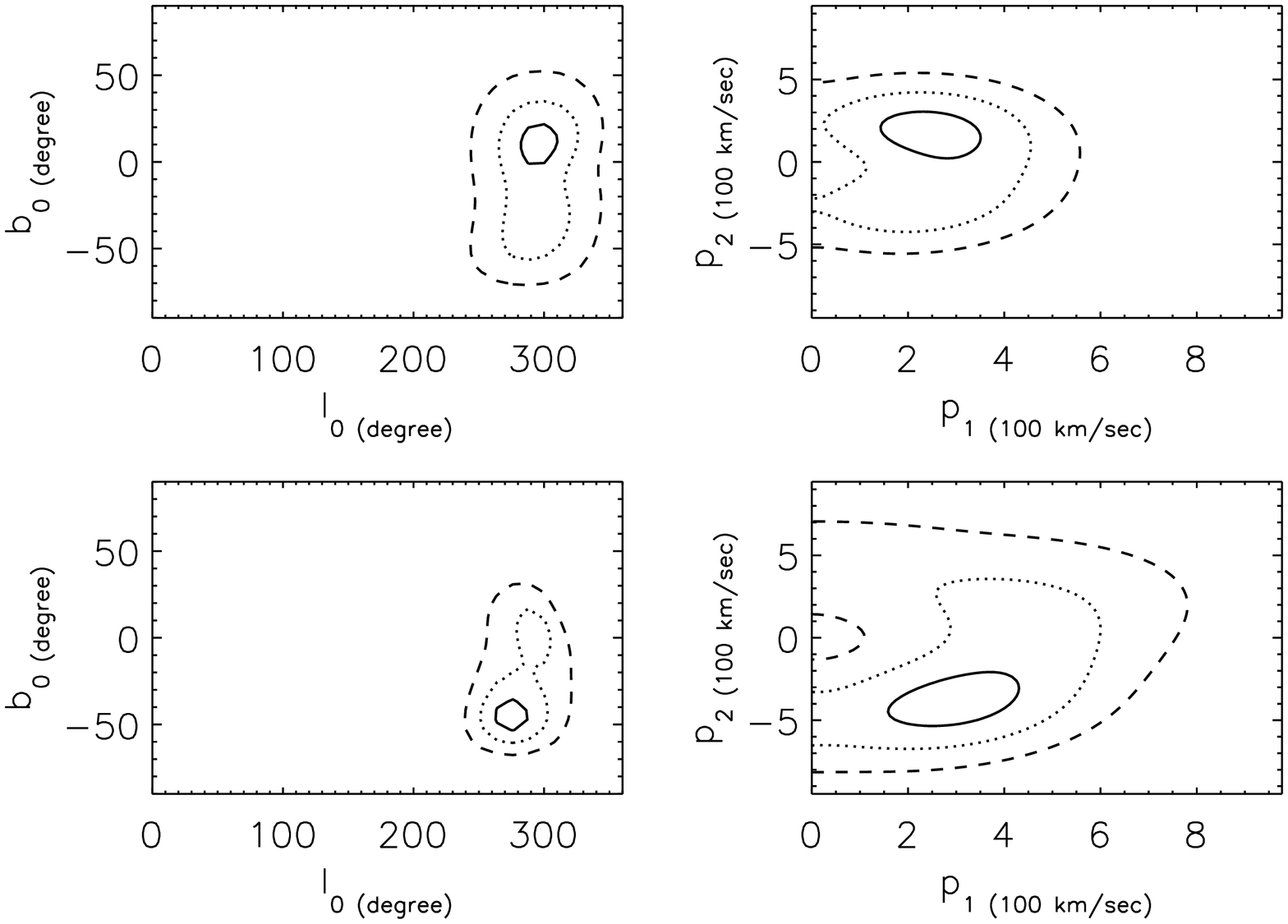}\\
\includegraphics[angle=90,scale=0.45]{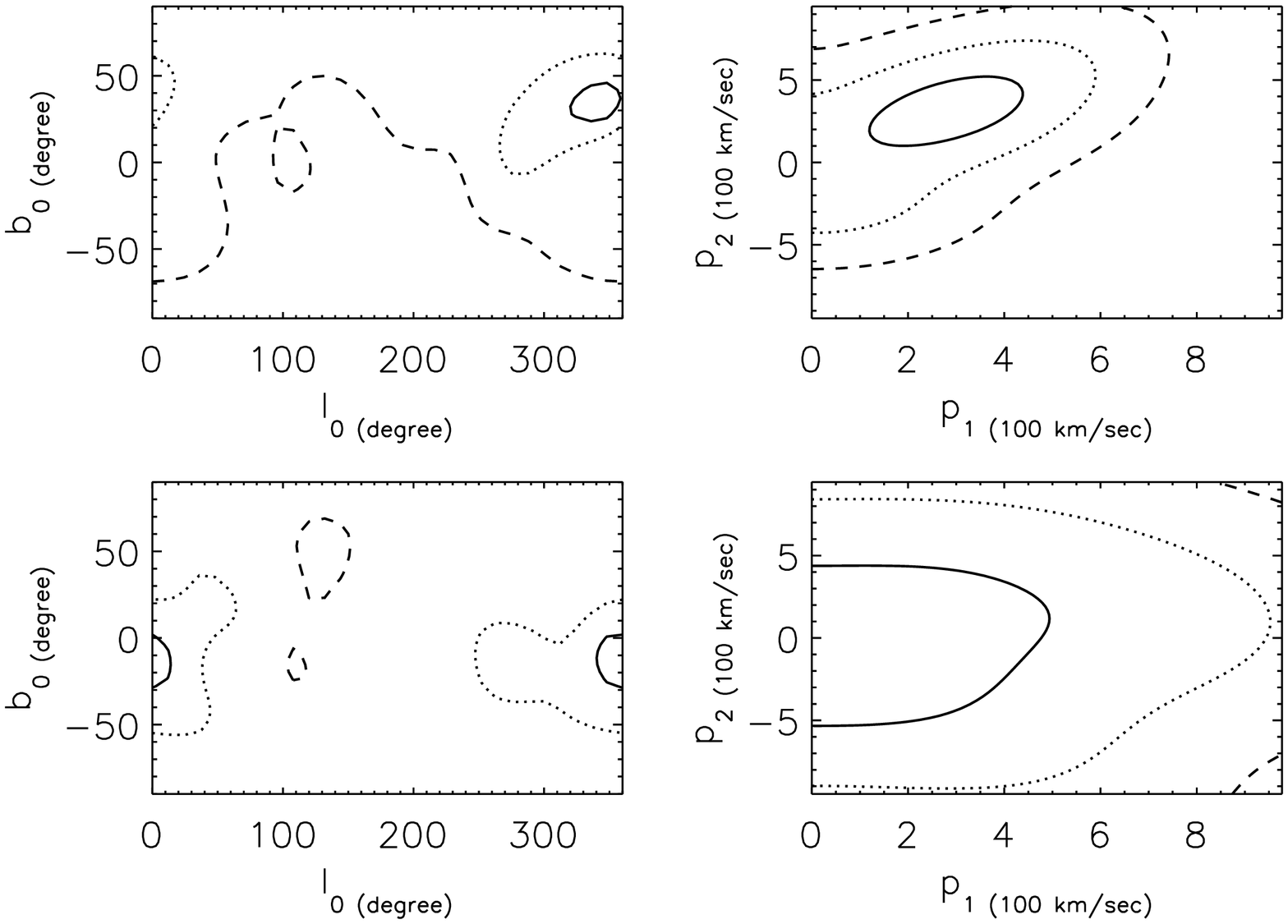}

\caption{The error contours of the parameters for the dipole plus quadruple fits using the MLM data. 
The left column shows the axis of symmetry $(l_0, b_0)$, and the right column shows the fitted 
coefficients of the dipole  and quadruple components.  From the top to the bottom, the panels show the 
fits for the entire sample, sample $A$, sample $B$, and sample $C$. For the contours of $(l_0, b_0)$, 
the probability distributions are marginalized over $p_1$ and $p_2$; for those of $p_1$ and $p_2$, they 
are marginalized over $l_0$ and $b_0$. The solid, dotted, and dashed lines are 1-, 2-, and 3-$\sigma$ confidence levels.
}
\label{Fig:fits12mlm}

\end{figure}

\begin{figure}
\epsscale{1.}
\includegraphics[angle=90,scale=0.45]{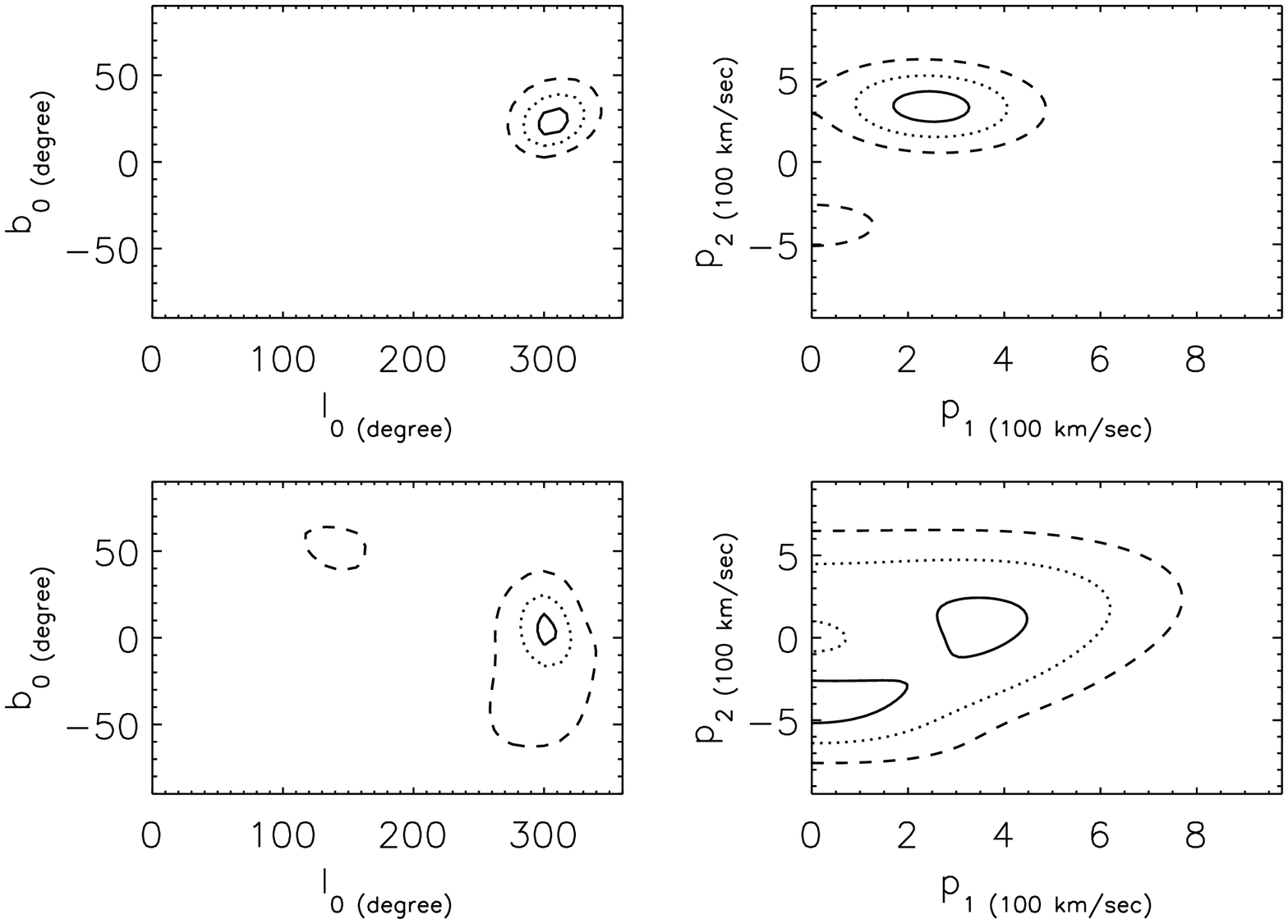}\\
\includegraphics[angle=90,scale=0.45]{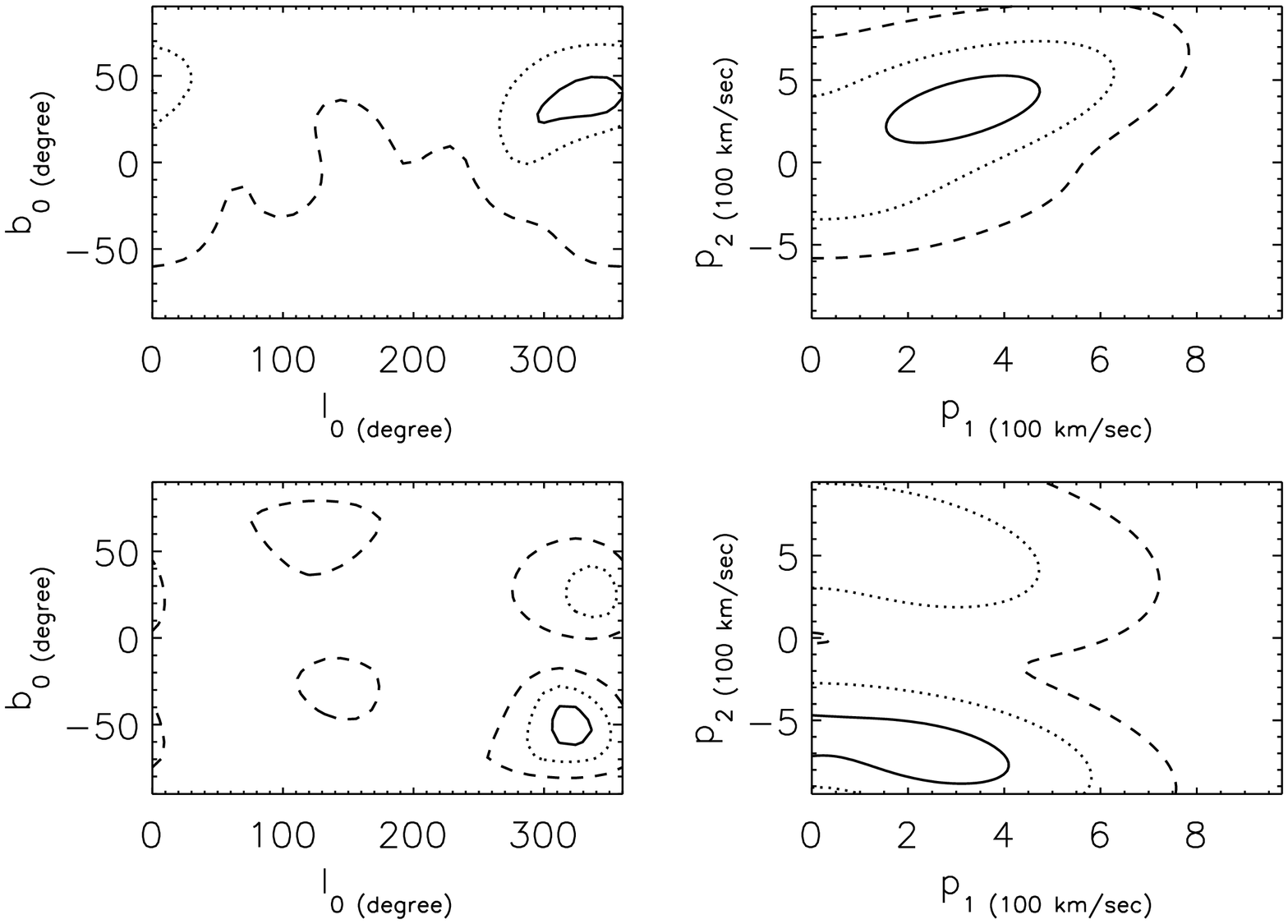}

\caption{Same as Figure~\ref{Fig:fits12mlm}, but for the CMAGIC method.
}
\label{Fig:fits12cmagic}

\end{figure}

\clearpage

In the reference frame where the polar direction is the symmetry axis, the model of the velocity field is then
\begin{equation}
V_{pec} \ =  \ p_1\sin{b'} + p_2 (3\sin^2{b'}-1)/2 
\label{Eq:quad}
\end{equation}
where $b' $  is the latitude in the rotated reference frame and has a value in the range from $-\pi/2$ to $\pi/2$. Just like the dipole model, the dipole plus quadruple model given in Equation~(\ref{Eq:quad})  has also four free parameters: $p_1$, $p_2$, and the direction of the symmetry axis $(l_0, b_0)$. 

The results of the fits are shown in Figures~\ref{Fig:fits12mlm}, \ref{Fig:fits12cmagic}, and in Table~\ref{Tab:fits}. As the figures show, the apex direction is at $(l_0,b_0)\ \sim \ (312^{\rm o}, 24^{\rm o})$ for the full sample using the CMAGIC method. The MLM method gives results that are consistent with the CMAGIC method for the full sample, but with larger errors. We found also that for Sample $A$, the flow can be equally well described by a pure dipole ($p_2\ = \ 0$) or a pure quadruple ($p_1 \ = \ 0$) with the CMAGIC method, but the significances of both the dipole and the quadruple field are less than $2-\sigma$ for both MLM and CMAGIC. For the intermediate redshift sample $B$, CMAGIC and MLM again give results that are in remarkable agreement; both methods show 
appreciable $\chi^2$ improvement as compared to the pure dipole model. For sample $C$, only the CMAGIC method is able to reveal the smooth component of the peculiar velocity field. At less than $2-\sigma$ level, the flow of sample $C$ shares the same apex direction and flow velocity as of sample $A$ and $B$, although the best fit models are significantly different. The sample $C$ of MLM method is unable to provide constraints on the flow field due to the larger associated errors.

 \subsection{A Bipolar Flow Field}
 
 It is found from \S\ref{S:dip} and \S\ref{S:quad} that there exists a coherent 
smooth streaming flow which requires a quadruple component. The exact functional form 
of this flow is unknown and the parametrization of the flow can be qualitatively 
different based on different model assumptions. For example, at $z$ below 0.01, a quadruple can provide 
an equally good fit as a dipole, but with the apex direction different. This is 
obviously a severe problem imposed by the lack of SNe in the zone of avoidance. Given 
this uncertainty,  we explored further another functional form
\begin{equation}
   V_{pec} \ = \ p_1 \cos [2(b'+p_2)],
 \label{Eq:mquad}
\end{equation}
where the parameter $p_2$ is a phase offset such that when it is equal to zero the 
above equation 
gives a pure quadruple flow. The above equation reduces to a pure quadruple field 
when $p_2\ = \ 0$.

\begin{figure}
\epsscale{1.}
\includegraphics[angle=90,scale=0.45]{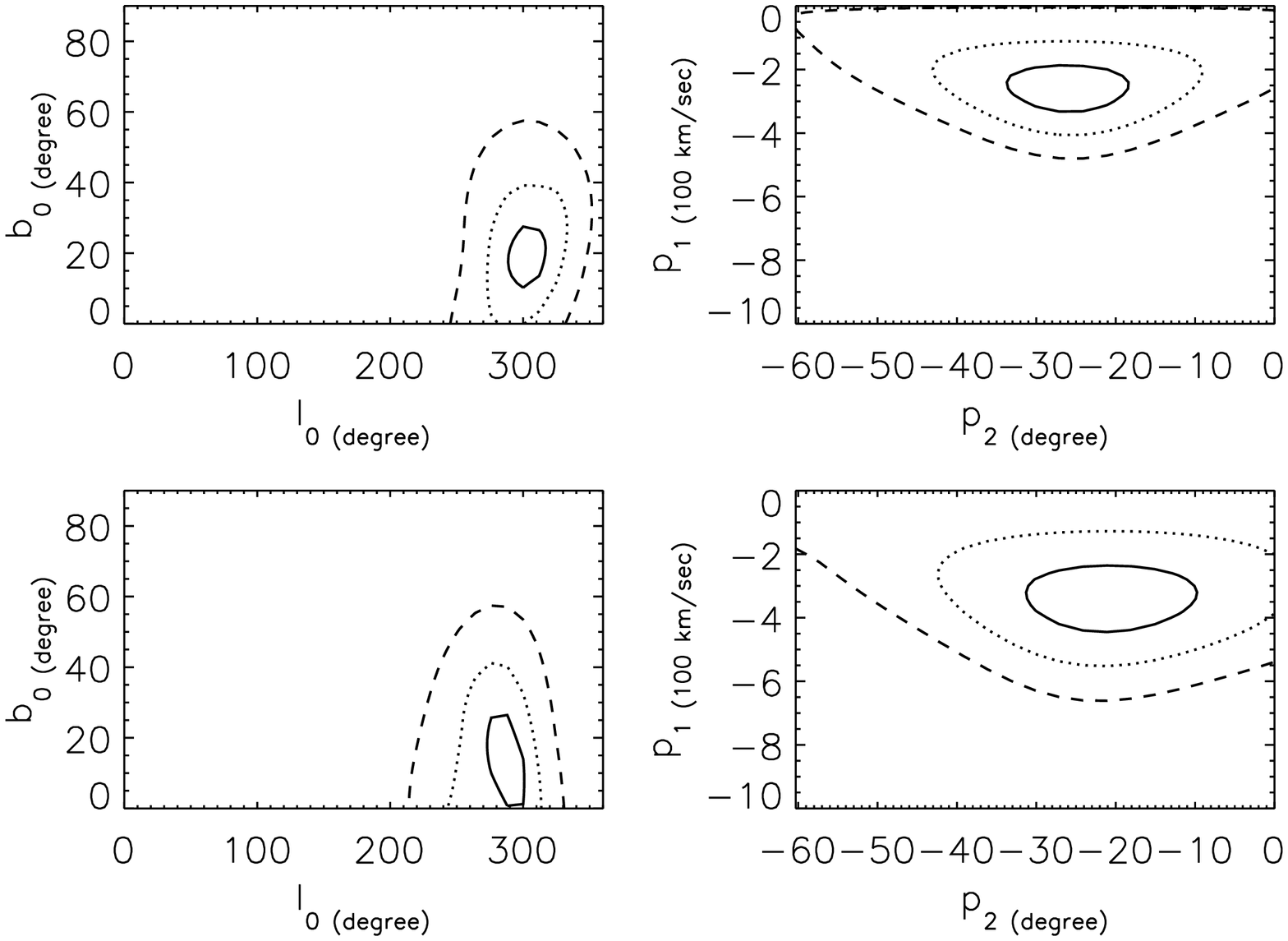}\\
\includegraphics[angle=90,scale=0.45]{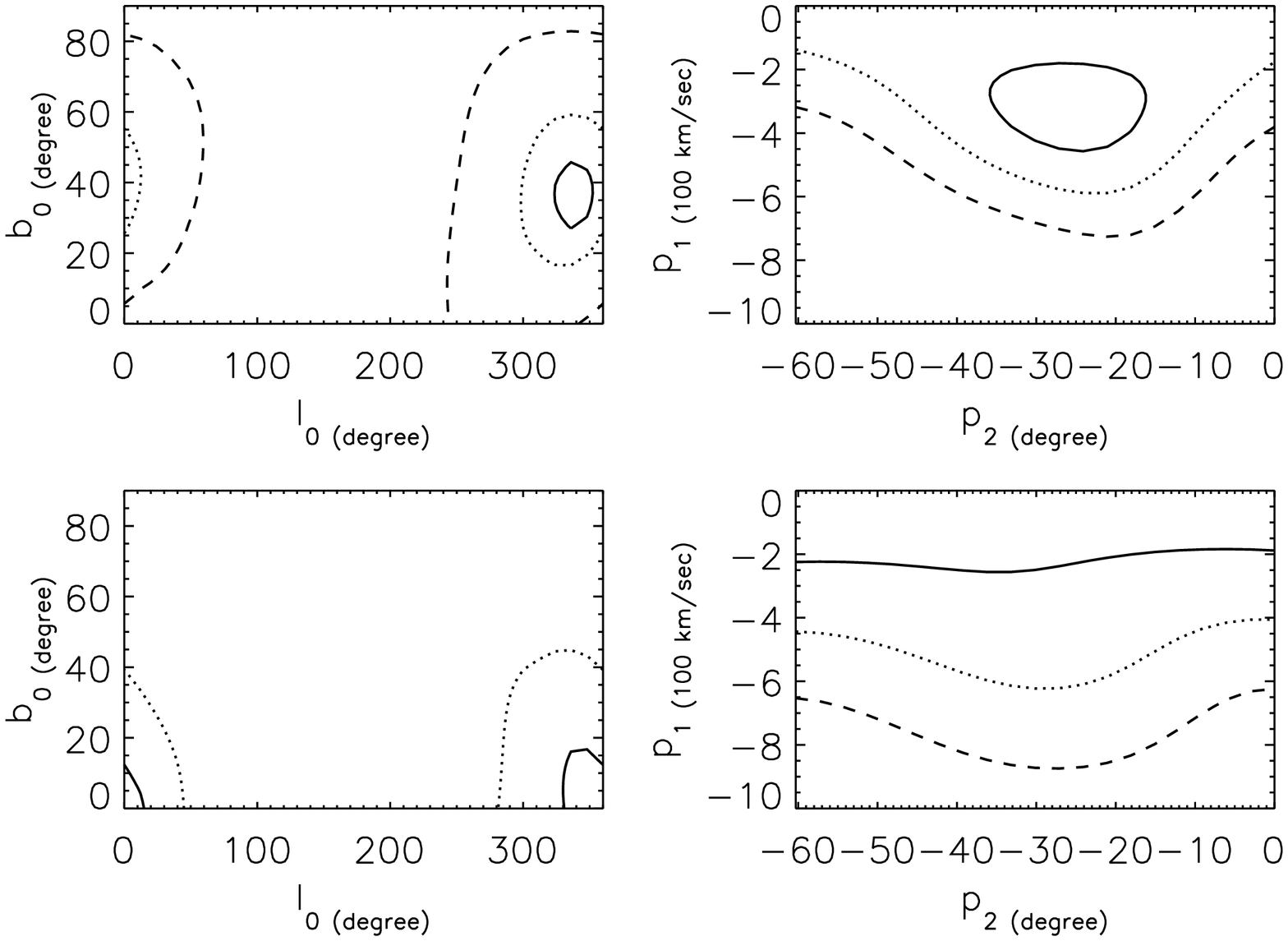}

\caption{The error contours of the parameters for Equation~\ref{Eq:mquad} using the MLM of SNIa. The left column 
shows the axis of symmetry $(l_0, b_0)$, and the right column shows the fitted coefficients of 
the velocity $p_1$ and the phase offset $p_2$. From the top to the bottom, the panels show the fits 
to the entire 
sample, sample $A$, sample $B$, and sample $C$. For the contours of $(l_0, b_0)$, the probability 
distributions are marginalized over $p_1$ and $p_2$; for those of $p_1$ and $p_2$, they are 
marginalized over $l_0$ and $b_0$. The $\chi^2$/DoF were normalized to unity to make these contours.
The solid, dotted, and dashed lines show 1-, 2-, and 3-$\sigma$ confidence levels.
}
\label{Fig:mqmlm}
\end{figure}

\begin{figure}
\epsscale{1.}
\includegraphics[angle=90,scale=0.45]{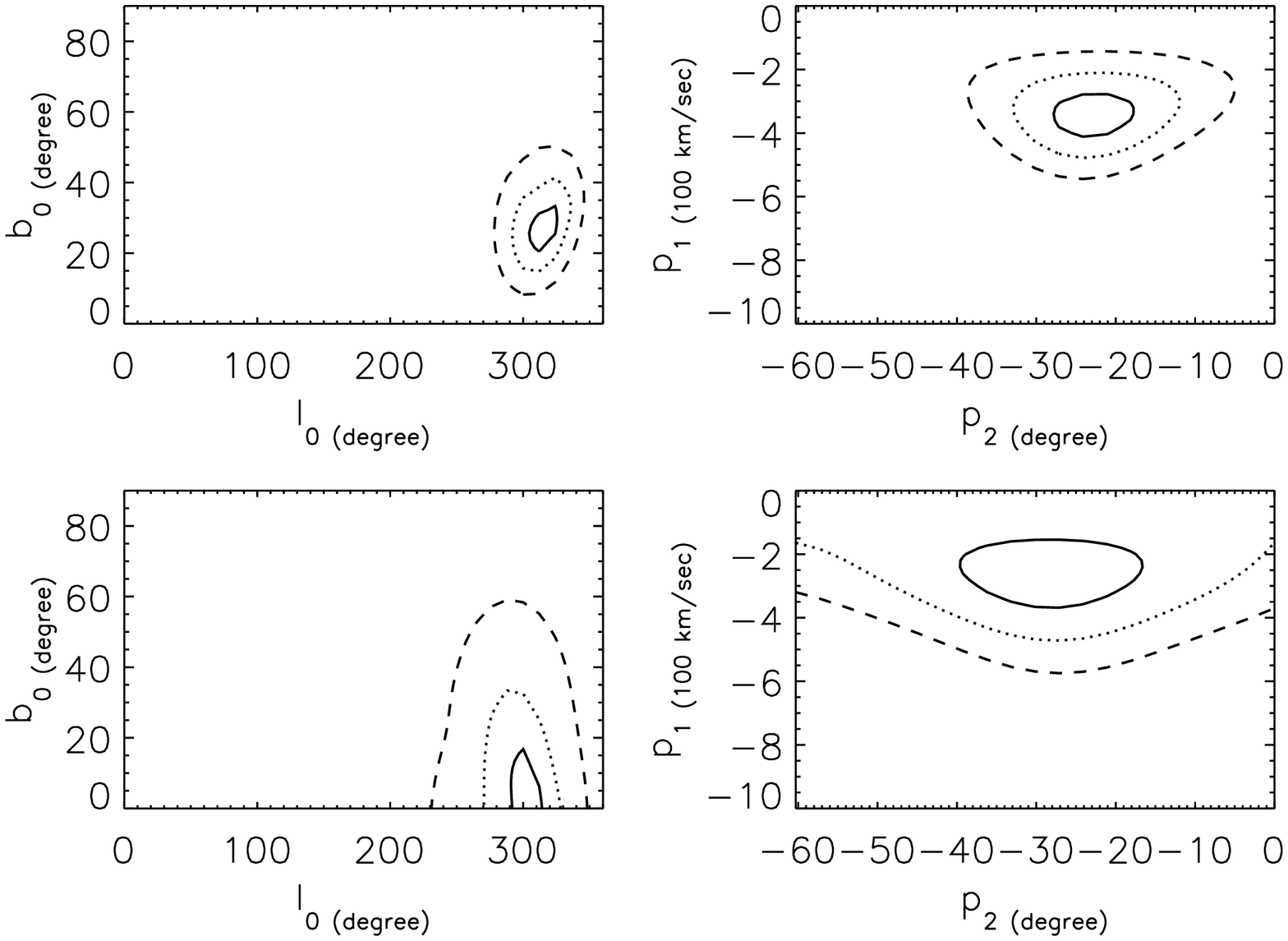}\\
\includegraphics[angle=90,scale=0.45]{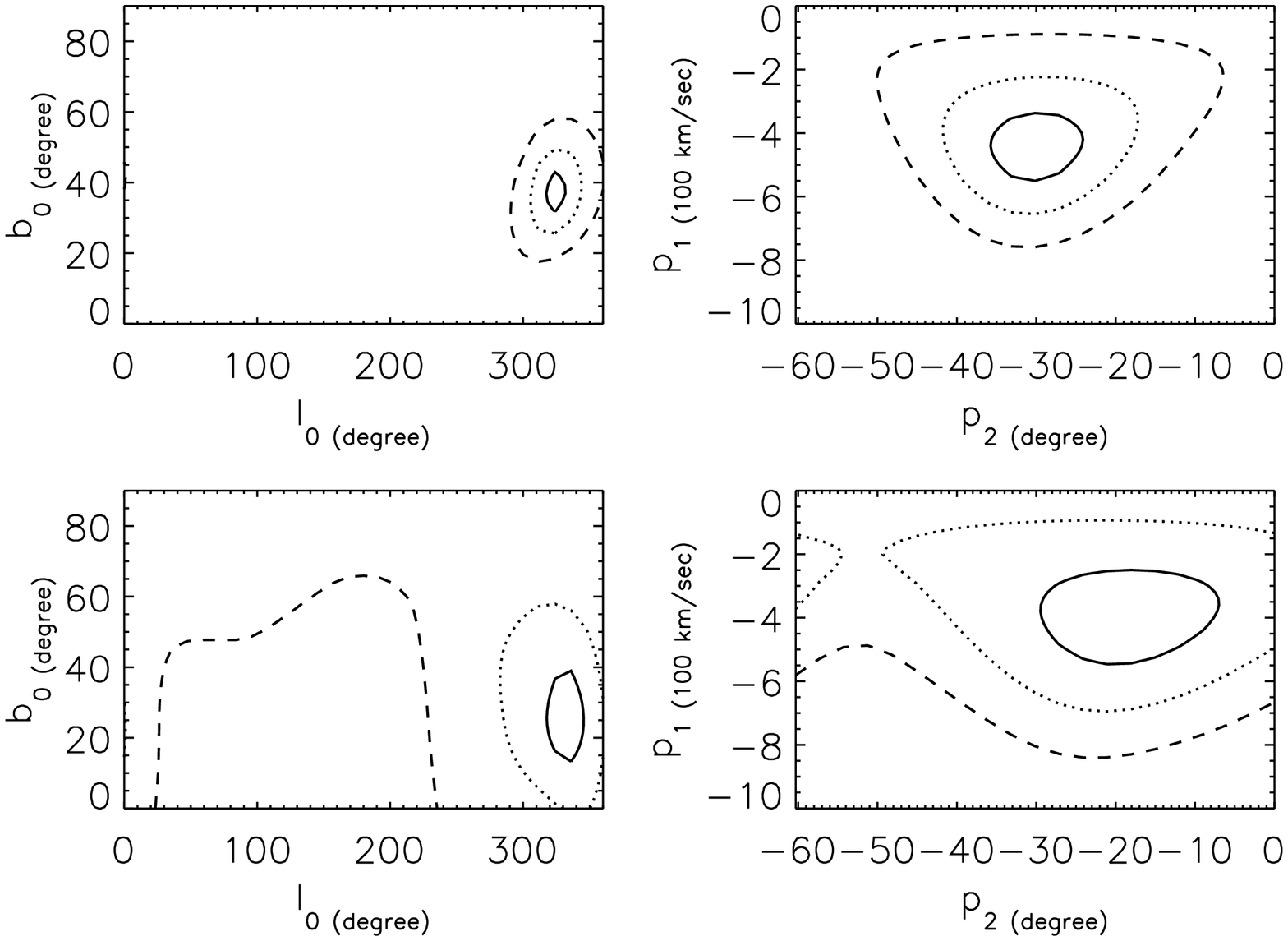}

\caption{Same as Figure~\ref{Fig:mqmlm}, but for the CMAGIC method and equation \ref{Eq:mquad}.
It is seen from these figures that the peculiar velocity fields for samples $A$, $B$, and $C$ share the same symmetry axis.}
\label{Fig:mqcmagic}
\end{figure}

The marginalized error contours of the fits are shown in Figure~\ref{Fig:mqmlm} 
and \ref{Fig:mqcmagic}, for MLM and CMAGIC, repsectively. 
The optimal fit to the CMAGIC distances gives 
\begin{equation}
V_{pec} \ = \ -340.0\{^{+62.4}_{-71.3}\}\cos[2(b'+24^{\rm o}.1\{^{+6.4}_{-3.7})\}]\  {\rm km/sec}
\label{Eq:mquad0}
\end{equation}
 for the full sample, where the numbers inside the curly parentheses are the errors 
of the preceding numbers. To within 2-$\sigma$, the above equation, deduced from the 
full sample, applies also to all the sub-samples $A$, $B$, and $C$ and to  both MLM and 
CMAGIC. By removing this 
velocity field, the reduction of $\chi^{2}$ is from  90 to 62 for the CMAGIC Hubble diagram. The same operation with MLM results in a reduction in $\chi^{2}$ from 78 to 63. Both fits are statistically significant.

\begin{figure}[tbh]
\epsscale{1.}
\includegraphics[angle=90,scale=0.3]{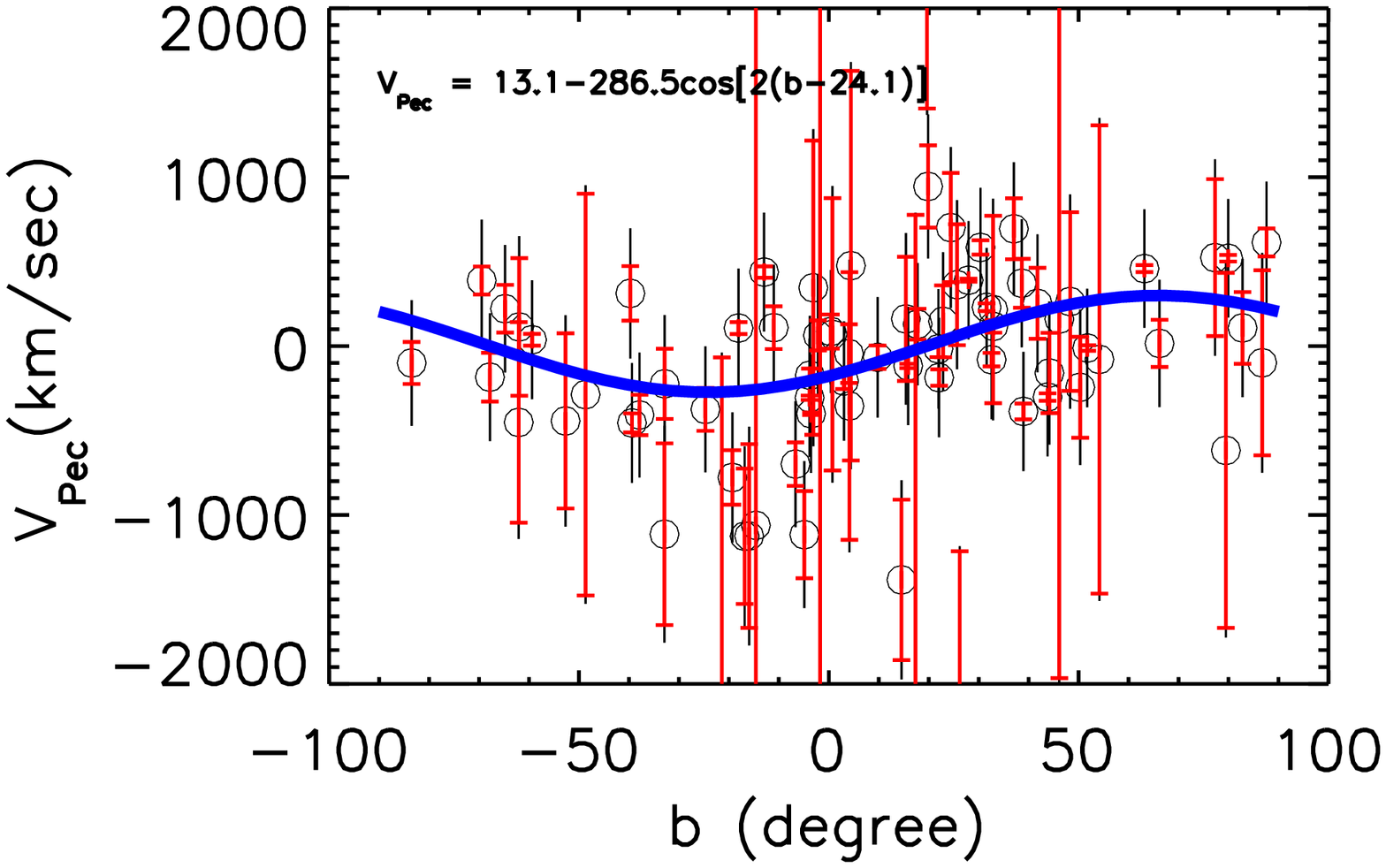}
\includegraphics[angle=90,scale=0.3]{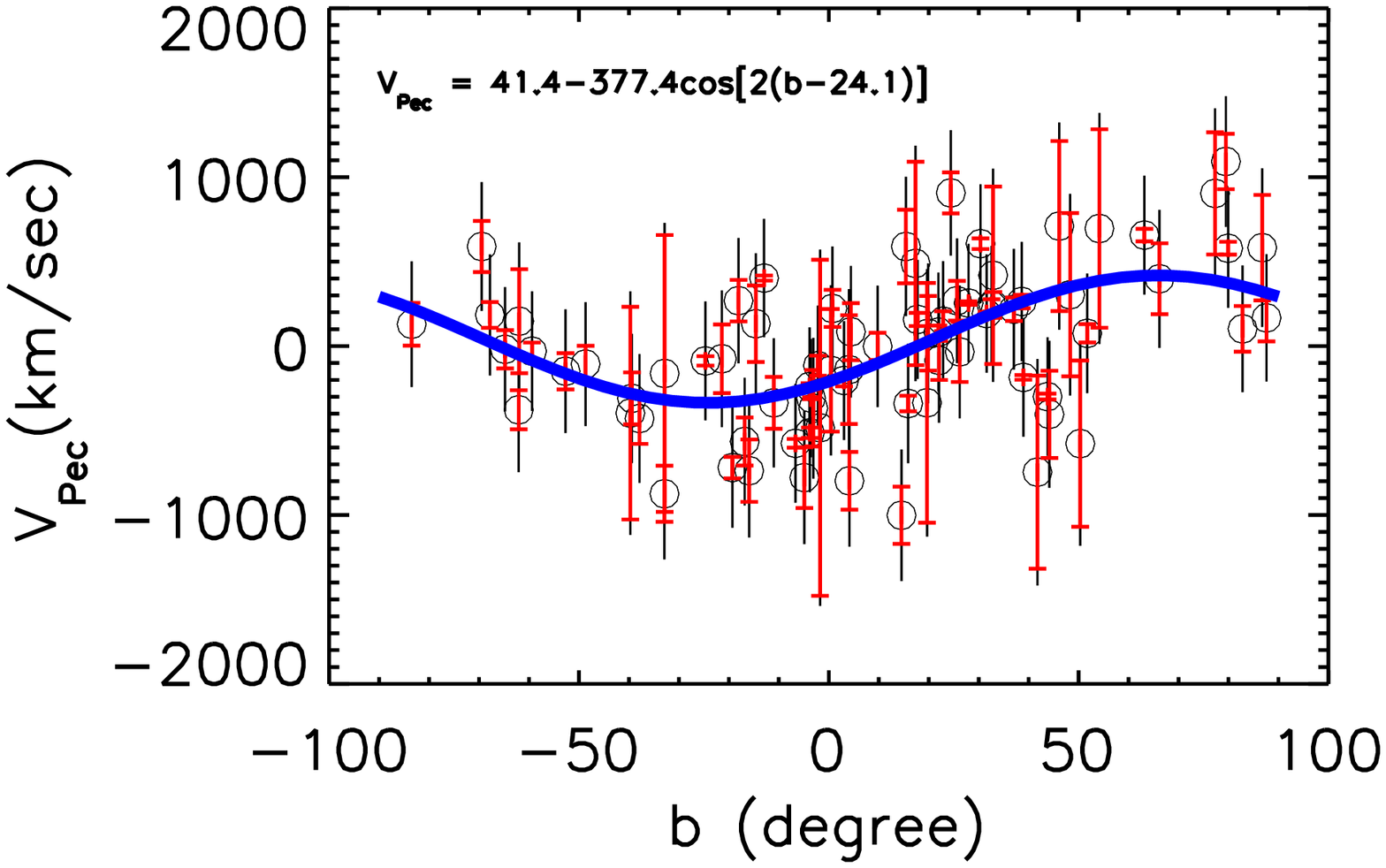}\\
\includegraphics[angle=90,scale=0.3]{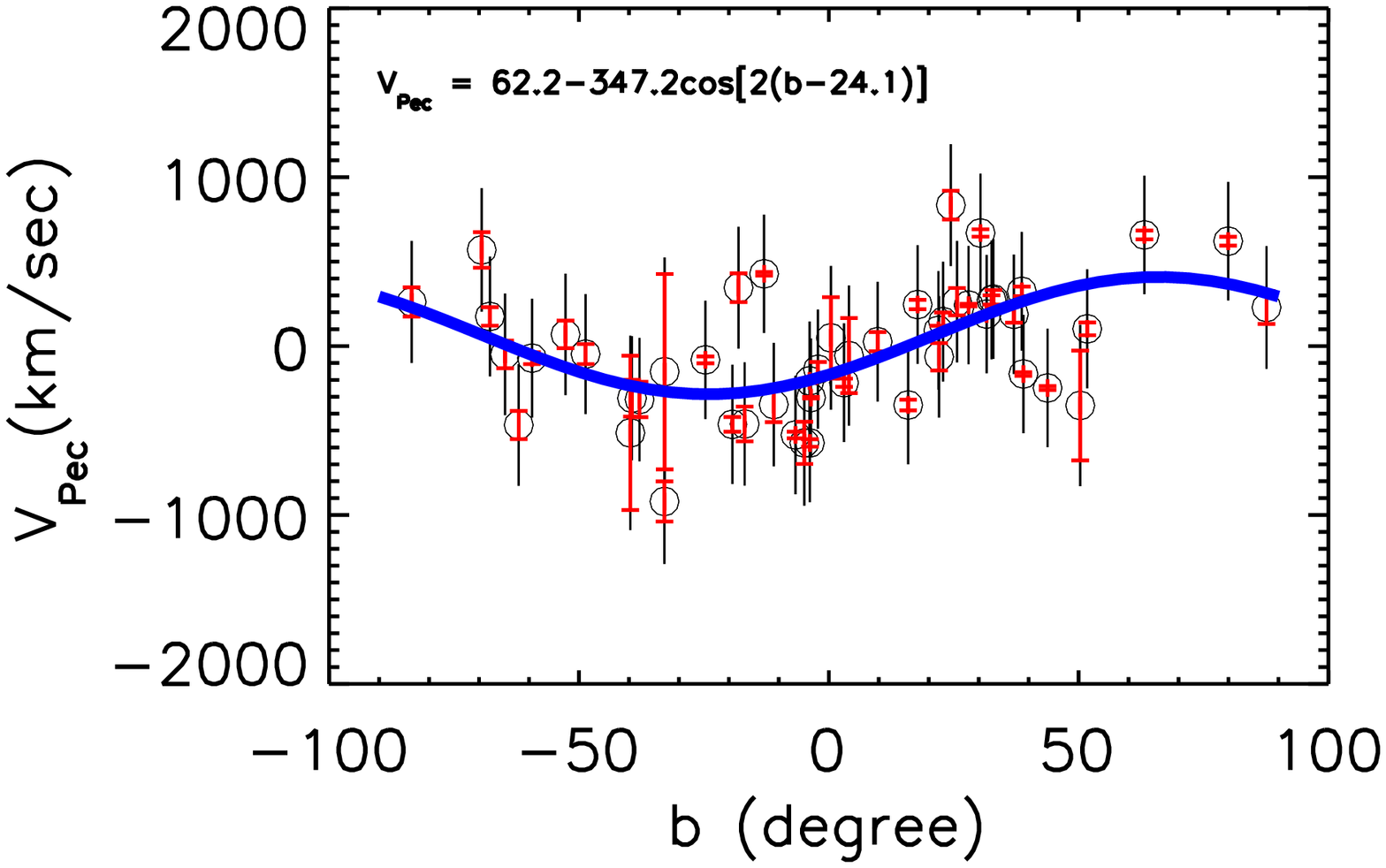}
\includegraphics[angle=90,scale=0.3]{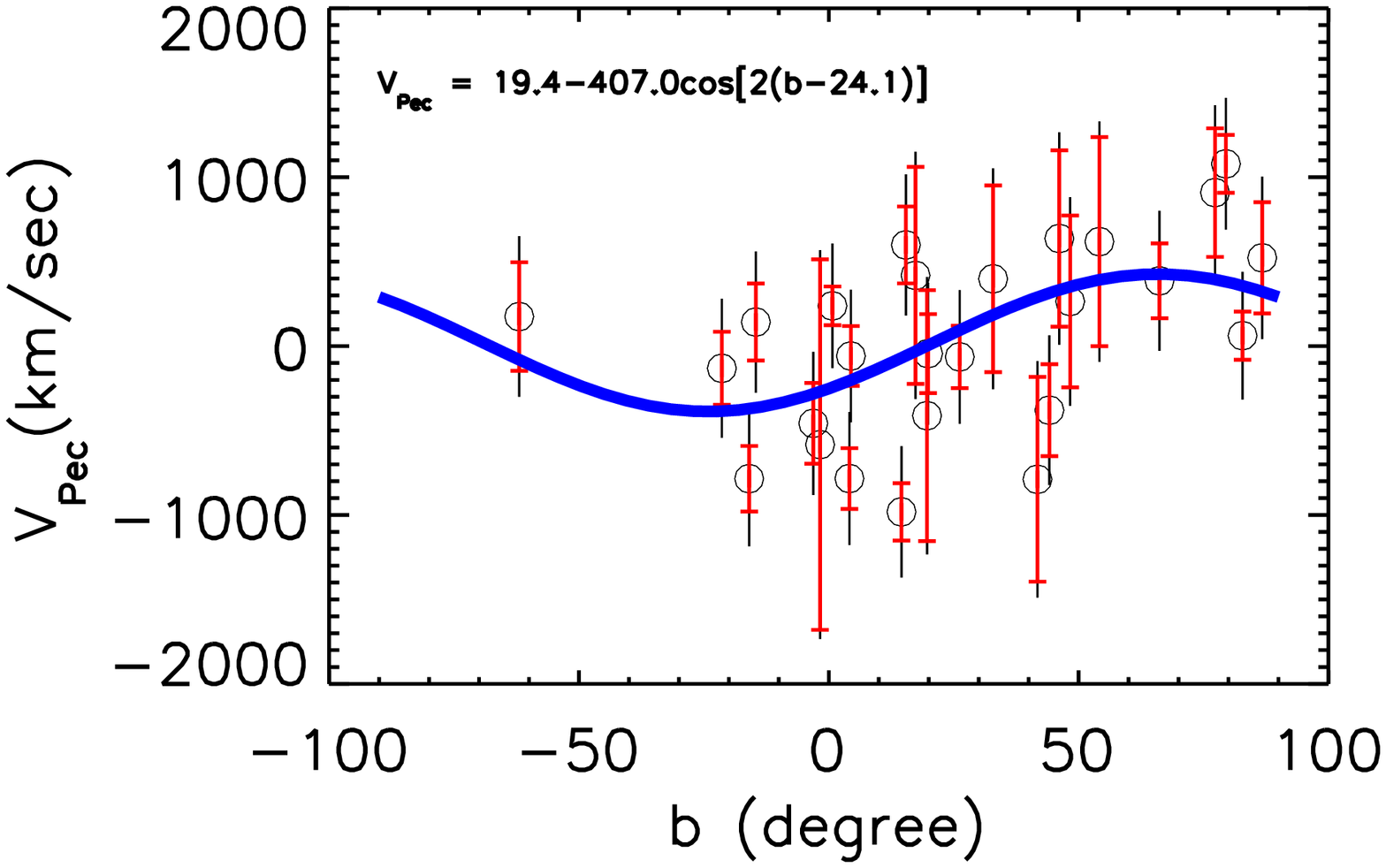}

\caption{
The residuals plotted against $b'$, the latitude measured with the polar direction at $(l_0,b_0)\ = \ (312^{\rm o}.0, 25^{\rm o}.7)$. The smooth solid lines show the fit $V_{pec} = 312.0 \cos[2(b'+24^{\rm o}.1)]$ km/sec,  as deduced from the $\chi^2$ fit to the peculiar velocity field of the full sample using the CMAGIC method. 
(a, upper left panel) shows the residual for the full sample of the conventional MLM method, (b, upper right panel) 
shows the same figure for the CMAGIC method. The residuals for the CMAGIC method using 
SNe below z = 0.025 and above z = 0.025 are shown in (c, lower left) and (d, lower right), respectively. No significant 
differences are identified in these fits, suggesting that the quadruple field extends at least 
beyond z = 0.025, possibly to z = 0.05. The hatted errors on the data points include only the 
observational errors, whereas the unhatted error bars include a 350 km/sec thermal peculiar velocity.}
\label{Fig:rotated0}
\end{figure}

\begin{figure}[tbh]
\epsscale{1.}
\includegraphics[angle=90,scale=0.3]{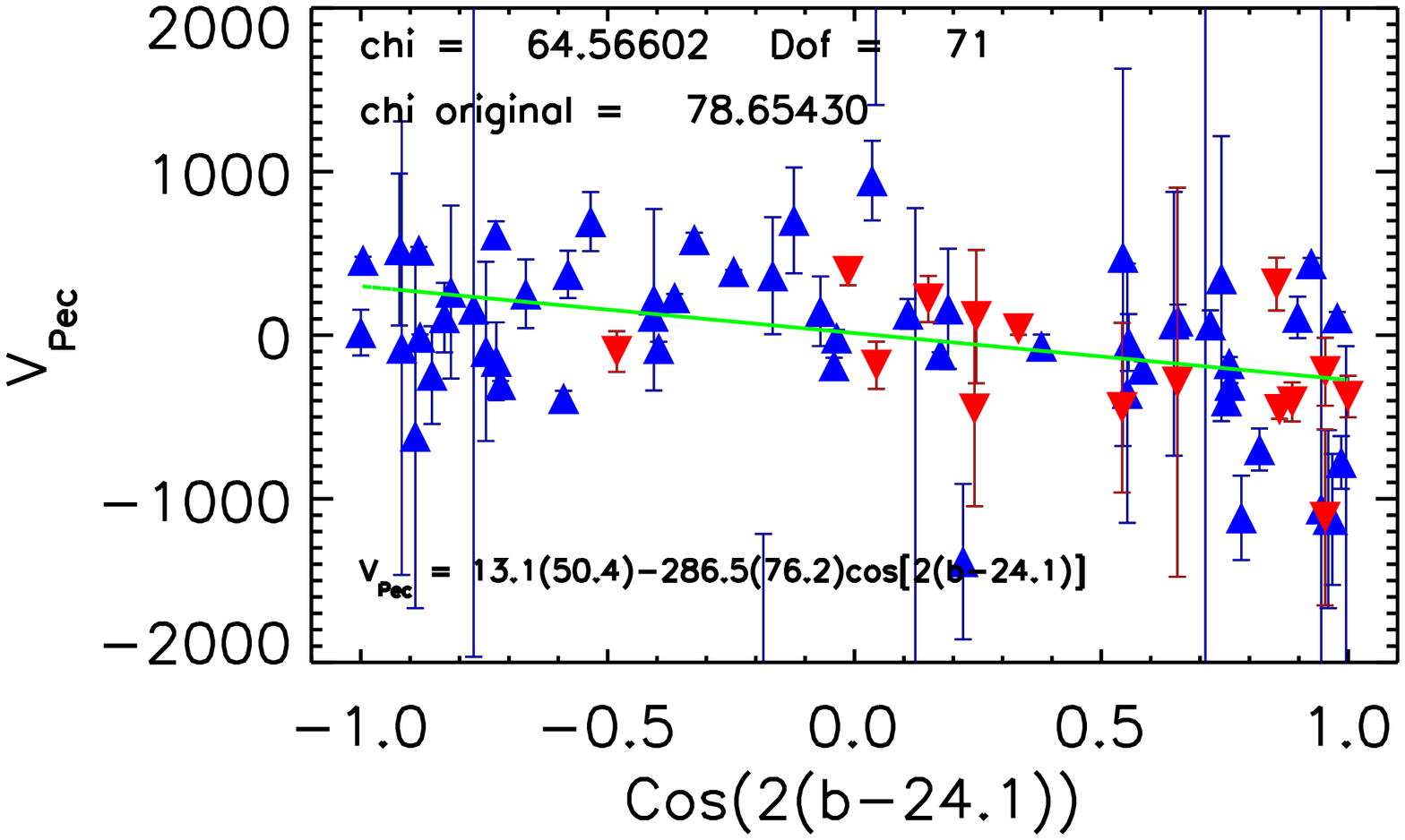}
\includegraphics[angle=90,scale=0.3]{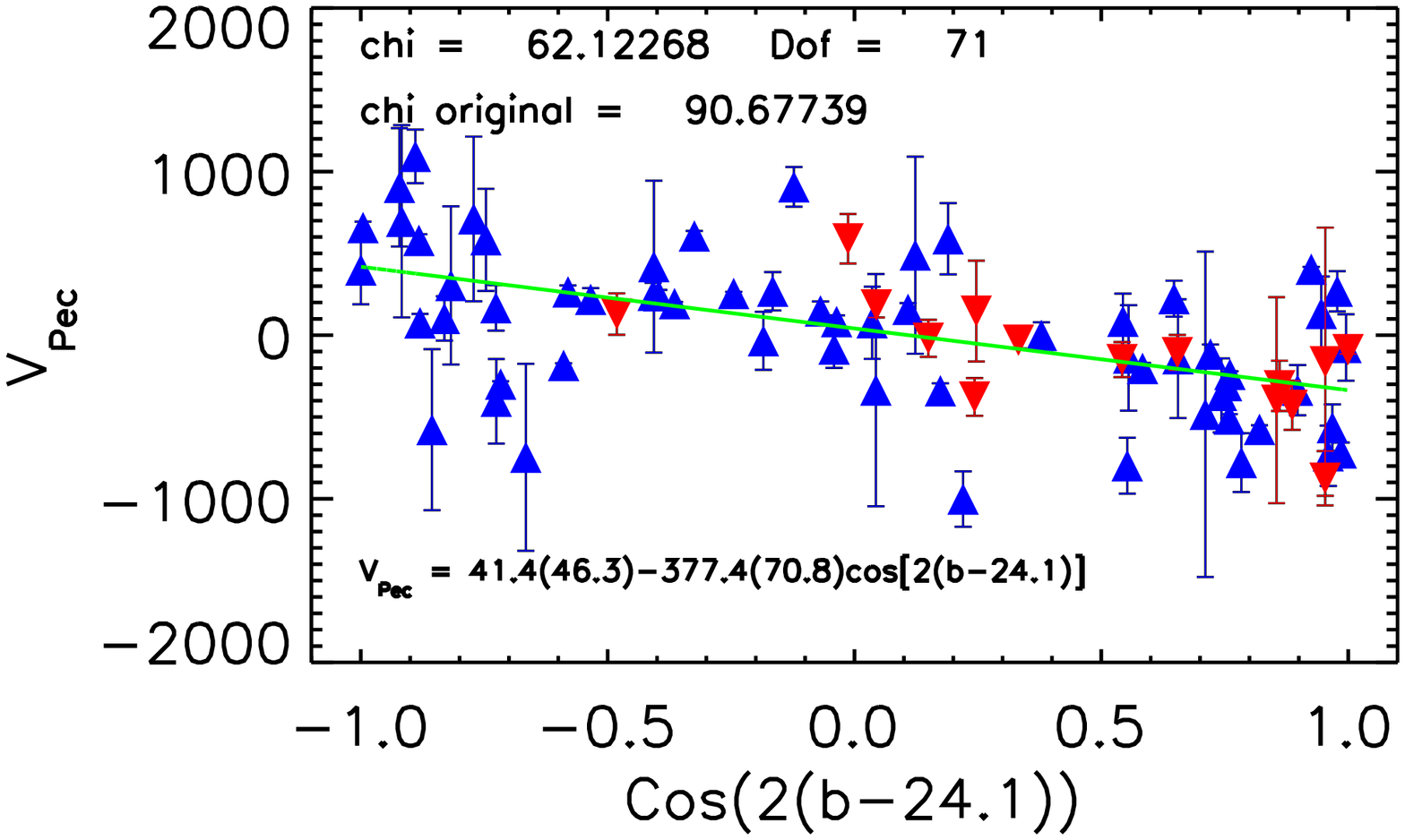}\\
\includegraphics[angle=90,scale=0.3]{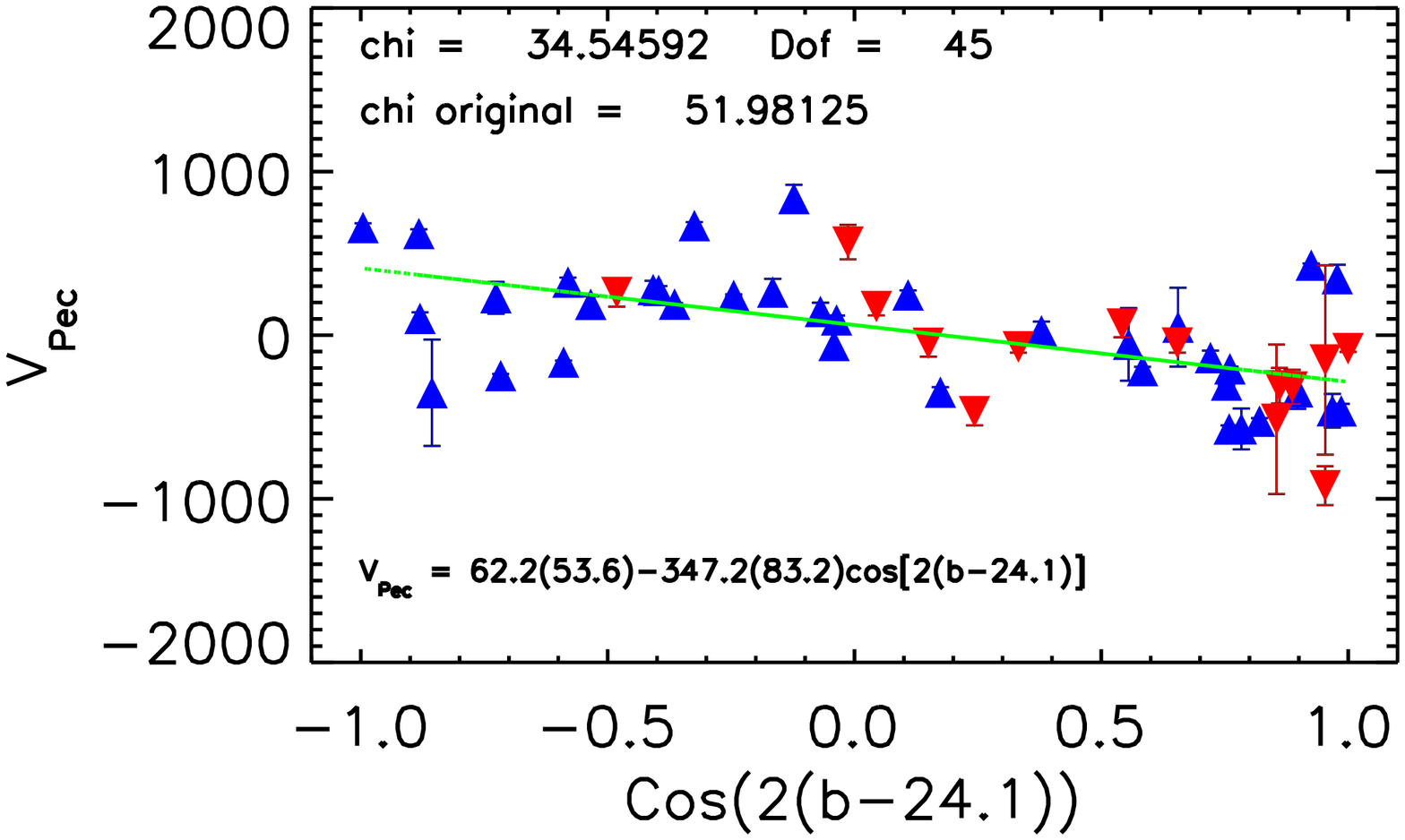}
\includegraphics[angle=90,scale=0.3]{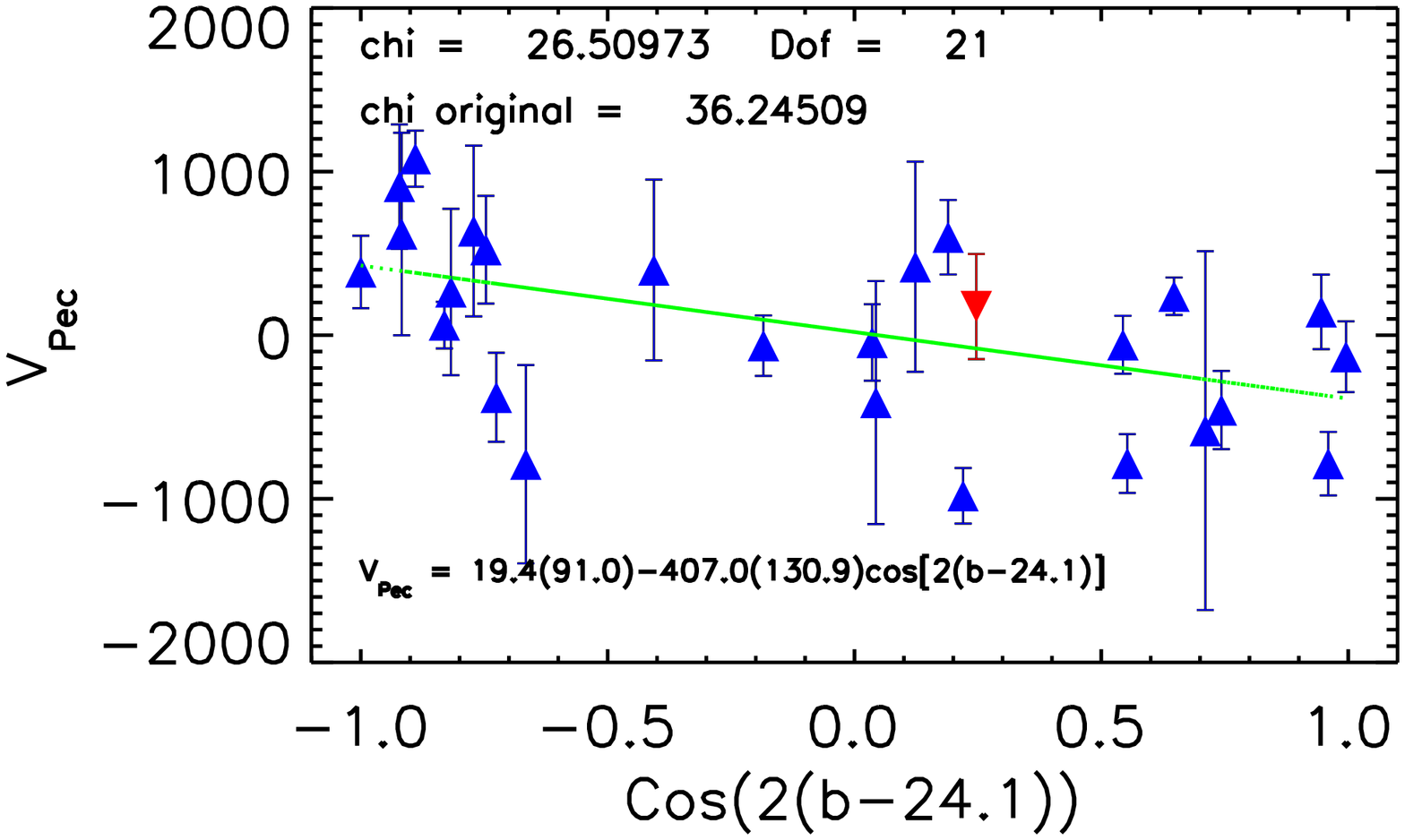}

\caption{
The same as Figure~\ref{Fig:rotated0}, but the horizontal axis is now $\cos[2(b'+p_2)]$. The data points are expected to follow a straight line with the adopted model. The upwards triangles are for data with $b'\ \ge\ -24^{\rm o}.1$, the downward triangles are for data with
$b'\ <\ -24^{\rm o}.1$.}
\label{Fig:rotated}
\end{figure}

These model fits can be better illustrated in figures with the peculiar velocity 
plotted versus  the latitude in the reference frame with the polar direction rotated to 
$(l_0, b_0)$. To a good approximation, we can choose the direction 
$(l_0,\ b_0)\ = \  (312^{\rm o}.0, 25^{\rm o}.7) $ to be 
the same for all the sub-samples. The fitting can be particularly simple as the a linear fit can be employed if the dependent variable is taken to be $\cos[2(b'+p_2)]$. The results are shown in 
Figures~\ref{Fig:rotated0} and \ref{Fig:rotated}. Here we have again allowed for a monopole component, so the fitting formula is $V_{pec} \ = \ c_0 + c_1\cos[2(b'+p_2)]$. The fitted parameters are shown in Figures~\ref{Fig:rotated0} and \ref{Fig:rotated} and are tabulated in Table~\ref{Tab:fitsbip}. The MLM method and CMAGIC give overall consistent results, but the CMAGIC method 
shows much stronger signal especially at $z$ above 0.025.  These figures illustrate the need for 
 more SNe at redshift above 0.025 to determine precisely the smooth peculiar velocity 
component and its redshift space extension. In the rotated reference frame, 
the peculiar velocity can also be plotted as a function of 
$\cos[2(b'+p_2)]$ (see Figure~\ref{Fig:rotated}). A linear relation is expected if 
Equation~(\ref{Eq:mquad}) is an 
acceptable description of the velocity field. This is indeed the case as shown in Figure~\ref{Fig:rotated}. Figure~\ref{Fig:rotated0} (c) and (d) shows the peculiar velocities 
from the CMAGIC method for SNe above and 
below z = 0.025, respectively.
These figures illustrate again that the local streaming motion extends very deep in 
redshift space, 
to at least z = 0.025 (105 Mpc/$h_{71}$) and, to a lower significance level, even 
the whole data set.

Sample $C$ has only one supernovae at $b'$ below $-30^{\rm o}$. It would be very important to fill up 
this region. However, it is remarkable that the data points do agree well with an expansion law that 
describes reasonably well sample $A$ and $B$ (see Figure~\ref{Fig:rotated0}(d) and Figure~\ref{Fig:rotated}(d)). This is 
an extraordinary property as the distance scales covered by these SNe is nearly 500 Mpc in diameter. 

\begin{deluxetable}{rrrrr|rrrrr}
\tabletypesize{\scriptsize}\tablecolumns{9}\tablewidth{0pt}\tablecaption{Fixed Axis Axially Symmetric Fits to the Peculiar Velocity Field}
\tablehead{
\colhead{} & \multicolumn{4}{c}{MLM} & \multicolumn{4}{c}{CMAGIC}  & \colhead{} \nl
\cline{2-5} \cline{6-9} \\
\colhead{z range}                          & \colhead{$c_0\ (\sigma)$\tablenotemark{*}}                    & \colhead{$c_1\ (\sigma)$\tablenotemark{*}}                    &\colhead{$\chi^2_o$\tablenotemark{\dag}}                  & \colhead{$\chi^2$\tablenotemark{\ddag}}                     & \colhead{$c_0\ (\sigma)$\tablenotemark{*}}                    & \colhead{$c_1\ (\sigma)$\tablenotemark{*}}                    &\colhead{$\chi^2_o$\tablenotemark{\dag}}                    & \colhead{$\chi^2$\tablenotemark{\ddag}}                    & \colhead{DoF}                \cr}
\startdata
 0 - 0.08 & 13.1 (50.4)  & -286.5 (76.2 )& 78.7 & 64.6 & 41.4 (46.3) & -377.4 (70.8) & 90.7 & 62.1 & 71 \\
 0 - 0.025 & 40.8 (55.1) & -326.1 (84.2) & 49.3 & 34.3 & 62.2 (53.4) & -347.2 (83.7) & 52.0 & 34.5 & 45 \\
 0.025 - 0.08 & -05.4 (177.5) & -145.5 (253.4) & 18.1 & 17.6 & 19.4 (91.0)    & -407 (130.9)     & 36.2 & 26.5 & 21 \\
\enddata
\tablenotetext{\dag, \ddag}{$\chi_o^2$ and $\chi$ are $\chi^2$ before and after the fits.}
\tablenotetext{*}{In units of km/sec.}
\label{Tab:fitsbip}
\end{deluxetable}

\clearpage

From Table~\ref{Tab:fitsbip}, we again did not find any significant monopole component at redshift above and below 0.05,  when the symmetry axis is fixed to $(l_0,b_0)\ = \ (312^{\rm o}.0, 25^{\rm o}.7)$.

\subsection{Monte-Carlo Test of the Significance of the Streaming Velocity Field}

\begin{figure}[tbh]
\epsscale{1.}
\includegraphics[angle=00,scale=0.8]{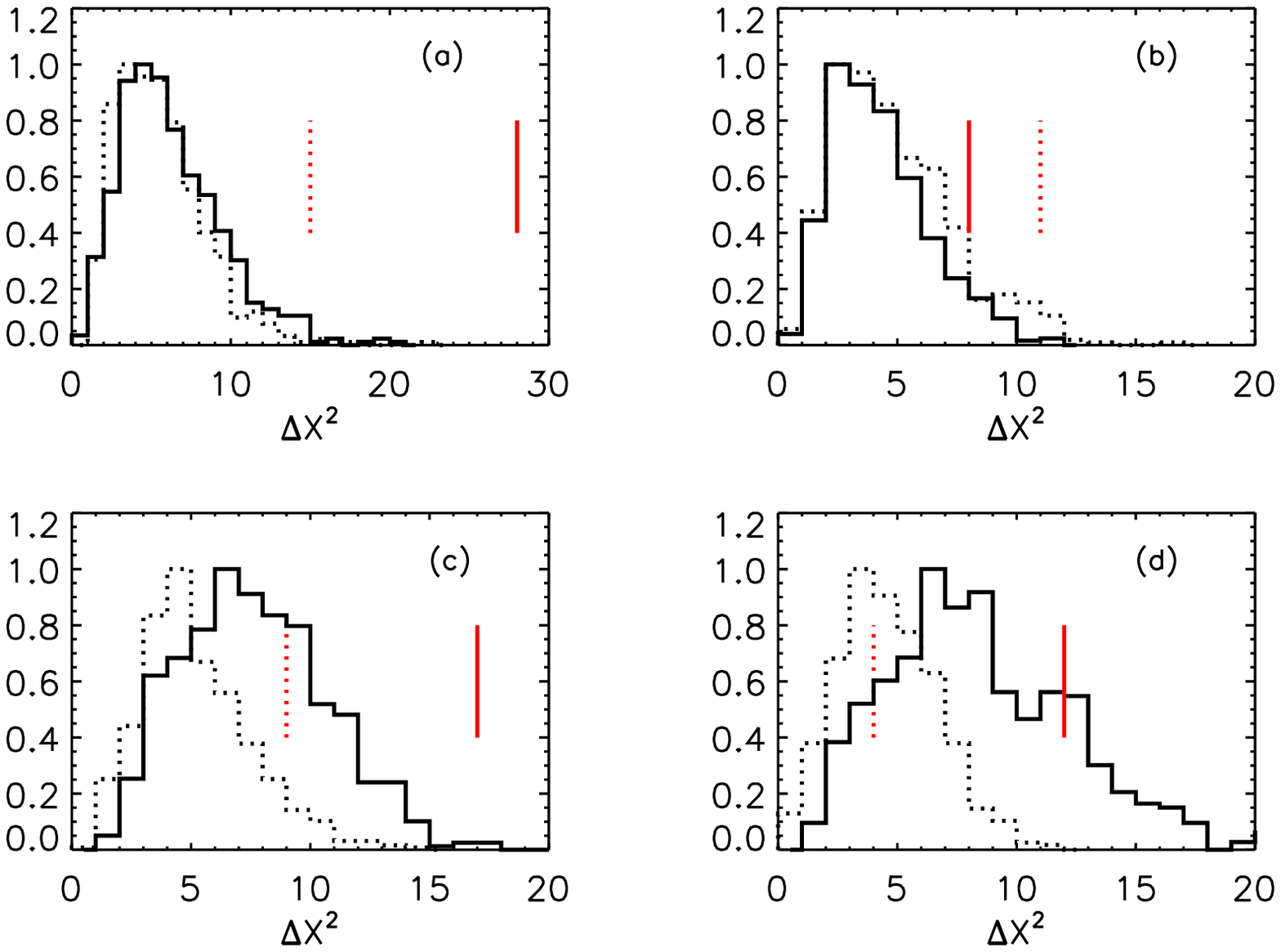}
\caption{Monte-Carlo simulations of the significance of the streaming velocity field. The figures show the histogram normalized to peak value of 1 of $\chi^2$ reductions from the simulation. The solid lines show the results for CMAGIC, and the dotted lines for MLM. The vertical straight lines show the $\chi^2$ reduction by fitting a streaming pattern as given in equation~\ref{Eq:mquad} to the MLM data (dotted) and CMAGIC data (solid).  From top left clockwise, the panels show the results for the entire sample, sample $A$, $B$, and $C$, respectively. A total of 1000 Monte-Carlo runs were performed for the full sample and each sub-sample. }
\label{Fig:mc}
\end{figure}

The uneven distribution of SNe on the sky is a worry of systematic errors that may lead to a false streaming velocity pattern.  Extensive Monte-Carlo test is employed to assess the robustness of the streaming velocity field.  We use the reduction of the $\chi^2$ as a measure of the goodness  of the streaming pattern. In each Monte-Carlo realization, we keep the redshift, and $(l, b)$ coordinates of each SN the same as were observed, but randomly assign a new $V_{pec}$ drawn from the observed peculiar velocities, with no repetition. The Monte-Carlo sample is fitted by the assumed streaming velocity models (Equations~(\ref{Eq:dip}), (\ref{Eq:quad}), and (\ref{Eq:mquad})) and the reduction of $\chi^2$ is recorded. Figure~\ref{Fig:mc} shows the results for the streaming motion model given in equation~(\ref{Eq:mquad}). 

From Figure~\ref{Fig:mc}, we can see that the streaming pattern is significant at larger than 99\% level for the full sample and for sample $C$, but is moderately significant at above 1 to 2-$\sigma$ level for the individual samples $A$ and $B$. 

Note that the Monte-Carlo test were done independently for different sub-samples, so the joint probability of having a streaming motion pointing in the same direction by the various Monte-Carlo sample is significantly lower. This suggests that the observed streaming velocity field is highly significant.

To test the effect of the uneven sampling on the errors of the parameter estimates, we have also applied bootstrapping simulations to study the parameter distribution. Only parameter space around the optimal fits given in Table~\ref{Tab:fits} were searched for in the Monte-Carlo simulation. In this approach, the data are resampled 2000 times by redrawing SNe randomly from the original data set, allowing for repetition. Each sample is analyzed in the same way as the original observations. This generates 2000 Monte-Carlo realizations of the original observations. We have carried through the error terms in the bootstrap simulation; thus for each Monte-Carlo realization the resulting model fits have different errors associated with them. To calculate the bootstrap confidence level, we have weighted the fitting parameters by the inverse-squared of their corresponding errors to construct their probability distribution. The resulting constraints on model parameters are found to be in broad agreement with those from the $\chi^{2}$ fits shown in Table~\ref{Tab:fits}. The bootstrap test suggests that the smooth streaming pattern is significant at more than 3-$\sigma$ level. The monopole term is consistent with zero from these analyses. This is in contradiction to the monopole field propose in \cite{Zahavi:1998, Jha:2007}, but is consistent with the conclusions of the previous sections of this paper.

   \begin{figure}[tbh]

\epsscale{1.}
\includegraphics[angle=90,scale=0.3]{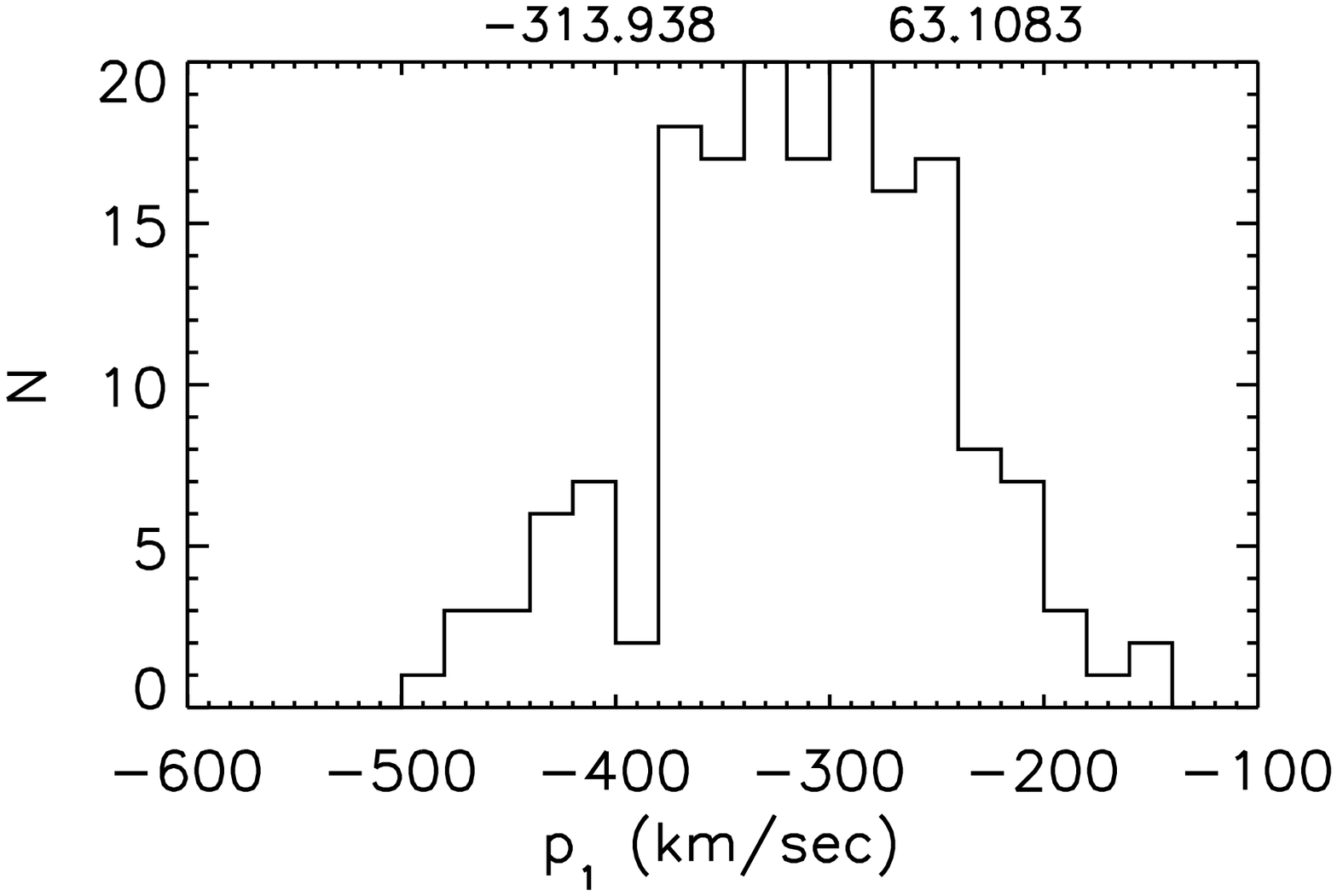}
\includegraphics[angle=90,scale=0.3]{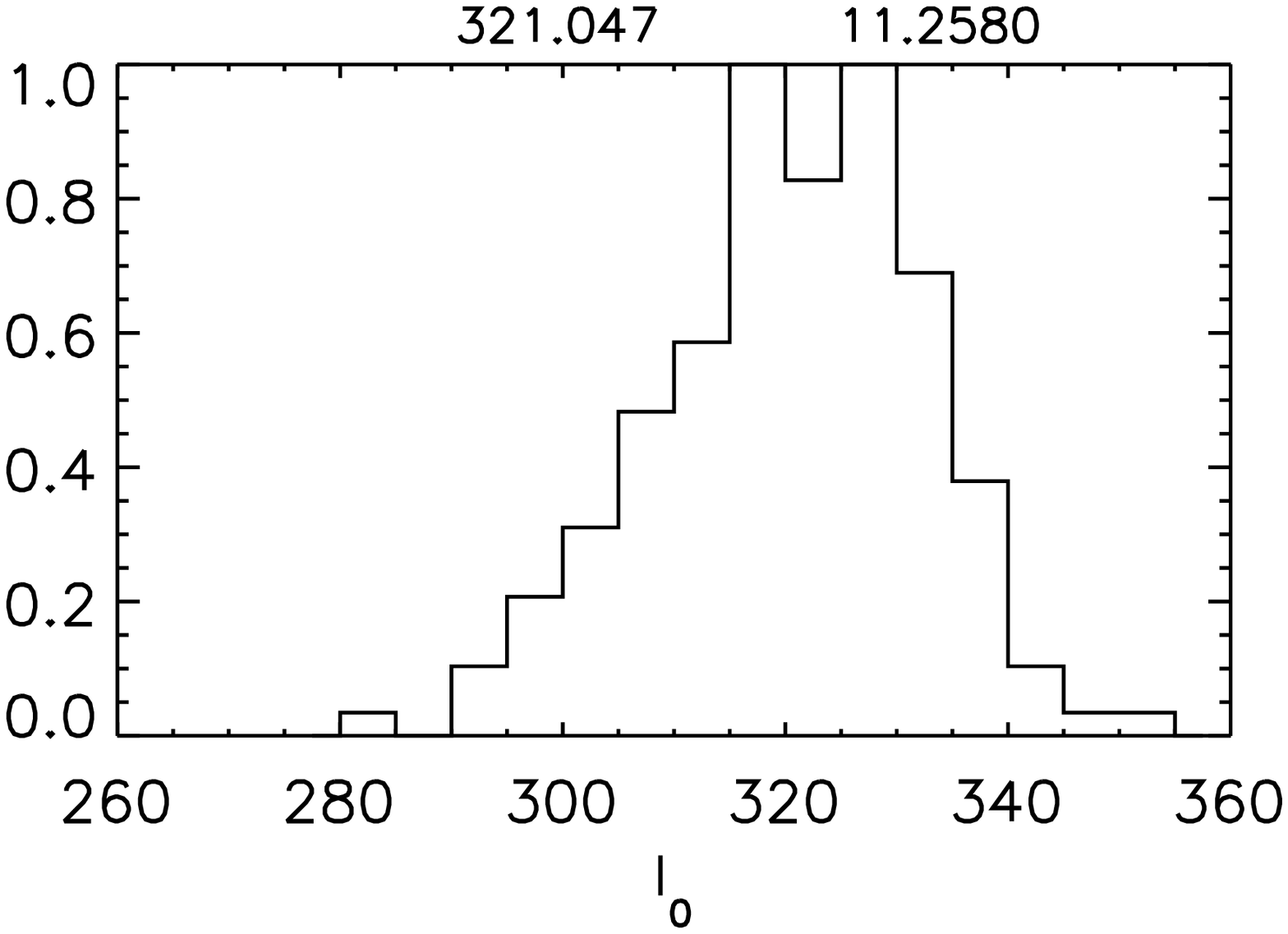}\\
\includegraphics[angle=90,scale=0.3]{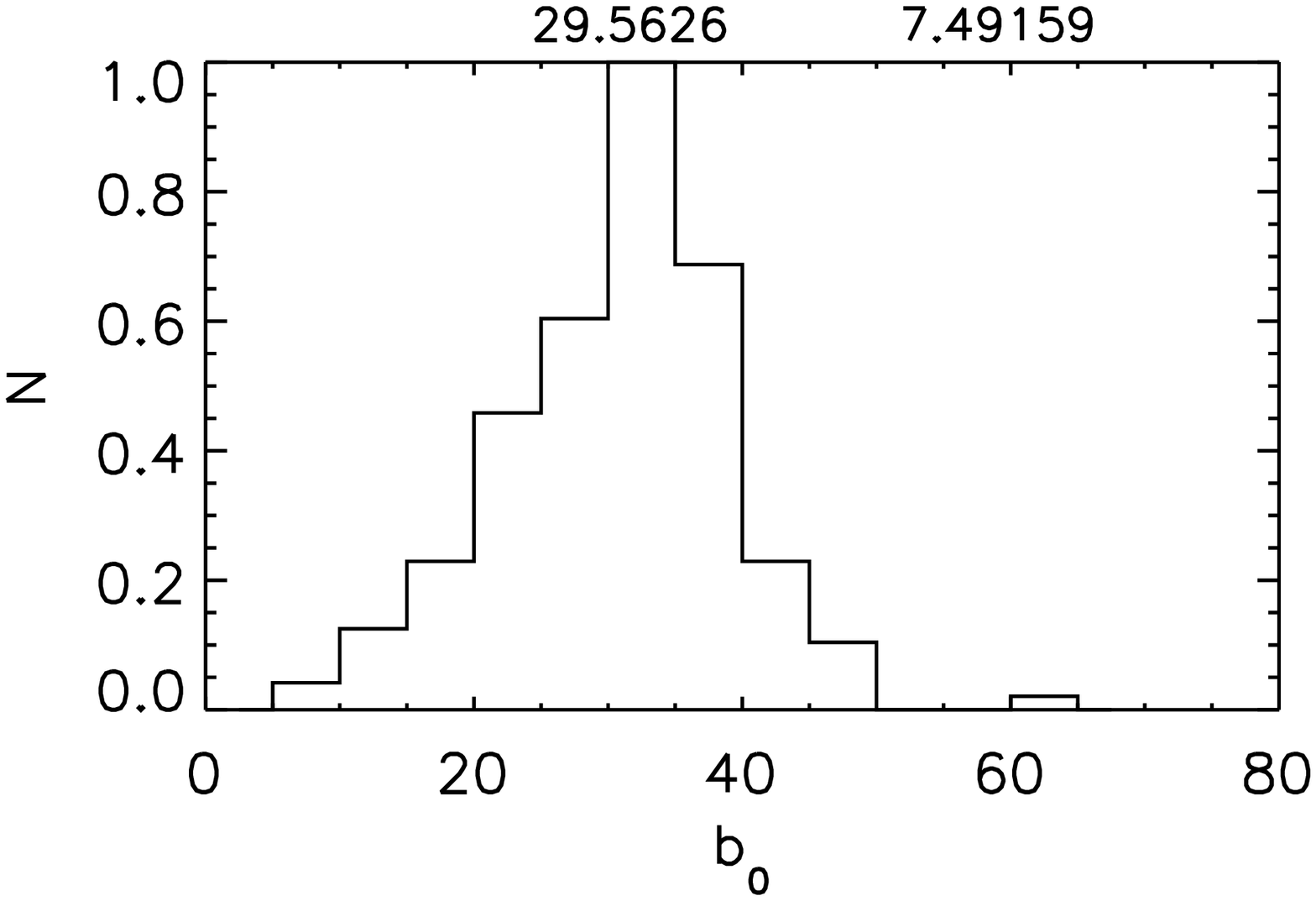}
\includegraphics[angle=90,scale=0.3]{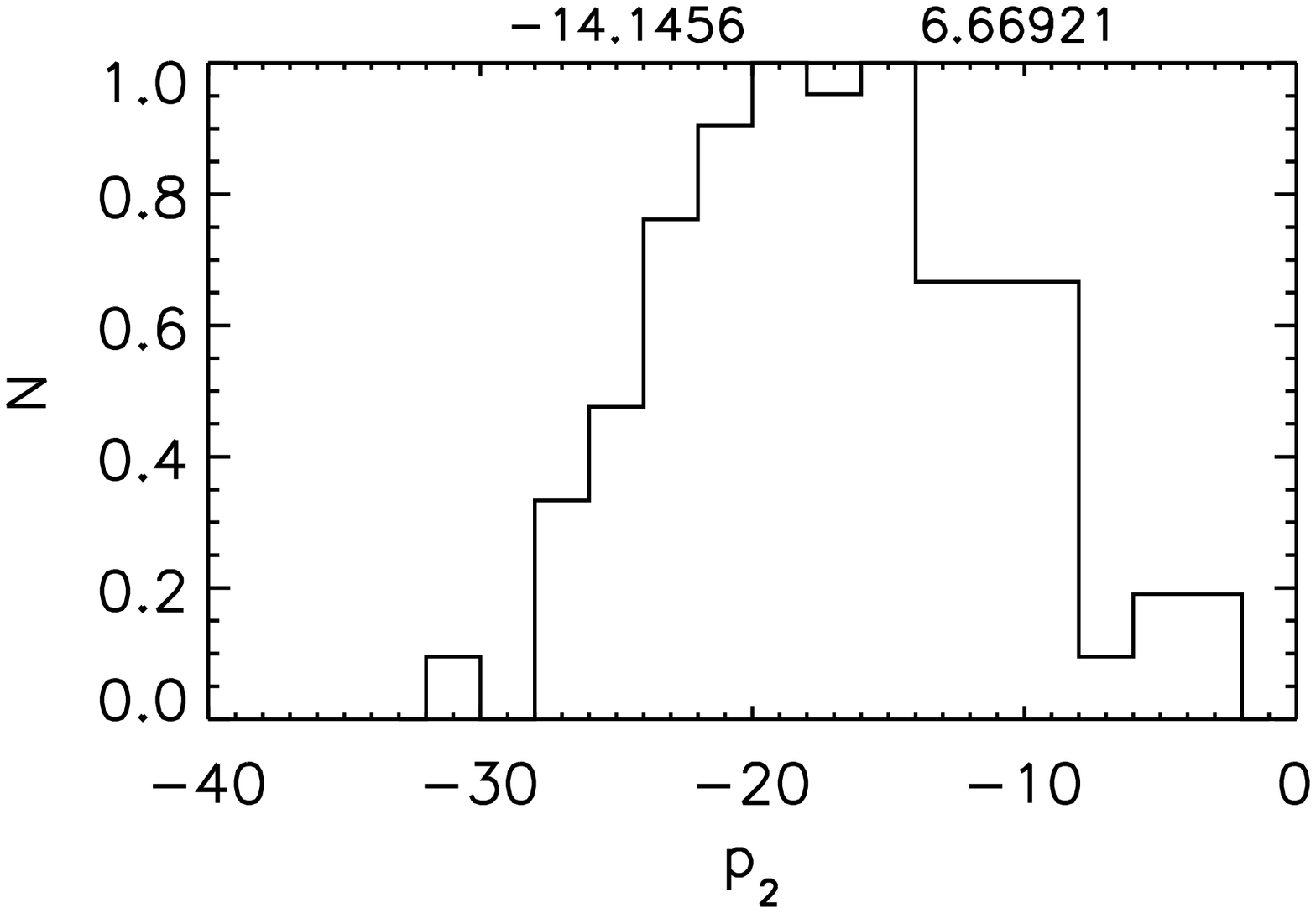}

\caption{Monte-Carlo bootstrap simulation showing the error levels of the parameters for equation~\ref{Eq:quad}. The numbers on top of each panel gives the mean and standard deviation of the corresponding parameter, for the full sample of the CMAGIC data. The errors are in good agreement with the results tabulated in Table~\ref{Tab:fits}, and in Figure~\ref{Fig:mqcmagic}. From top-left clockwise, the figures show the errors of $p_1$, $l_0$, $b_0$, and $p_2$, respectively.
}
\label{Fig:mcboot}
\end{figure}

\subsection{Systematic Errors That May Lead to the Observed Pattern}

The results derived here, although statistically significant, may subject to many systematic errors. First, the data were acquired by different groups on different telescopes which studies supernovae only in certain part of the sky. The observations from different groups may have systematic errors as large as a few percent. Even for data taken by the same group with the same  instrument, seasonality may introduce patterns across the sky as different parts of the sky are are accessible to the observer only in different seasons. To assess these effects, we studied the magnitude residuals as a function of the different research groups that acquired the original data. No dependence on the observers were identified.  We conclude that the systematic errors introduced by different groups is unlikely the course of the streaming field, although it is difficult to rule out this possibility. 

Another source of systematic error is seasonality of the observations. Although never being reported to be the case, data taken in the winter may be systematically different from those taken in the summer, due to atmospheric changes or due to temperature changes that may affect the performance of the observations. However, no obvious pattern is observed, suggesting that this again is an unlikely origin for the large scale streaming field.

We should stress that the streaming field is {\it not} to be due to the analysis method we used to homogenizing the supernova distances. Neither the MLM or CMAGIC method used in this study has a sensitivity to the position of the supernovae on the sky, and thus can not by itself introduce systematic correlations of the magnitude residuals on the sky.

Sampling errors are certainly present and important.  The presence of the zone of avoidance makes it difficult to observe supernovae close to the galactic plane. This causes ambiguities in decomposing the clustering patterns of the magnitude residuals on the sky into spherical harmonics. The dipole field for sample $A$, for example, is found to be pointing toward the suspected great attractor located behind the zone of avoidance. It can, however, be equally well modeled by a quadruple field pointing toward the center of Shapley supercluster (see Figure~\ref{Fig:fits12cmagic}). More SNe are needed to disentangle this ambiguity. 

Systematic errors may also arise from  corrections for Galactic dust extinction. The observed pattern may be the result of a similar systematic pattern in the dust extinction maps. The amplitude of the systematic error of the extinction map needs to be around 0.02 mag to explain the amplitude of the streaming field for the sub-sample z at $z \  > \ 0.025$. However, if systematic errors of Galactic extinction is the cause, we expect the amplitude of the deduced coherent flow to be inversely proportional to the redshift of the supernova. The analyses here for SNe below and above $z\ =\ 0.025$ give quite consistent expansion velocities at different redshifts. It thus seems that errors in Galactic extinction is not a favored explanation of the large scale streaming field. For the same reason, the observed velocity field is not due to an angular dependent Hubble constant.

\section{The Streaming Motion Corrected Hubble Diagram}

\label{S:NEWH}

\begin{figure}[tbh]
\epsscale{1.}
\includegraphics[angle=90,scale=0.5]{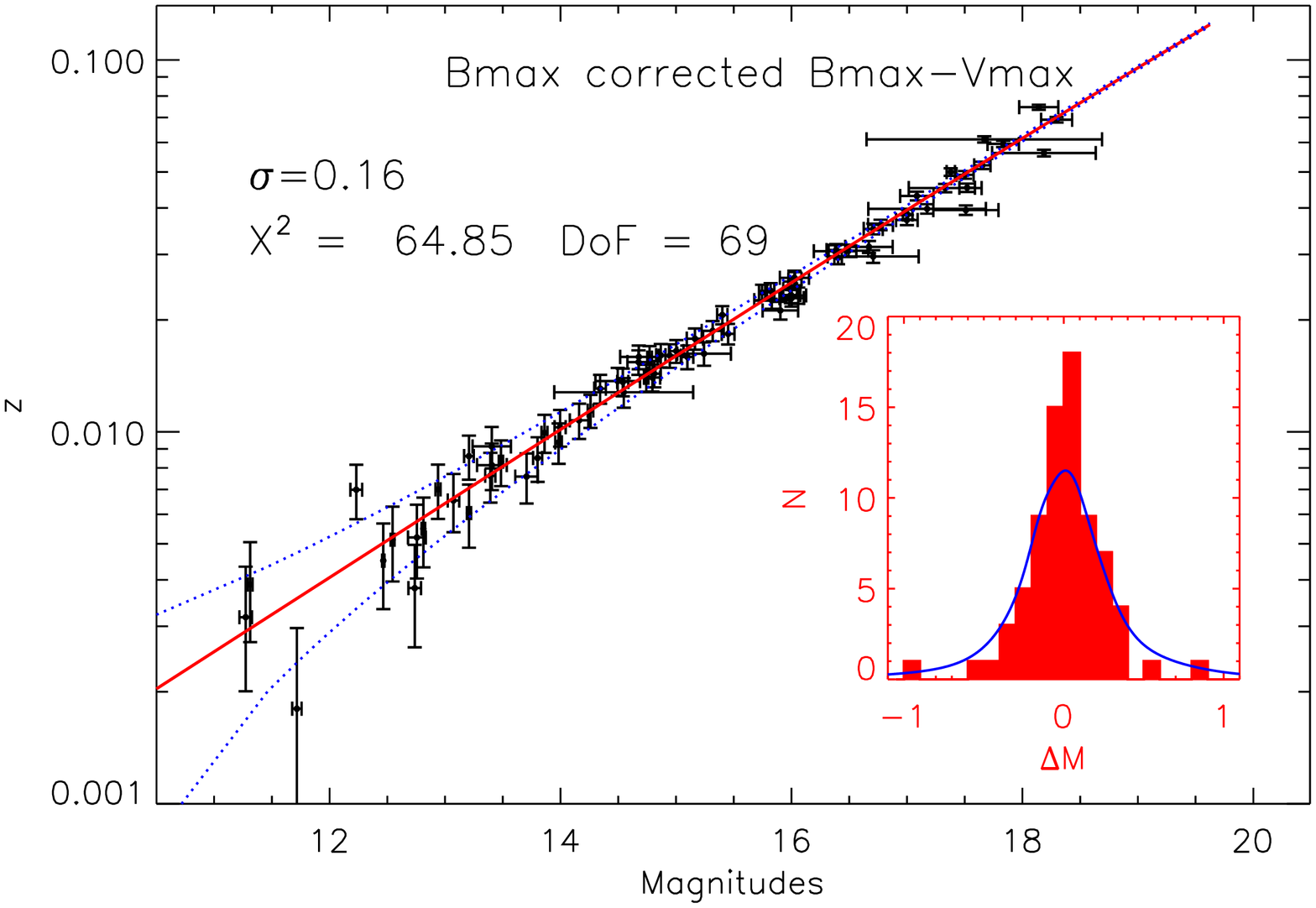}

\includegraphics[angle=90,scale=0.5]{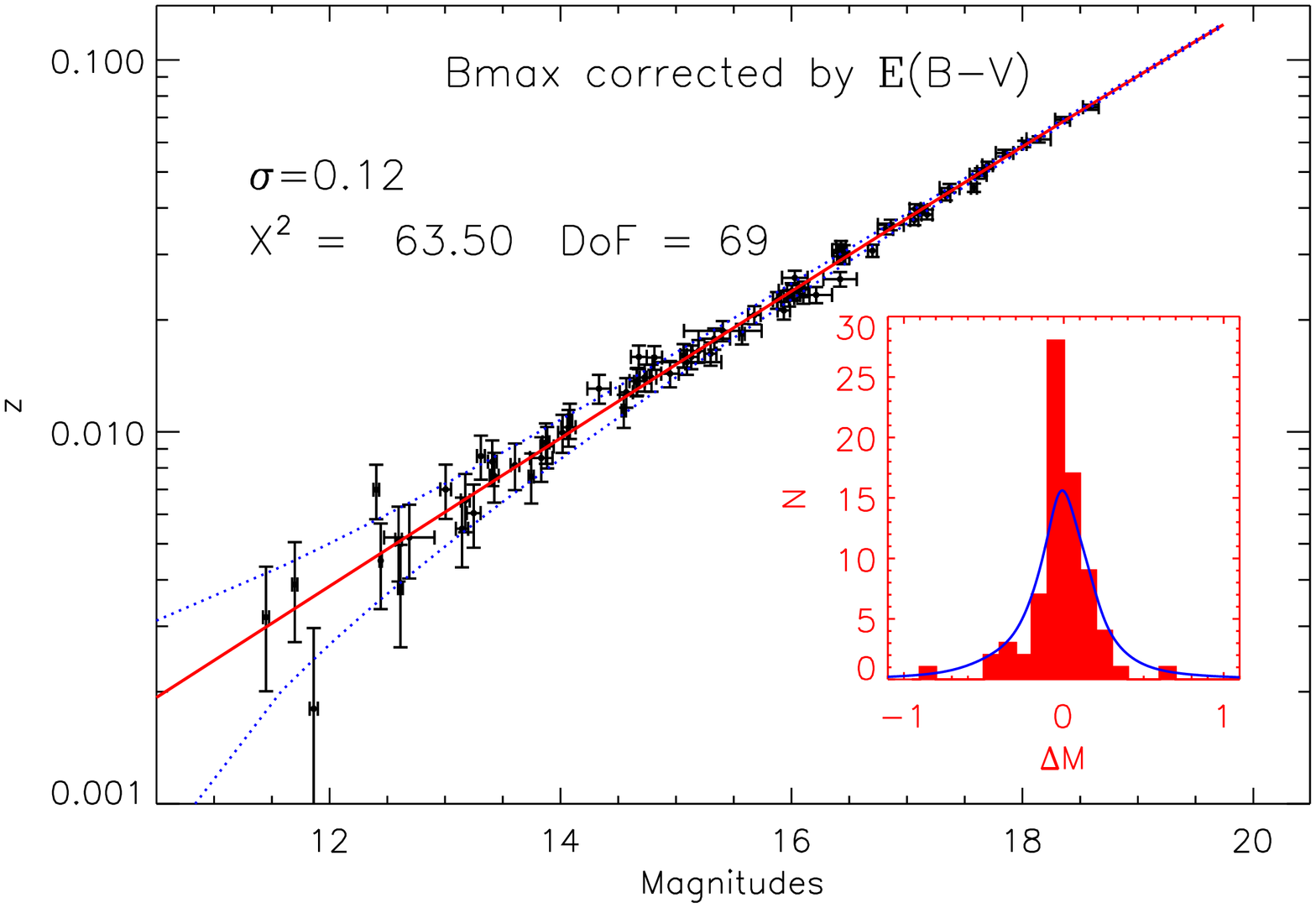}
\caption{Hubble diagrams after corrections of the coherent flow for the conventional method (a, upper panel) and the CMAGIC (b, lower panel). The Hubble diagrams shown here also have smaller magnitude dispersions as compared to those in Figure~\ref{Fig:hubble}. A dramatic decrease of $\chi^2$ is found for the CMAGIC Hubble diagram as compared to Figure~\ref{Fig:hubble}.
}
\label{Fig:newhubble}
\end{figure}

We can now apply the correction of the streaming motion pattern and construct a new Hubble diagram. We will use Equation~(\ref{Eq:mquad0}) for all of the SNe, with the coordinate transformation given as
\begin{equation}
 b'\ =\ \sin^{-1} [\cos b_0 \cos b \cos(l_0-l) + \sin b_0 \sin b],  
\label{Eq:coord}
\end{equation}
where ($l_0, \ b_0$) = ($312^{\rm o}.0, 25.7^{\rm o}$).
After correction of the coherent flow given in equation~(\ref{Eq:mquad0}), the new Hubble diagram is shown in Figure~\ref{Fig:newhubble}. As expected, the streaming motion correction dramatically improves the fits to the Hubble flow.  Both the $\chi^{2}$/DoF and the magnitude dispersions around the best fit Hubble line are significantly reduced. Most noticeably, the $\chi^2$ of the CMAGIC Hubble diagram decreased from 90 to 63, for 69 degrees of freedom, this makes the CMAGIC distances nicely agree with the a smooth Hubble flow. The magnitude dispersions on the Hubble diagram becomes 0.12 mag, the smaller ever for a sample of SNIa of this size. The correction of the streaming motion improves also the MLM Hubble diagram, at a more moderate level.

\clearpage
\begin{figure}[htbp]
\includegraphics[angle=90,scale=0.31]{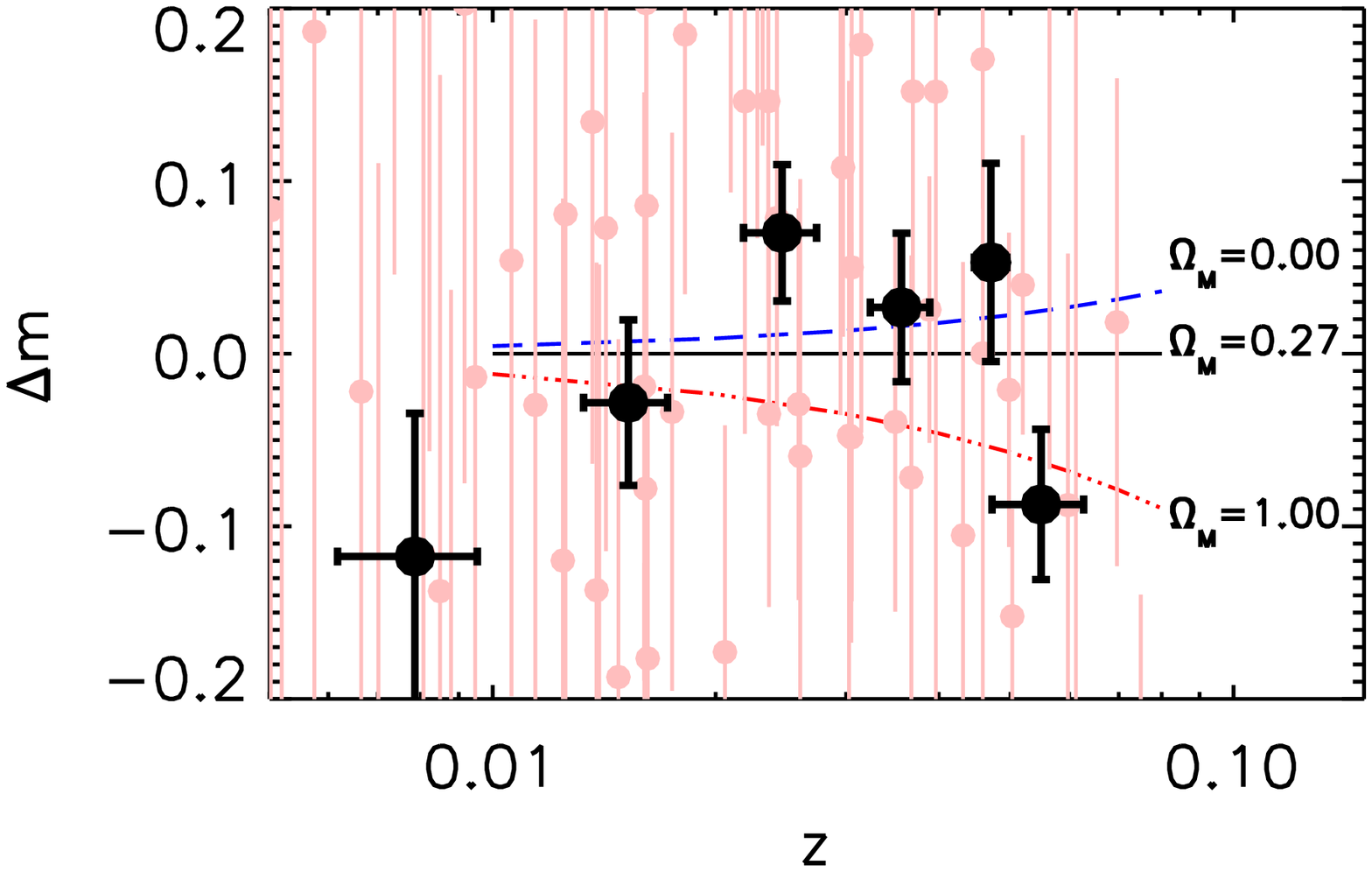}
\includegraphics[angle=90,scale=0.31]{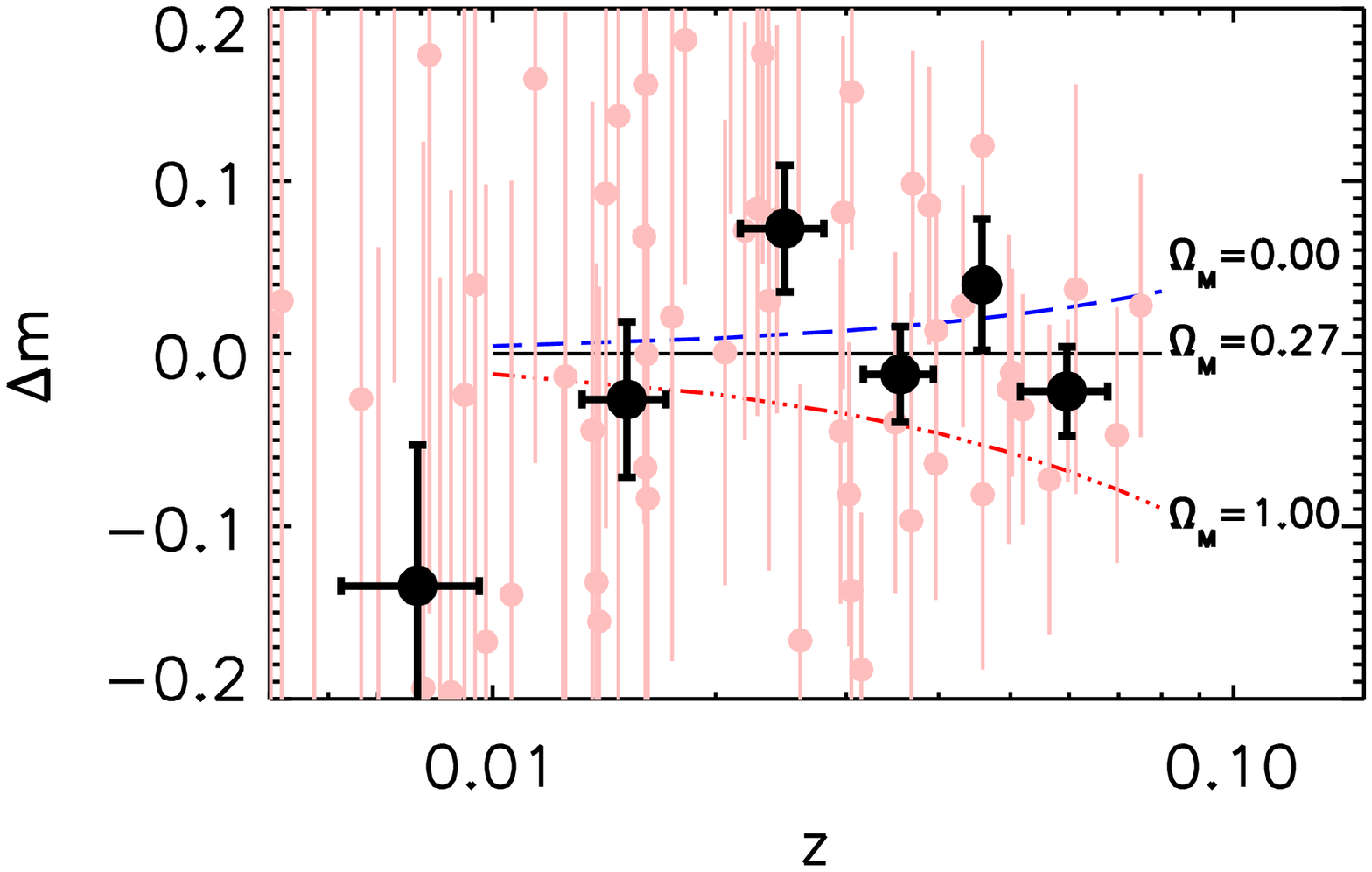}
\caption{The residuals of the coherent flow corrected distance moduli after the subtraction of  the smooth Hubble flow as a function of redshift. The light dots show the residuals of individual SNe. The heavy dots show the average of data points in redshift bins (from left to right): 0 to 0.01, 0.01 to 0.02, 0.02  to 0.03, 0.03 to 0.05, and 0.05 to 0.08. The solid lines show the Hubble line of a flat universe with $\Omega_{M}\ = \ 0.27$, the dashed lines with $\Omega_{M}\ = \ 0$, and the dash-dotted lines with $\Omega_{M}\ = \ 1.0.$ The panels
are for the MLM (a, left) and CMAGIC (b, right) methods.
}
\label{Fig:residual}
\end{figure}

To see the redshift dependence of the streaming field, we have plotted the magnitude residuals versus 
observed redshifts (see Figure~\ref{Fig:residual} for the streaming velocity corrected Hubble diagram. 
The individual SN was grouped into smaller redshift bins [0, 0.01), [0.01, 0.02), [0.02, 0.03), [0.03, 0.05), and [0.05, 0.08). The weighted averages of the SNe in the different redshift bins are shown also in  Figure~\ref{Fig:residual}. With the assumed thermal velocity of 350 km/sec, Figure~\ref{Fig:residual} shows the data at different redshift are consistent with the assumed $\Omega_m\ = \ 0.3$ and $\Omega_\Lambda\ = \ 0.7$ cosmology. An overall parabolic distribution of the average residual is indeed apparent in the data. We will defer the significance of such a pattern to future studies involving larger data set, but want to remark that at least in the redshift range below 0.01, the 350 km/sec is an over estimate of the thermal peculiar velocity as we found in \S\ref{S:FITS}; it is highly likely that a redshift dependent streaming motion will emerge from nearby SNIa observations. Two different cosmology models are also over-plotted Figure~\ref{Fig:residual}  which illustrates that it is potentially possible to study the nature of the dark matter and dark energy content of the Universe with SNe at redshifts as low as below 0.1.

\section{Discussions}

\label{S:DIS}

The primary goal of this study is to reveal the large scale streaming motion of the local Universe. Earlier efforts based on distances derived from nearby galaxies and clusters of galaxies have shown results in rough agreement with the current analysis. At redshift below 0.02,  \cite{Giovanelli:1998} found a dipole flow (in the CMB frame) of 200$\pm65$ km/sec toward (l, b) =  (295$^{\rm o}$, 25$^{\rm o}$) from an I-band
Tully-Fisher survey. \cite{Dekel:1999} found a velocity 370$\pm$110km/sec toward (l, b) = (306$^{\rm o}$, 10$^{\rm o}$) from the Mark III velocity compilation. At higher redshift, \cite{Lauer:1994} found
a 689$\pm$178 km/sec bulk motion (toward (l, b) = $(343^{\rm o},\  52^{\rm o})$ for an all sky survey of 119 clusters of galaxies to z $\sim$ 0.03. \cite{Hudson:1999} reported results from the Streaming Motions of Abell Clusters (SMAC) project, and found a bulk flow of amplitude 630$\pm$200 km/sec toward (l, b) = (260$\pm15^{\rm o}$, -1$\pm12^{\rm o}$). More recently, \cite{Hudson:2004} analyzed 56 clusters within 120 h$^{-1}$ Mpc in the SMAC sample and identified a bulk flow of 687 $\pm$ 203 km/sec toward (l, b) = (260$^{\rm o}\pm13^{\rm o}$, 0$^{\rm o}\pm11^{\rm o}$), and found that flow does not drop off significantly with depth. 

The amplitude of these velocity fields are consistent with what we found in this study. The apex direction of \cite{Giovanelli:1998} and \cite{Dekel:1999} agree with the apex deduced from this study with Equation~(\ref{Eq:mquad}). Due to the poor sky coverage of supernovae in sample $C$, the apex from the supernova data are quite sensitive to the underlying models. These differences may not be alarming considering the fact that so far most studies of large scale streaming motions have assumed dipole flows (cf. \cite{Haugboelle:2006}). There are no strong physical reasons  for using only  a dipole field for the coherent flow, especially at very large scales. It would be interesting to see if fits using different model functions would bring the  results derived from galaxy and cluster observations into better agreement with the supernova observations.
 
The fact that the velocity field we derived points to the location of Shapley concentration at all of the redshift bins is a direct confirmation that Shapley supercluster contributes significantly to the streaming motion of the local Universe. The detection of a high-amplitude coherent flow on such a large scale argues for excess mass density fluctuation power at wavelengths about 60-100/h Mpc, relative to the predictions of currently popular cosmological models \citep{Hudson:2004}. 

There may be a physical relation between the quadruple field revealed here and the quadruple fields identified from CMB maps \citep{Tegmark:2003,  Bennett:2003}. The directions of CMB dipole, quadruple, and octopole are found to be aligned roughly in the same direction. Very likely these phenomena can be understood in a coherent picture. For example, it has been suggested that the CMB dipole may be caused by interaction of CMB photons with the local structures \citep{Rakic:2006, Inoue:2006}.  A comprehensive full sky nearby supernova survey is needed to fully address these questions observationally. Such a survey can be performed on telescopes of large field of view, but not necessarily large telescopes. Future projects such as PANSTARRS and LSST provides some interesting scientific opportunity in this regard. However, for SNe below redshift of around 0.05, the supernova may be saturated with LSST. An array of small telescopes such as the system described in \cite{York:2006} is a powerful tool for mapping the local universe at $z$ below 0.05.

An important question to ask is the level of the intrinsic magnitude dispersions of SNIa after standardization procedures are applied. This  has been addressed in a previous paper by \cite{Wang:2006}. We revisit this issue in light of the new understandings of the peculiar velocity field. We will consider only SNe at $z \ >\ 0.01$ to reduce the effect of the thermal component of the peculiar velocities. We then deduce the required amount of intrinsic dispersion in order to achieve a fit  with reduced $\chi^{2}$ of 1 for an assumed value of thermal component of the peculiar velocity.  The conventional MLM method does not have sufficient statistical power to probe the intrinsic dispersion; for most reasonable values of the thermal component of the peculiar velocity, the amount of intrinsic dispersion needed is zero for MLM. For the CMAGIC method, this exercise yields values for the intrinsic dispersion of 0.074 mag, 0.069 mag, 0.062 mag, 0.050 mag, and 0.029 mag for thermal peculiar velocities of 100, 150, 200, 250, and 300 km/sec, respectively. These are remarkably low values, especially when considering the fact that the data set assembled  in this study came from diverse sources with unknown systematic errors.  For the thermal velocity of about 250-300 km/sec as we deduced from sample $A$, the intrinsic dispersion is found to be at the level of 0.03-0.05 magnitudes. This argues favorably that SNIa are indeed excellent standard candles, with the potential to reach distance precisions of around 2\%. These numbers also suggest that the thermal velocities must be very low if the the intrinsic dispersions of the distance modulus deduced from CMAGIC is larger than 0.07 mag, {\it i. e.}, the local streaming motion flow superimposed on the overall Hubble expansion, must be very smooth at the redshift range from 0 to about 0.08.

The reason that CMAGIC out performs MLM is perhaps related to the fact that CMAGIC weighs 
heavily on data about 2 weeks past optical maximum, whereas MLM uses data around optical maximum. 
At optical maximum, SNIa show more diverse behavior than at around day 15 past maximum. Spectroscopically 
peculiar SNIa such as SN~1991T are hardly distinguishable from normal SNIa at day 15 but the differences
are obvious at around maximum. Furthermore,  the polarization around optical maximum decreases 
to practically 
undetectable at day 15 
\citep{Wang:1996, Wang:2003, Wang:20062004dt, Leonard:2005, Wang:2007}. All these effects may have contributed.

Finally, we want to remark the effect of the streaming motion on the determination of cosmological parameters 
using SNIa at z $\sim$ 1. The large scale flow pattern implies that there could be correlated systematic 
errors at more than 0.02 magnitudes out to redshift of 0.1. A sample of supernova evenly distributed on 
the sky is needed to average out the effect of the large scale correlated flow if no streaming motion 
correction is performed.


\input{ms.bbl}
\end{document}

%% file: BmaxBmaxVmaxBipolar.tex
\begin{deluxetable}{lllllllllrrrrllrr}\tabletypesize{\scriptsize}\rotate\tablecolumns{14}\tablewidth{0pt}\tablecaption{MLM Parameters of the Supernovae}\tablehead{\colhead{SN}              & \colhead{z$_{cmb}$}                  & \colhead{Mag}                    & \colhead{$\sigma$}                    & \colhead{$\Delta m_{15}$}                    & \colhead{$\sigma$}                    & \colhead{$Color$}                    & \colhead{$\sigma$}                    & \colhead{$\mu_0$}                    & \colhead{$\Delta \mu$}               & \colhead{$\sigma$}                   & \colhead{$V_{Pec}$ (km/sec)}               & \colhead{$\sigma$ (km/dec)}                   & \colhead{RA (2000.0)}                         & \colhead{Dec (2000.0)}                        & \colhead{$b\prime$}                        & \cr\colhead{(1)}              & \colhead{(2)}                  & \colhead{(3)}              & \colhead{(4)}           & \colhead{(5)}              & \colhead{(6)}           & \colhead{(7)}         & \colhead{(8)}           & \colhead{(9)}           & \colhead{(10)}       &                 \colhead{(11)}       &                 \colhead{(12)}       &                 \colhead{(13)}       &                 \colhead{(14)}       &                 \colhead{(15)}       &                 \colhead{(16)}                        \cr}\startdata
1989B&         0.00359&    12.23& 0.02& 1.25& 0.02& 0.35& 0.03&30.39&-0.52& 0.06&  229.72&   23.19&11 20 13.91&$+ 13$  00 19.20&   31.59&\\
1990af&         0.04992&    17.77& 0.01& 1.59& 0.02& 0.04& 0.02&36.56&-0.14& 0.04&  944.88&  250.71&21 34 58.12&$-62$  44  7.38&   19.88&\\
1990N&         0.00447&    12.54& 0.03& 1.13& 0.02&-0.09& 0.03&31.94& 0.55& 0.06& -386.85&   48.37&12 42 56.70&$+ 13$  15 23.69&   38.95&\\
1990O&         0.03060&    16.22& 0.04& 0.93& 0.03&-0.03& 0.04&35.57&-0.04& 0.09&  160.87&  378.63&17 15 35.92&$+ 16$  19 25.80&   15.43&\\
1990T&         0.03967&    17.26& 0.09& 1.24& 0.03&-0.14& 0.13&36.72& 0.53& 0.31&-3317.41& 2166.39&19 59  2.28&$-56$  15 30.00&   26.21&\\
1990Y&         0.03870&    17.63& 0.31& 1.15& 0.05& 0.17& 0.32&36.32& 0.19& 0.57&-1061.54& 3382.79&03 37 22.64&$-33$  02 40.09&  -14.60&\\
1991ag&         0.01388&    14.45& 0.05& 0.89& 0.03& 0.03& 0.09&33.67&-0.20& 0.21&  363.48&  368.60&20 00  8.66&$-55$  22  3.38&   25.67&\\
1991S&         0.05606&    17.76& 0.24& 1.30& 0.05&-0.22& 0.26&37.42& 0.45& 0.50&-3880.42& 4797.62&10 29 27.79&$+ 22$  00 46.41&   17.36&\\
1991U&         0.03243&    16.49& 0.10& 1.17& 0.04&-0.09& 0.11&35.87& 0.13& 0.23& -617.58& 1082.72&13 23 22.20&$-26$  06 28.70&   79.49&\\
1992A&         0.00594&    12.62& 0.02& 1.31& 0.01& 0.11& 0.02&31.39&-0.61& 0.06&  438.21&   34.56&03 36 27.40&$-34$  57 31.59&  -12.95&\\
1992ae&         0.07461&    18.53& 0.08& 1.24& 0.07& 0.12& 0.09&37.30&-0.31& 0.19& 3011.66& 1653.42&21 28 17.66&$-61$  33  0.00&   19.66&\\
1992ag&         0.02700&    16.25& 0.05& 1.07& 0.05& 0.10& 0.06&35.19&-0.15& 0.14&  523.87&  478.32&13 24 10.12&$-23$  52 39.30&   77.31&\\
1992al&         0.01350&    14.48& 0.02& 1.21& 0.02&-0.05& 0.02&33.73&-0.07& 0.05&  129.53&   94.19&20 45 56.45&$-51$  23 40.00&   17.79&\\
1992au&         0.06035&    18.35& 0.55& 1.44& 0.07& 0.19& 0.61&36.82&-0.31& 1.14& 2434.49& 8609.03&00 10 40.27&$-49$  56 43.31&   -1.75&\\
1992bc&         0.01960&    15.11& 0.02& 0.75& 0.01&-0.03& 0.02&34.58&-0.04& 0.05&  109.09&  130.38&03 05 17.28&$-39$  33 39.69&  -11.02&\\
1992bg&         0.03648&    16.75& 0.04& 1.14& 0.03&-0.02& 0.05&35.96&-0.04& 0.12&  216.53&  571.38&07 41 56.53&$-62$  31  8.81&   32.89&\\
1992bh&         0.04509&    17.61& 0.03& 0.97& 0.06& 0.07& 0.03&36.69& 0.21& 0.07&-1384.05&  489.30&04 59 27.55&$-58$  49 44.19&   14.55&\\
1992bk&         0.05885&    18.26& 0.06& 1.63& 0.03& 0.05& 0.07&37.02&-0.06& 0.15&  475.81& 1187.83&03 43  1.90&$-53$  37 56.81&    4.43&\\
1992bl&         0.04223&    17.41& 0.06& 1.50& 0.04& 0.03& 0.07&36.27&-0.06& 0.16&  345.25&  897.23&23 15 13.25&$-44$  44 34.50&   -3.10&\\
1992bo&         0.01723&    15.75& 0.02& 1.63& 0.02&-0.01& 0.03&34.64& 0.30& 0.06& -778.01&  166.24&01 21 58.44&$-34$  12 43.50&  -19.29&\\
1992J&         0.04612&    17.83& 0.11& 1.53& 0.12& 0.10& 0.16&36.50&-0.03& 0.35&  159.88& 2189.69&10 09  0.30&$-26$  38 24.41&   46.14&\\
1992P&         0.02649&    16.07& 0.02& 1.25& 0.04&-0.02& 0.02&35.22&-0.07& 0.06&  252.86&  216.55&12 42 48.95&$+ 10$  21 37.50&   41.77&\\
1993ag&         0.05004&    17.80& 0.03& 1.27& 0.03& 0.08& 0.04&36.67&-0.04& 0.08&  263.66&  544.50&10 03 35.00&$-35$  27 47.59&   48.32&\\
1993ah&         0.02850&    16.41& 0.14& 1.46& 0.04&-0.21& 0.19&35.94& 0.49& 0.43&-2149.05& 2144.05&23 51 50.27&$-27$  57 47.00&  -21.40&\\
1993B&         0.07008&    18.44& 0.06& 1.31& 0.04& 0.01& 0.07&37.48& 0.01& 0.15&  -79.17& 1428.12&10 34 51.38&$-34$  26 30.00&   54.16&\\
1993H&         0.02513&    16.71& 0.02& 1.56& 0.02& 0.18& 0.03&35.14&-0.03& 0.06&  107.52&  218.93&13 52 50.34&$-30$  42 23.31&   82.81&\\
1993O&         0.05293&    17.62& 0.03& 1.27& 0.03&-0.06& 0.03&36.85& 0.01& 0.08&  -98.95&  564.51&13 31  7.88&$-33$  12 50.50&   86.75&\\
1994D&         0.00261&    11.76& 0.01& 1.57& 0.01&-0.10& 0.02&30.93& 0.71& 0.04& -301.80&   22.37&12 34  2.37&$+ 07$  42  4.70&   43.79&\\
1994M&         0.02431&    16.32& 0.03& 1.41& 0.03& 0.06& 0.04&35.17& 0.07& 0.09& -244.09&  307.86&12 31  8.61&$+ 00$  36 19.90&   50.33&\\
1994Q&         0.02987&    16.52& 0.08& 0.96& 0.04& 0.09& 0.09&35.54&-0.02& 0.20&   69.60&  830.10&16 49 51.06&$+ 40$  25 55.53&    0.73&\\
1994S&         0.01611&    14.79& 0.02& 0.88& 0.07&-0.01& 0.04&34.12&-0.07& 0.10&  146.34&  219.04&12 31 21.82&$+ 29$  08  4.19&   22.88&\\
1995ak&         0.02198&    16.06& 0.05& 1.48& 0.05& 0.00& 0.07&35.02& 0.14& 0.16& -443.22&  533.96&02 45 48.83&$+ 03$  13 50.10&  -52.69&\\
1995al&         0.00588&    13.32& 0.02& 0.96& 0.01& 0.13& 0.02&32.22& 0.24& 0.05& -206.55&   49.02&09 50 55.97&$+ 33$  33  9.41&    3.05&\\
1995bd&         0.01443&    15.22& 0.02& 0.73& 0.03& 0.31& 0.05&33.79&-0.16& 0.09&  312.09&  166.42&04 45 21.24&$+ 11$  04  2.50&  -39.72&\\
1995D&         0.00766&    13.23& 0.02& 0.97& 0.01&-0.03& 0.02&32.58& 0.02& 0.05&  -16.29&   50.52&09 40 54.79&$+ 05$  08 26.60&   21.94&\\
1996bl&         0.03485&    16.68& 0.03& 0.84& 0.04& 0.05& 0.04&35.87&-0.02& 0.09&  113.36&  419.58&00 36 17.97&$+ 11$  23 40.50&  -61.96&\\
1996bo&         0.01632&    15.84& 0.03& 1.16& 0.03& 0.32& 0.03&34.12&-0.10& 0.07&  221.28&  145.48&01 48 22.86&$+ 11$  31 15.10&  -64.80&\\
1996C&         0.03007&    16.62& 0.08& 0.94& 0.04& 0.09& 0.09&35.65& 0.08& 0.19& -354.78&  816.04&13 50 48.59&$+ 49$  19  7.09&    4.13&\\
1996X&         0.00800&    12.99& 0.01& 1.28& 0.01&-0.02& 0.01&32.13&-0.53& 0.02&  519.91&   19.76&13 18  1.12&$-26$  50 45.31&   79.94&\\
1996Z&         0.00974&    14.41& 0.05& 1.05& 0.04& 0.42& 0.08&32.50&-0.59& 0.18&  694.31&  185.91&09 36 44.82&$-21$  08 51.70&   37.05&\\
1997bp&         0.00944&    13.91& 0.01& 1.19& 0.01& 0.15& 0.01&32.64&-0.38& 0.02&  458.71&   21.38&12 46 53.75&$-11$  38 33.20&   63.18&\\
1997bq&         0.00960&    14.44& 0.03& 1.16& 0.02& 0.10& 0.04&33.32& 0.27& 0.09& -374.69&  131.13&10 17  5.33&$+ 73$  23  2.12&  -24.69&\\
1997cw&         0.01595&    15.86& 0.07& 0.99& 0.03& 0.28& 0.11&34.37& 0.20& 0.25& -451.71&  611.57&00 25 17.27&$+ 12$  53  6.20&  -62.06&\\
1997E&         0.01333&    15.11& 0.02& 1.38& 0.03& 0.05& 0.03&33.99& 0.21& 0.06& -408.07&  123.10&06 47 38.10&$+ 74$  29 51.00&  -37.88&\\
1998aq&         0.00426&    12.31& 0.01& 1.17& 0.01&-0.11& 0.01&31.75& 0.47& 0.02& -305.78&   13.42&11 56 25.87&$+ 55$  07 43.19&   -3.85&\\
1998bu&         0.00416&    12.13& 0.01& 1.01& 0.01& 0.35& 0.01&30.42&-0.81& 0.02&  389.92&    9.15&10 46 46.01&$+ 11$  50  7.50&   27.98&\\
1998dk&         0.01204&    14.84& 0.32& 1.30& 0.07& 0.07& 0.35&33.72& 0.17& 0.67& -287.18& 1224.73&00 14 32.16&$ 00$  44 10.90&  -48.68&\\
1998ec&         0.02012&    16.31& 0.05& 1.05& 0.03& 0.17& 0.07&35.05& 0.37& 0.17&-1113.86&  553.99&06 53  6.11&$+ 50$  02 22.09&  -32.88&\\
1998es&         0.00957&    13.84& 0.01& 0.74& 0.01& 0.08& 0.01&33.02&-0.03& 0.03&   37.49&   36.04&01 37 17.60&$+ 05$  52 50.60&  -59.40&\\
1998V&         0.01717&    15.10& 0.03& 1.10& 0.02&-0.03& 0.04&34.35& 0.02& 0.07&  -44.96&  178.95&18 22 37.41&$+ 15$  42  8.41&    4.04&\\
1999aa&         0.01525&    14.74& 0.02& 0.81& 0.02& 0.02& 0.02&34.04&-0.03& 0.04&   63.09&   91.66&08 27 42.03&$+ 21$  29 14.80&   -2.17&\\
1999ac&         0.00982&    14.09& 0.01& 1.23& 0.03& 0.01& 0.01&33.16& 0.06& 0.03&  -80.55&   41.33&16 07 15.02&$+ 07$  58 20.40&   32.58&\\
1999aw&         0.03924&    16.69& 0.02& 0.78& 0.02&-0.05& 0.02&36.19& 0.03& 0.04& -159.23&  244.86&11 01 36.37&$-06$  06 31.60&   44.16&\\
1999dk&         0.01395&    14.82& 0.03& 1.00& 0.03& 0.03& 0.03&33.97& 0.09& 0.07& -183.78&  148.41&01 31 26.92&$+ 14$  17  5.70&  -67.83&\\
1999dq&         0.01308&    14.41& 0.02& 0.94& 0.03& 0.06& 0.02&33.51&-0.23& 0.05&  388.27&   85.65&02 33 59.68&$+ 20$  58 30.41&  -69.48&\\
1999ee&         0.01055&    14.85& 0.01& 0.90& 0.01& 0.29& 0.01&33.38& 0.12& 0.03& -172.99&   40.38&22 16 10.00&$-36$  50 39.69&   -3.79&\\
1999ek&         0.01760&    15.57& 0.01& 1.08& 0.01& 0.11& 0.06&34.48& 0.09& 0.08& -223.52&  214.10&05 36 31.60&$+ 16$  38 17.80&  -32.86&\\
1999gh&         0.00879&    14.51& 0.07& 1.67& 0.03& 0.31& 0.07&32.53&-0.33& 0.14&  371.78&  149.36&09 44 19.75&$-21$  16 25.00&   38.64&\\
2000ca&         0.02452&    15.58& 0.01& 0.93& 0.01&-0.03& 0.01&34.93&-0.19& 0.03&  613.89&   85.22&13 35 22.98&$-34$  09 37.00&   87.61&\\
2000cf&         0.03603&    17.09& 0.05& 1.31& 0.04&-0.02& 0.05&36.19& 0.21& 0.10&-1124.49&  560.47&15 52 56.19&$+ 65$  56 12.69&  -15.94&\\
2000cn&         0.02321&    16.60& 0.01& 1.58& 0.02& 0.20& 0.01&34.97&-0.03& 0.03&   86.30&  103.97&17 57 40.42&$+ 27$  49 58.09&    0.43&\\
2000dk&         0.01645&    15.35& 0.02& 1.43& 0.03& 0.02& 0.03&34.28& 0.04& 0.06& -100.00&  128.02&01 07 23.52&$+ 32$  24 23.19&  -83.49&\\
2000E&         0.00422&    12.78& 0.01& 1.03& 0.02& 0.03& 0.04&31.93& 0.67& 0.07& -455.01&   58.20&20 37 13.77&$+ 66$  05 50.19&  -39.39&\\
2000fa&         0.02179&    15.75& 0.05& 1.09& 0.02&-0.10& 0.06&35.20& 0.34& 0.12&-1125.79&  412.65&07 15 29.88&$+ 23$  25 42.41&  -16.84&\\
2001ba&         0.03053&    16.19& 0.01& 1.06& 0.01&-0.08& 0.01&35.60&-0.00& 0.03&   15.85&  144.11&11 38  1.76&$-32$  19 51.00&   66.20&\\
2001bt&         0.01445&    15.27& 0.01& 1.29& 0.01& 0.17& 0.01&33.89&-0.06& 0.02&  116.43&   38.54&19 13 46.75&$-59$  17 22.81&   32.94&\\
2001el&         0.00365&    12.79& 0.00& 1.16& 0.00& 0.12& 0.01&31.62& 0.68& 0.01& -400.54&    7.76&03 44 30.57&$-44$  38 23.69&   -3.55&\\
2001V&         0.01604&    14.61& 0.03& 0.73& 0.03& 0.06& 0.07&33.84&-0.34& 0.18&  701.75&  332.88&11 57 24.93&$+ 25$  12  9.00&   24.42&\\
2002bo&         0.00529&    13.94& 0.01& 1.24& 0.01& 0.42& 0.01&31.91& 0.15& 0.02& -117.46&   13.09&10 18  6.51&$+ 21$  49 41.70&   15.89&\\
2002el&         0.02238&    16.15& 0.03& 1.37& 0.02&-0.03& 0.03&35.25& 0.33& 0.07&-1116.58&  265.68&20 56 30.09&$-18$  33 34.30&   -4.89&\\
2002er&         0.00855&    14.26& 0.01& 1.26& 0.01& 0.15& 0.02&32.96& 0.15& 0.04& -186.42&   49.59&17 11 29.88&$+ 07$  59 44.80&   22.08&\\
2003du&         0.00665&    13.50& 0.04& 1.15& 0.03&-0.09& 0.05&32.91& 0.65& 0.11& -698.57&  132.64&14 34 35.80&$+ 59$  20  3.81&   -6.66&\\
2004eo&         0.01473&    15.08& 0.00& 1.33& 0.00& 0.07& 0.01&33.94&-0.05& 0.02&  106.82&   36.45&20 32 54.19&$+ 09$  55 42.70&  -18.07&\\
2004S&         0.00986&    14.24& 0.02& 1.18& 0.01& 0.08& 0.02&33.16& 0.05& 0.05&  -66.36&   71.79&06 45 43.50&$-31$  13 52.50&    9.78&\\
2005am&         0.00897&    13.69& 0.01& 1.54& 0.02& 0.09& 0.02&32.38&-0.53& 0.04&  584.00&   43.53&09 16 12.47&$-16$  18 16.00&   30.36&\\
2005cf&         0.00704&    13.20& 0.00& 1.14& 0.00&-0.01& 0.01&32.39& 0.01& 0.02&  -11.42&   16.69&15 21 32.20&$-07$  24 47.50&   51.73&\\
\enddata
\label{Tab:mlm}
\end{deluxetable}

%% file: BmaxmathcalEBVBipolar.tex
\begin{deluxetable}{lllllllllrrrrllrr}\tabletypesize{\scriptsize}\rotate\tablecolumns{14}\tablewidth{0pt}\tablecaption{CMAGIC Parameters of the Supernovae}\tablehead{\colhead{SN}              & \colhead{z$_{cmb}$}                  & \colhead{Mag}                    & \colhead{$\sigma$}                    & \colhead{$\Delta m_{15}$}                    & \colhead{$\sigma$}                    & \colhead{$Color$}                    & \colhead{$\sigma$}                    & \colhead{$\mu_0$}                    & \colhead{$\Delta \mu$}               & \colhead{$\sigma$}                   & \colhead{$V_{Pec}$ (km/sec)}               & \colhead{$\sigma$ (km/dec)}                   & \colhead{RA (2000.0)}                         & \colhead{Dec (2000.0)}                        & \colhead{$b\prime$}                        & \cr\colhead{(1)}              & \colhead{(2)}                  & \colhead{(3)}              & \colhead{(4)}           & \colhead{(5)}              & \colhead{(6)}           & \colhead{(7)}         & \colhead{(8)}           & \colhead{(9)}           & \colhead{(10)}       &                 \colhead{(11)}       &                 \colhead{(12)}       &                 \colhead{(13)}       &                 \colhead{(14)}       &                 \colhead{(15)}       &                 \colhead{(16)}                        \cr}\startdata
1989B&         0.00359&    12.59& 0.02& 1.25& 0.02& 0.39& 0.02&30.48&-0.43& 0.03&  192.42&   10.81&11 20 13.91&$+ 13$  00 19.20&   31.59&\\
1990af&         0.04992&    18.98& 0.03& 1.59& 0.02&-0.01& 0.03&36.69&-0.01& 0.03&   75.54&  230.29&21 34 58.12&$-62$  44  7.38&   19.88&\\
1990N&         0.00447&    13.83& 0.01& 1.13& 0.02&-0.05& 0.02&31.67& 0.28& 0.02& -185.38&   14.51&12 42 56.70&$+ 13$  15 23.69&   38.95&\\
1990O&         0.03060&    17.57& 0.01& 0.93& 0.03&-0.07& 0.02&35.46&-0.14& 0.06&  589.60&  227.05&17 15 35.92&$+ 16$  19 25.80&   15.43&\\
1990T&         0.03967&    18.38& 0.00& 1.24& 0.03& 0.04& 0.04&36.19& 0.01& 0.03&  -34.41&  185.27&19 59  2.28&$-56$  15 30.00&   26.21&\\
1990Y&         0.03870&    18.22& 0.06& 1.15& 0.05& 0.29& 0.15&36.11&-0.02& 0.04&  132.32&  235.90&03 37 22.64&$-33$  02 40.09&  -14.60&\\
1991ag&         0.01388&    15.80& 0.05& 0.89& 0.03&-0.08& 0.05&33.72&-0.14& 0.07&  268.57&  122.44&20 00  8.66&$-55$  22  3.38&   25.67&\\
1991S&         0.05606&    19.07& 0.02& 1.30& 0.05&-0.08& 0.17&36.90&-0.06& 0.08&  489.33&  628.09&10 29 27.79&$+ 22$  00 46.41&   17.36&\\
1991U&         0.03243&    17.64& 0.02& 1.17& 0.04& 0.02& 0.05&35.48&-0.26& 0.04& 1092.84&  170.85&13 23 22.20&$-26$  06 28.70&   79.49&\\
1992A&         0.00594&    13.65& 0.02& 1.31& 0.01& 0.08& 0.03&31.45&-0.56& 0.03&  402.61&   15.97&03 36 27.40&$-34$  57 31.59&  -12.95&\\
1992ae&         0.07461&    19.81& 0.05& 1.24& 0.07&-0.06& 0.10&37.65& 0.03& 0.07& -335.62&  739.98&21 28 17.66&$-61$  33  0.00&   19.66&\\
1992ag&         0.02700&    17.22& 0.10& 1.07& 0.05& 0.12& 0.08&35.07&-0.26& 0.11&  903.74&  377.14&13 24 10.12&$-23$  52 39.30&   77.31&\\
1992al&         0.01350&    15.92& 0.01& 1.21& 0.02&-0.11& 0.02&33.72&-0.09& 0.02&  157.58&   40.70&20 45 56.45&$-51$  23 40.00&   17.79&\\
1992au&         0.06035&    19.45& 0.05& 1.44& 0.07& 0.06& 0.36&37.19& 0.06& 0.12& -483.97& 1036.97&00 10 40.27&$-49$  56 43.31&   -1.75&\\
1992bc&         0.01960&    16.81& 0.04& 0.75& 0.01&-0.25& 0.05&34.74& 0.12& 0.06& -336.03&  159.89&03 05 17.28&$-39$  33 39.69&  -11.02&\\
1992bg&         0.03648&    18.09& 0.02& 1.14& 0.03&-0.06& 0.03&35.91&-0.08& 0.11&  419.01&  547.94&07 41 56.53&$-62$  31  8.81&   32.89&\\
1992bh&         0.04509&    18.73& 0.02& 0.97& 0.06& 0.04& 0.02&36.63& 0.16& 0.03&-1001.76&  176.62&04 59 27.55&$-58$  49 44.19&   14.55&\\
1992bk&         0.05885&    19.38& 0.01& 1.63& 0.03& 0.05& 0.04&37.06&-0.01& 0.02&   84.65&  176.44&03 43  1.90&$-53$  37 56.81&    4.43&\\
1992bl&         0.04223&    18.66& 0.05& 1.50& 0.04&-0.02& 0.05&36.39& 0.06& 0.04& -367.83&  235.79&23 15 13.25&$-44$  44 34.50&   -3.10&\\
1992bo&         0.01723&    16.92& 0.02& 1.63& 0.02& 0.01& 0.03&34.62& 0.28& 0.02& -719.47&   66.34&01 21 58.44&$-34$  12 43.50&  -19.29&\\
1992J&         0.04612&    18.67& 0.11& 1.53& 0.12& 0.18& 0.10&36.41&-0.11& 0.09&  710.27&  524.77&10 09  0.30&$-26$  38 24.41&   46.14&\\
1992P&         0.02649&    17.83& 0.43& 1.25& 0.04&-0.21& 0.27&35.48& 0.19& 0.15& -746.38&  596.00&12 42 48.95&$+ 10$  21 37.50&   41.77&\\
1993ag&         0.05004&    18.86& 0.03& 1.27& 0.03& 0.07& 0.04&36.66&-0.04& 0.07&  303.88&  504.07&10 03 35.00&$-35$  27 47.59&   48.32&\\
1993ah&         0.02850&    17.70& 0.03& 1.46& 0.04&-0.06& 0.10&35.47& 0.02& 0.05&  -75.12&  211.84&23 51 50.27&$-27$  57 47.00&  -21.40&\\
1993B&         0.07008&    19.61& 0.05& 1.31& 0.04& 0.01& 0.07&37.40&-0.07& 0.07&  695.94&  612.54&10 34 51.38&$-34$  26 30.00&   54.16&\\
1993H&         0.02513&    17.40& 0.02& 1.56& 0.02& 0.25& 0.02&35.14&-0.03& 0.04&  101.81&  141.08&13 52 50.34&$-30$  42 23.31&   82.81&\\
1993O&         0.05293&    18.96& 0.03& 1.27& 0.03&-0.07& 0.06&36.75&-0.08& 0.05&  582.89&  325.53&13 31  7.88&$-33$  12 50.50&   86.75&\\
1994D&         0.00261&    13.22& 0.05& 1.57& 0.01&-0.13& 0.05&30.92& 0.70& 0.04& -299.14&   18.15&12 34  2.37&$+ 07$  42  4.70&   43.79&\\
1994M&         0.02431&    17.50& 0.17& 1.41& 0.03& 0.01& 0.22&35.26& 0.17& 0.14& -577.43&  513.17&12 31  8.61&$+ 00$  36 19.90&   50.33&\\
1994Q&         0.02987&    17.60& 0.03& 0.96& 0.04& 0.06& 0.05&35.50&-0.05& 0.03&  223.04&  114.42&16 49 51.06&$+ 40$  25 55.53&    0.73&\\
1994S&         0.01611&    16.21& 0.01& 0.88& 0.07&-0.11& 0.02&34.12&-0.07& 0.03&  147.84&   61.47&12 31 21.82&$+ 29$  08  4.19&   22.88&\\
1995ak&         0.02198&    17.18& 0.02& 1.48& 0.05& 0.04& 0.04&34.92& 0.05& 0.04& -148.92&  111.91&02 45 48.83&$+ 03$  13 50.10&  -52.69&\\
1995al&         0.00588&    14.32& 0.04& 0.96& 0.01& 0.10& 0.04&32.22& 0.24& 0.04& -203.81&   36.20&09 50 55.97&$+ 33$  33  9.41&    3.05&\\
1995bd&         0.01443&    16.15& 0.05& 0.73& 0.03& 0.12& 0.06&34.14& 0.19& 0.30& -396.86&  656.67&04 45 21.24&$+ 11$  04  2.50&  -39.72&\\
1995D&         0.00766&    14.61& 0.02& 0.97& 0.01&-0.08& 0.03&32.48&-0.08& 0.04&   83.27&   39.81&09 40 54.79&$+ 05$  08 26.60&   21.94&\\
1996bl&         0.03485&    17.93& 0.02& 0.84& 0.04&-0.03& 0.03&35.87&-0.03& 0.07&  147.37&  320.88&00 36 17.97&$+ 11$  23 40.50&  -61.96&\\
1996bo&         0.01632&    16.34& 0.02& 1.16& 0.03& 0.33& 0.02&34.23& 0.01& 0.05&  -19.10&  118.05&01 48 22.86&$+ 11$  31 15.10&  -64.80&\\
1996C&         0.03007&    17.85& 0.04& 0.94& 0.04&-0.02& 0.05&35.75& 0.18& 0.04& -797.53&  178.15&13 50 48.59&$+ 49$  19  7.09&    4.13&\\
1996X&         0.00800&    14.26& 0.02& 1.28& 0.01&-0.04& 0.03&32.06&-0.60& 0.05&  579.17&   39.65&13 18  1.12&$-26$  50 45.31&   79.94&\\
1996Z&         0.00974&    14.90& 0.01& 1.05& 0.04& 0.28& 0.03&32.92&-0.17& 0.06&  218.22&   72.94&09 36 44.82&$-21$  08 51.70&   37.05&\\
1997bp&         0.00944&    14.64& 0.04& 1.19& 0.01& 0.25& 0.02&32.45&-0.57& 0.04&  656.95&   38.63&12 46 53.75&$-11$  38 33.20&   63.18&\\
1997bq&         0.00960&    15.28& 0.02& 1.16& 0.02& 0.18& 0.02&33.12& 0.07& 0.02&  -87.73&   29.02&10 17  5.33&$+ 73$  23  2.12&  -24.69&\\
1997cw&         0.01595&    16.49& 0.03& 0.99& 0.03& 0.31& 0.05&34.33& 0.16& 0.05& -377.28&  120.63&00 25 17.27&$+ 12$  53  6.20&  -62.06&\\
1997E&         0.01333&    16.23& 0.01& 1.38& 0.03& 0.04& 0.03&34.00& 0.22& 0.08& -428.68&  157.00&06 47 38.10&$+ 74$  29 51.00&  -37.88&\\
1998aq&         0.00426&    13.85& 0.03& 1.17& 0.01&-0.16& 0.03&31.66& 0.37& 0.03& -239.64&   20.33&11 56 25.87&$+ 55$  07 43.19&   -3.85&\\
1998bu&         0.00416&    12.85& 0.02& 1.01& 0.01& 0.24& 0.02&30.74&-0.50& 0.02&  254.75&   10.41&10 46 46.01&$+ 11$  50  7.50&   27.98&\\
1998dk&         0.01204&    15.86& 0.00& 1.30& 0.07& 0.11& 0.20&33.61& 0.06& 0.07& -106.50&  112.57&00 14 32.16&$ 00$  44 10.90&  -48.68&\\
1998ec&         0.02012&    17.08& 0.02& 1.05& 0.03& 0.21& 0.03&34.97& 0.29& 0.05& -874.46&  173.10&06 53  6.11&$+ 50$  02 22.09&  -32.88&\\
1998es&         0.00957&    15.10& 0.03& 0.74& 0.01&-0.04& 0.09&33.07& 0.02& 0.04&  -30.06&   52.77&01 37 17.60&$+ 05$  52 50.60&  -59.40&\\
1998V&         0.01717&    16.65& 0.25& 1.10& 0.02&-0.12& 0.30&34.39& 0.06& 0.14& -139.19&  335.96&18 22 37.41&$+ 15$  42  8.41&    4.04&\\
1999aa&         0.01525&    16.16& 0.01& 0.81& 0.02&-0.13& 0.02&34.12& 0.05& 0.03& -115.88&   61.86&08 27 42.03&$+ 21$  29 14.80&   -2.17&\\
1999ac&         0.00982&    15.05& 0.00& 1.23& 0.03& 0.09& 0.01&32.92&-0.19& 0.03&  241.03&   37.28&16 07 15.02&$+ 07$  58 20.40&   32.58&\\
1999aw&         0.03924&    18.24& 0.02& 0.78& 0.02&-0.20& 0.05&36.24& 0.07& 0.05& -404.16&  269.47&11 01 36.37&$-06$  06 31.60&   44.16&\\
1999dk&         0.01395&    15.88& 0.03& 1.00& 0.03& 0.06& 0.03&33.78&-0.10& 0.04&  184.53&   79.48&01 31 26.92&$+ 14$  17  5.70&  -67.83&\\
1999dq&         0.01308&    15.47& 0.09& 0.94& 0.03& 0.06& 0.06&33.38&-0.35& 0.10&  589.50&  157.41&02 33 59.68&$+ 20$  58 30.41&  -69.48&\\
1999ee&         0.01055&    15.65& 0.02& 0.90& 0.01& 0.19& 0.01&33.59& 0.33& 0.02& -513.26&   32.23&22 16 10.00&$-36$  50 39.69&   -3.79&\\
1999ek&         0.01760&    16.58& 0.04& 1.08& 0.01& 0.09& 0.06&34.45& 0.07& 0.34& -161.92&  854.34&05 36 31.60&$+ 16$  38 17.80&  -32.86&\\
1999gh&         0.00879&    14.92& 0.02& 1.67& 0.03& 0.39& 0.04&32.64&-0.23& 0.04&  263.87&   41.83&09 44 19.75&$-21$  16 25.00&   38.64&\\
2000ca&         0.02452&    17.28& 0.05& 0.93& 0.01&-0.20& 0.05&35.07&-0.05& 0.04&  167.06&  145.52&13 35 22.98&$-34$  09 37.00&   87.61&\\
2000cf&         0.03603&    18.33& 0.03& 1.31& 0.04&-0.02& 0.04&36.12& 0.14& 0.04& -737.69&  192.47&15 52 56.19&$+ 65$  56 12.69&  -15.94&\\
2000cn&         0.02321&    17.32& 0.04& 1.58& 0.02& 0.24& 0.09&35.04& 0.04& 0.12& -143.94&  378.10&17 57 40.42&$+ 27$  49 58.09&    0.43&\\
2000dk&         0.01645&    16.42& 0.04& 1.43& 0.03& 0.07& 0.06&34.18&-0.06& 0.06&  129.33&  131.96&01 07 23.52&$+ 32$  24 23.19&  -83.49&\\
2000E&         0.00422&    13.88& 0.00& 1.03& 0.02& 0.06& 0.04&31.74& 0.47& 0.22& -309.42&  160.15&20 37 13.77&$+ 66$  05 50.19&  -39.39&\\
2000fa&         0.02179&    17.28& 0.02& 1.09& 0.02&-0.12& 0.03&35.04& 0.18& 0.05& -565.02&  148.52&07 15 29.88&$+ 23$  25 42.41&  -16.84&\\
2001ba&         0.03053&    17.66& 0.03& 1.06& 0.01&-0.13& 0.05&35.51&-0.10& 0.05&  398.44&  218.97&11 38  1.76&$-32$  19 51.00&   66.20&\\
2001bt&         0.01445&    16.01& 0.02& 1.29& 0.01& 0.23& 0.02&33.83&-0.13& 0.04&  243.21&   80.00&19 13 46.75&$-59$  17 22.81&   32.94&\\
2001el&         0.00365&    13.62& 0.01& 1.16& 0.00& 0.18& 0.01&31.48& 0.54& 0.01& -308.31&    6.27&03 44 30.57&$-44$  38 23.69&   -3.55&\\
2001V&         0.01604&    15.77& 0.06& 0.73& 0.03& 0.02& 0.08&33.73&-0.45& 0.07&  906.79&  126.76&11 57 24.93&$+ 25$  12  9.00&   24.42&\\
2002bo&         0.00529&    14.41& 0.08& 1.24& 0.01& 0.40& 0.05&32.18& 0.42& 0.05& -340.00&   47.96&10 18  6.51&$+ 21$  49 41.70&   15.89&\\
2002el&         0.02238&    17.38& 0.02& 1.37& 0.02&-0.01& 0.03&35.15& 0.24& 0.05& -778.70&  187.50&20 56 30.09&$-18$  33 34.30&   -4.89&\\
2002er&         0.00855&    15.06& 0.04& 1.26& 0.01& 0.20& 0.03&32.87& 0.07& 0.10&  -84.75&  120.45&17 11 29.88&$+ 07$  59 44.80&   22.08&\\
2003du&         0.00665&    14.97& 0.02& 1.15& 0.03&-0.13& 0.03&32.80& 0.55& 0.02& -575.72&   26.64&14 34 35.80&$+ 59$  20  3.81&   -6.66&\\
2004eo&         0.01473&    16.06& 0.02& 1.33& 0.00& 0.10& 0.02&33.86&-0.14& 0.07&  268.80&  128.20&20 32 54.19&$+ 09$  55 42.70&  -18.07&\\
2004S&         0.00986&    15.27& 0.01& 1.18& 0.01& 0.08& 0.02&33.12& 0.00& 0.06&   -1.81&   84.18&06 45 43.50&$-31$  13 52.50&    9.78&\\
2005am&         0.00897&    14.62& 0.01& 1.54& 0.02& 0.14& 0.01&32.35&-0.55& 0.03&  606.13&   32.94&09 16 12.47&$-16$  18 16.00&   30.36&\\
2005cf&         0.00704&    14.46& 0.01& 1.14& 0.00&-0.03& 0.02&32.30&-0.08& 0.06&   75.45&   56.78&15 21 32.20&$-07$  24 47.50&   51.73&\\
\enddata
\label{Tab:cmagic}
\end{deluxetable}